\documentclass[onecolumn,a4]{aa}
\usepackage{amsmath}
\usepackage{amssymb}
\usepackage{graphics}
\usepackage{hyperref}
\usepackage[varg]{txfonts}
\usepackage[utf8]{inputenc}

\begin{document}

\title{Star formation activity in Balmer break galaxies at $z$ $<$ 1.5}

\author{J. Díaz Tello \inst{1,6}\and C. Donzelli \inst{1}\and N. Padilla \inst{2}\and M. Akiyama \inst{4}\and N. Fujishiro \inst{3}\and T. Yoshikawa \inst{3}\and H. Hanami \inst{5}
}

\institute{IATE, Observatorio Astronómico de Córdoba, Universidad Nacional de Córdoba, Argentina.\\ \email{jdiazt@oac.uncor.edu}\and Centro de Astro-Ingeniería, Pontificia Universidad Católica de Chile, Chile.\and Koyama Astronomical Observatory, Kyoto Sangyo University, Japan.\and Astronomical Institute, Graduate School of Science, Tohoku University, Japan.\and Physics Section, Iwate University, Japan.\and Instituto de Astronomía-Ensenada, Universidad Nacional Autónoma de México, México
}

\abstract{}{
We present a spectroscopic study of the properties of 64 Balmer break galaxies that show signs of star formation. The studied sample of star-forming galaxies spans a redshift range from 0.094 to 1.475 with stellar masses in the range 10$^{8}-$10$^{12}$ $M_{\odot}$. The sample also includes eight broad emission line galaxies with redshifts between 1.5 $<z<$ 3.0. 
}{
We derived star formation rates (SFRs) from emission line luminosities and investigated the dependence of the SFR and specific SFR (SSFR) on the stellar mass and color. Furthermore, we investigated the evolution of these relations with the redshift.
}{
We found that the SFR correlates with the stellar mass; our data is consistent with previous results from other authors in that there is a break in the correlation, which reveals the presence of massive galaxies with lower SFR values (i.e., decreasing star formation). We also note an anticorrelation for the SSFR with the stellar mass. Again in this case, our data is also consistent with a break in the correlation, revealing the presence of massive star-forming galaxies with lower SSFR values, thereby increasing the anticorrelation. These results might suggest a characteristic mass ($M_{0}$) at which the red sequence could mostly be assembled. In addition, at a given stellar mass, high-redshift galaxies have on average higher SFR and SSFR values than local galaxies. Finally, we explored whether a similar trend could be observed with redshift in the SSFR$-(u-B)$ color diagram, and we hypothesize that a possible $(u-B)_{0}$ break color  may define a characteristic color for the formation of the red sequence.
}{}


\titlerunning{SF in BBGs at $z$ $<$ 1.5}

\maketitle
\section{Introduction}

A primary aim of studies on the formation and evolution of galaxies is to resolve how the stellar content of the present day galaxies became assembled over cosmic time, with the observed basic structural properties of local elliptical galaxies of different masses suggesting that a scenario of monolithic collapse could have formed these systems (\citeauthor{Chiosi,Romano,Kaviraj}). In this scenario, massive elliptical galaxies could have been the result of a violent burst of star formation at high redshift followed by passive evolution over time.  However, in the accepted hierarchical scenario, massive galaxies are assembled by mergers of lower mass units and gas accretion (\citeauthor{Davis,White}). The critical aspect in all these models is how and when the most massive present day galaxies were built and what type of evolution they had.

It has been shown that the global star formation rate (SFR) declines from z $\sim$ 1.5 (\citeauthor{Cowie,Lilly,Madau}). The inspection of the SFR and specific SFR (SSFR) in individual galaxies has revealed that the star formation migrates from massive systems at high redshift to low-mass systems at low redshift, an effect mentioned by Cowie et al. as downsizing, which has been confirmed by optic (\citeauthor{Bundy}), infrared (\citeauthor{HopkinsA}) and radio (\citeauthor{Perez}) observations. Moreover, it has been consistently shown that the SFR depends strongly on both the stellar mass and  redshift (\citeauthor{Brinchman2,Bauer,Diaz}).  

A fundamental problem is to understand better the link between the global decline of the SFR and the evolution of the mass assembly. While the $\Lambda$-CDM hierarchical model describes the dynamics of the Universe and the building in mass of galaxies, the quenching of star formation in massive systems is less well understood. Several mechanisms have been proposed to regulate star formation, such as variations in the gas accretion history (\citeauthor{Dutton}), strong and ubiquitous galactic outflows (\citeauthor{Dave}), and feedback processes on galactic scales, such as active galactic nuclei (AGNs) (\citeauthor{Bower}) and supernovas (\citeauthor{Cole}), among others.

It has become clear that there is a bimodal distribution of galaxy properties observed up to $z\sim$1$-$2 with a narrow red sequence dominated by passive galaxies and a blue cloud of star-forming galaxies (\citeauthor{Strateva,Kauffman,Driver}). This bimodality is also seen in the stellar mass function (SMF) of star-forming and passive galaxies (\citeauthor{Borch,Franceschini,Bundy}). Interestingly, the SMFs differ strongly from each other at all redshifts probed;  passive galaxies dominate the massive tail while star-forming galaxies dominate the opposite tail end. Furthermore, a downward evolution has been found in the mass value at which the populations intersect (the transition mass $M_{tr}$, \citeauthor{Bundy}), thus revealing signs of downsizing in the mean stellar mass of passive galaxies. Hopkins et al \citeyearpar{Hopkins} found that the transition mass values derived from quasar and merger luminosity functions also follow a similar trend, perhaps reflecting the characteristic mass on which the red sequence population is mostly assembled at a given redshift. This mass also increases with redshift and thus suggests that cosmic downsizing may apply to star formation as well as to red galaxy assembly.

Observing the properties of galaxies at different redshifts (i.e., epochs) allows us to study  their evolution as a function of time directly. In this study, we investigated the physical properties of a sample of 72 emission line Balmer break galaxies at redshift 0 $<z<$ 3 for which multiwavelength photometric and optical spectra were available. Balmer break galaxies are galaxies at high redshift that are selected using the position of the Balmer limit. It is known that these galaxies can be selected using a simple color-color band photometry diagram (\citeauthor{Daddi}). Furthermore, to achieve better statistic results in our analysis, we extended the sample by utilizing the sample galaxies of Díaz Tello et al. (\citeyear{Diaz}; 37 galaxies). This galaxy sample is part of a pilot project of a spectroscopic survey of galaxies in the Subaru XMM Deep Field (SXDF) up to z $\sim$ 3.0 that investigates the end of star formation in massive galaxies.

This paper is organized as follows: Section \ref{sample} describes the criteria used to select the sample; section \ref{obsred} mentions the spectrophotometric observations and explains the reduction process and  methodology used to analyze the data;  section \ref{phyprop} summarizes the final galaxy sample used in this work and gives details about its physical properties. In section \ref{nuc} we explore the presence of AGN activity  using the available methods for detecting this at $z>$ 1.0, and in section \ref{starfor}, we derive SFRs and study the evolution of the SFR$-$stellar mass, SSFR$-$stellar mass and SSFR$-$color relations in the redshift range 0 $<z<$ 1.5. Finally, in section \ref{Sum}, we summarize our results and conclusions.

Throughout this paper we have assumed a flat $\Lambda-$do\-mi\-na\-ted cosmology with $\Omega_{m}$= 0.28, $\Omega_{\Lambda}$= 0.72, and 
H$_{0}$= 70 km s$^{-1}$ Mpc.

\section{Sample selection}\label{sample}

The galaxies presented in this article come from a sample constructed to study star formation activity of massive galaxies in the redshift range $z$ = 0.1$-$3.0 (\citeauthor{Diaz}; hereafter D13). The chosen field, the SXDF (R.A. = 02:18:00, decl. = -05:00:00; \citeauthor{Furusawa}) has the advantage that deep photometric data is available in the bands \textit{u, B, R, i, z} (Subaru), \textit{J, H, K} (UKIRT), and in all MIR \textit{Spitzer} bands. The parent sample was selected using the $\lambda$3646 Balmer and $\lambda$4000 break features as tracers of redshift, as described by Daddi et al. \citeyearpar{Daddi}, utilizing the $BzK$ color$-$color diagram to select star-forming galaxies in a particular redshift range. Furthermore, we used two color$-$color diagrams to select star-forming galaxies in a lower redshift range than the $BzK$ diagram; these diagrams were presented in Hanami et al. (\citeyear{Hanami}; $uVi$ and $uRJ$ diagrams).

\begin{figure}
\begin{center}
\includegraphics[width=0.49\textwidth]{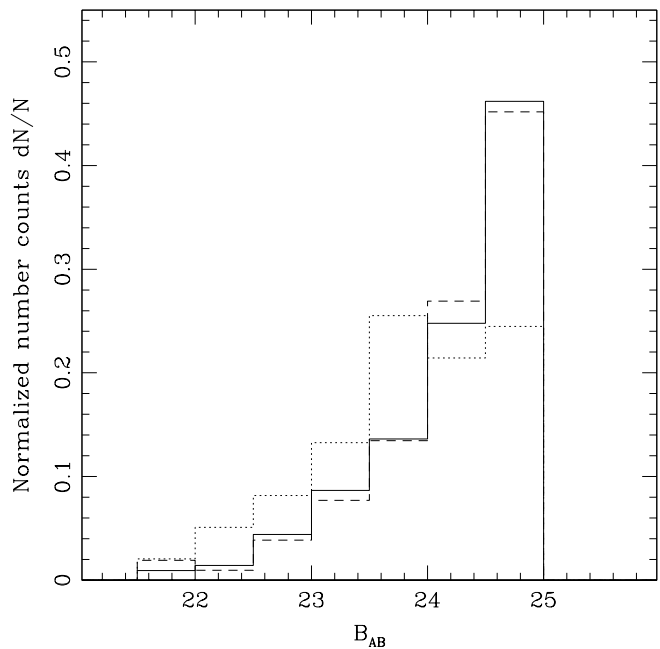}
\caption{Magnitude distribution of objects selected, observed, and presented. The solid line shows the normalized distribution of galaxies photometrically selected in the fields used for spectroscopy; the dashed line shows the distribution of the galaxies targeted for spectroscopy; and the dotted line shows the distribution of the galaxies presented in the paper.}\label{comp}
\end{center}
\end{figure}

A total of 3840 objects were photometrically selected by these color criteria. However, only 417 of these objects ($\sim$ 10 $\%$) could be included randomly in the four observed fields. Unexpectedly, parasite light problems affected both bright and faint galaxy spectra. Thus, only 132 spectra had the necessary signal-to-noise (S/N) ratio for calculating a redshift. Figure \ref{comp} shows the normalized magnitude distribution of objects selected in this research (solid line), comprised of galaxies observed among the selected objects (dashed line) and objects presented in this paper (dotted line).

\defcitealias{Diaz}{D13}

\begin{figure*}
\begin{center}
\includegraphics[width=0.33\textwidth]{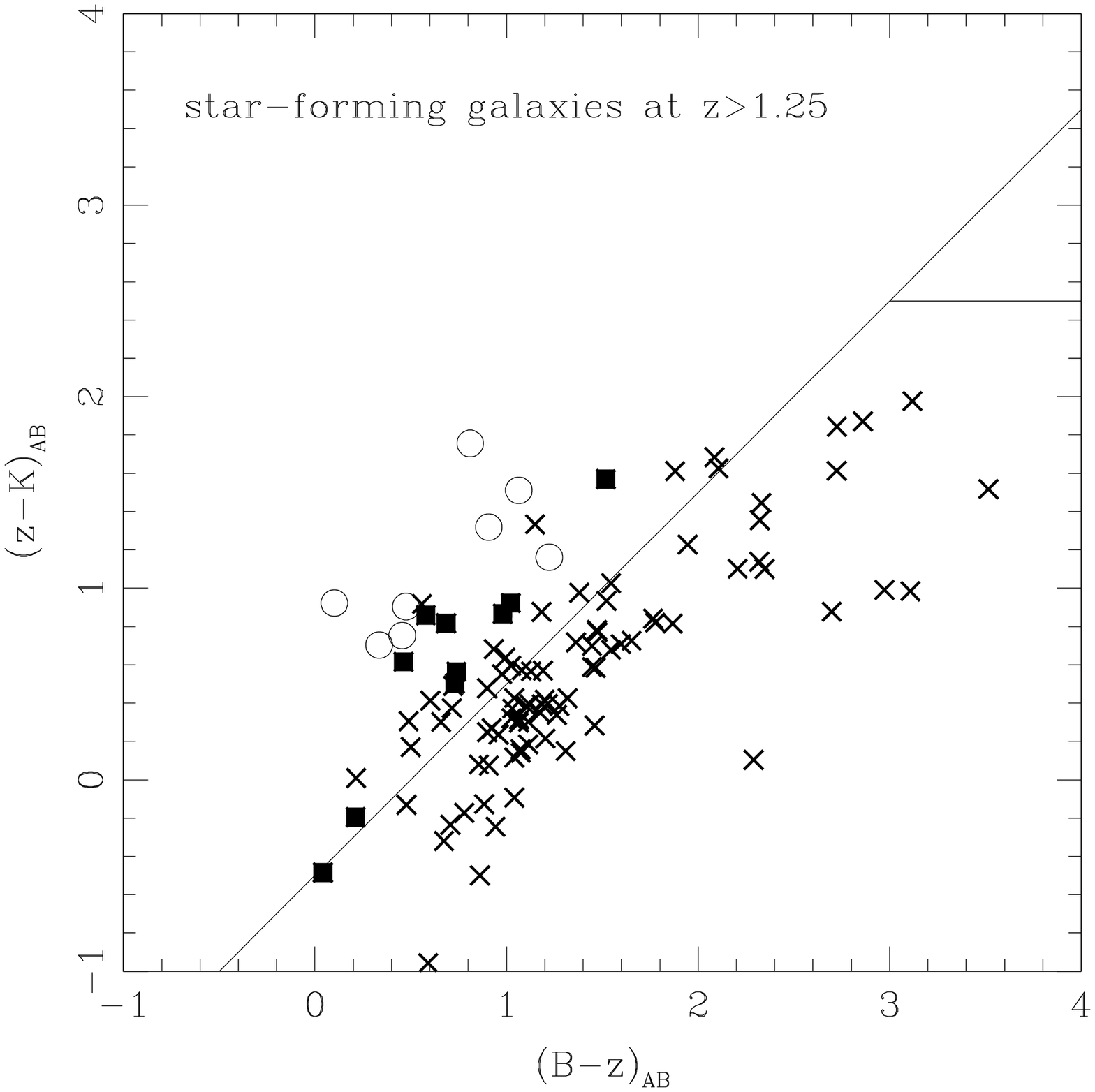}
\includegraphics[width=0.33\textwidth]{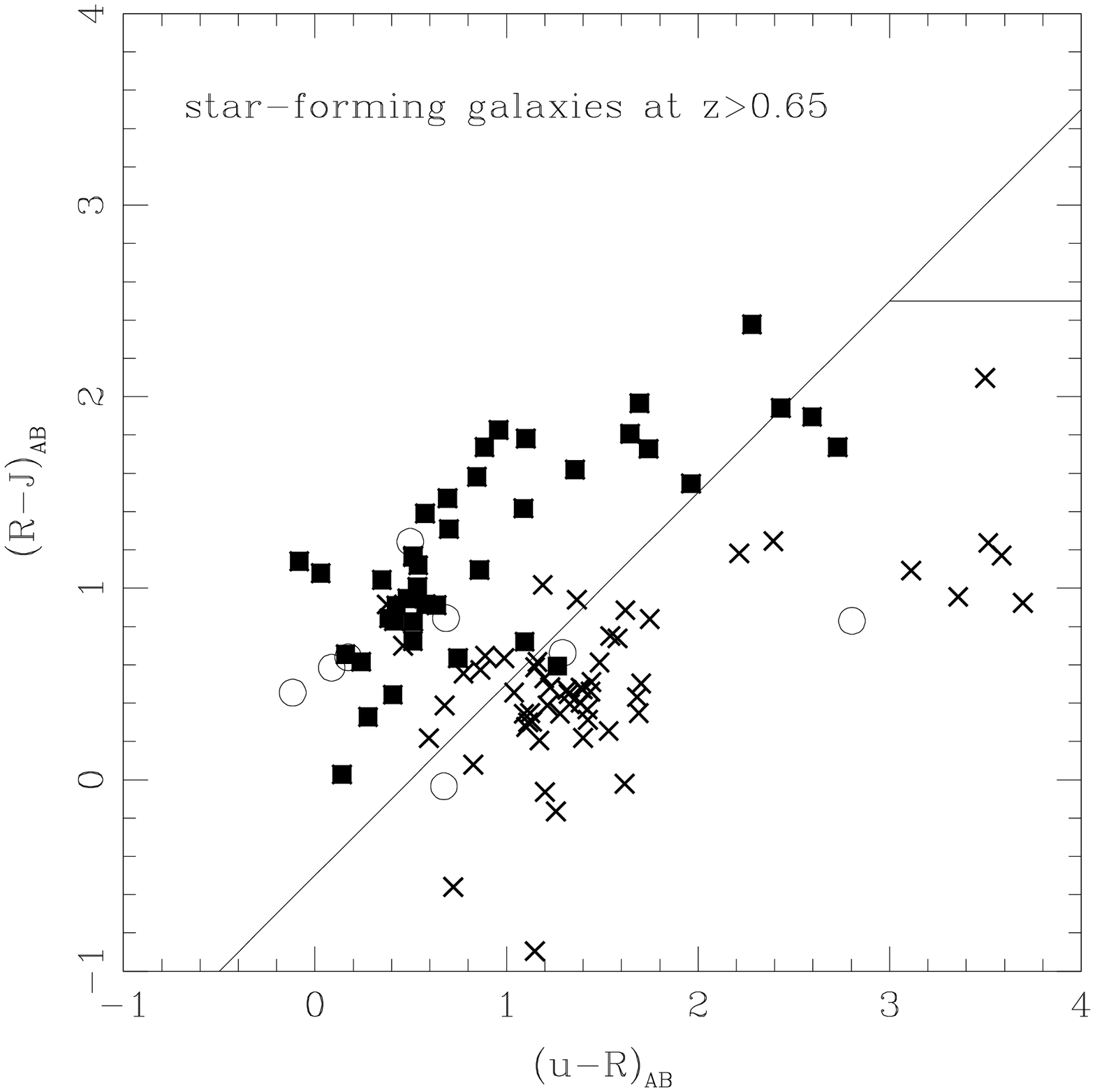}
\includegraphics[width=0.33\textwidth]{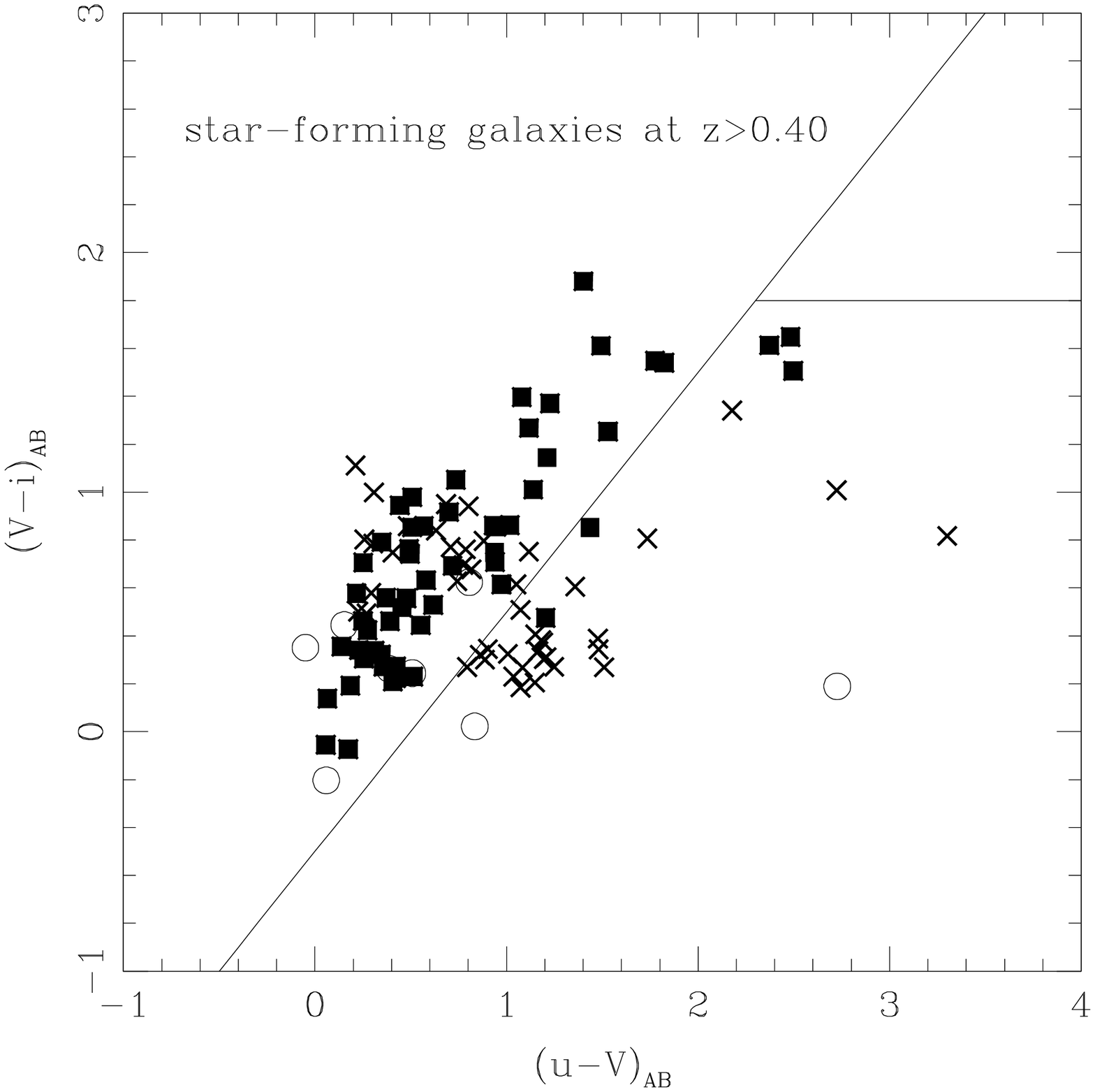}
\caption{Color-color diagrams $BzK$ (left), $uRJ$ (middle) and $uVi$ (right)  used to select our star-forming galaxies. In each panel, the galaxies whose spectroscopic redshifts satisfy the redshift domain of each color criteria utilized (indicated in each panel) are represented by black squares, while the remaining galaxies are represented by crosses. The circles are the galaxies whose spectra showed broad emission line features. These galaxies revealed spectroscopic redshift $z$ $>$ 2.0. The solid lines delimit the selection zone used by each criteria with the smallest region indicating the expected location of the passive galaxies.}\label{colcol}
\end{center}
\end{figure*}

Of these 132 spectra, we found galaxies that show the necessary emission lines to classify them using the BPT diagrams. These galaxies were studied in \citetalias{Diaz}.  We also study those galaxies that show only one of the following emission lines: [OII]$\lambda$3727 or H$\beta$ 4861 $\AA{}$ or H$\alpha$ 6563 $\AA{}$. Additionally, we also selected those galaxies that show broad emission lines in their spectra in order to explore AGN activity. Thus, our final sample is composed of 72 emission line galaxies, of which 64 show one narrow emission lines and eight have broad emission lines. Figure \ref{colcol} shows the color-color diagrams used together with the galaxies we present in this paper. The black squares are the galaxies whose spectroscopic redshifts satisfy the redshift domain of the color criteria used (indicated in each panel), while the crosses are the remaining galaxies with lower redshifts than the criteria used. The circles are galaxies whose spectra reveal broad emission lines. The solid lines delimit the selection zone used by each criteria with the smallest region indicating the expected location of  passive galaxies.

\section{Observations and reductions}\label{obsred}

As described in \citetalias{Diaz}, we obtained $u$-band data and optical spectroscopy for a more accurate study of the physical properties of the galaxies. 
The $u$-band data were obtained during 2006 September 16-17 and 19-20 (PI: N. Padilla) with MosaicII/Blanco Telescope at Cerro Tololo Inter American Observatory (CTIO), while the spectroscopic data were obtained from two set of observations. The first set was obtained during 2007 December 11-12 (PI: N. Padilla) with the Inamori-Magellan Areal Camera and Spectrograph (IMACS)/Magellan Telescope at Las Campanas Observatory, while the second set was observed in service mode during the second semester of 2008 with the Gemini Multi-Object Spectrograph (GMOS)/Gemini South. In both observations, the slit size adopted was 1" aperture, which is consistent with the average seeing of the run. We observed the central region of the SXDF, using a multi-slit mask of 227 slits in IMACS and three multi-slit mask with a total of 190 slits in GMOS. The IMACS spectra had a dispersion of 2.04 $\AA{}$ pixel$^{-1}$, while GMOS spectra had a dispersion of 3.59 $\AA{}$ pixel$^{-1}$. During the spectroscopic observations, standard star spectra were acquired to calibrate by flux the galaxy spectra. We obtained the spectra of Feige110 and Hilt600 stars for the IMACS
observations, while with GMOS only LTT2415 was used, and the same slit width utilized in the multislit observations (1") are adopted. The data reduction process is also described in \citetalias{Diaz}. Here, the more relevant aspects are mentioned.

\subsection{Emission line fluxes}

The flux calibration with the standard star data can recover the flux of the stellar object, but extended objects can be affected by flux loss. Therefore, prior to taking the emission line measurements, 1" aperture photometry was used to correct errors in the spectra due to this effect. The calculated aperture $B$-, $V$-, $R$-, $i$-, $z$-band fluxes were compared with the fluxes from the spectrum. Then, the average of the flux ratio between the aperture photometry and the spectrum, calculated for each filter, was used as a correction factor. Typical values of such corrections were of about 30$\%$, which are essential to obtain consistent results when comparing properties derived from the spectra with those derived from SED fitting (SFRs). Then, the spectrum of the underlying stellar population (Balmer stellar absorption) was subtracted to produce a pure emission line spectrum. Stellar template spectra were obtained from stellar population synthesis models (Section \ref{SED}). Finally, fluxes of the emission lines were measured using the \textit{splot} routine within IRAF. The flux limit in the emission lines was 4$\times$10$^{-18}$ erg s$^{-1}$cm$^{-2}$ at the 3$\sigma$ level, implying an equivalent-width limit of $\sim$4 $\AA{}$ . This flux corresponds, in 1" aperture, to a limit continuum level of $\sim$25 AB mag (fig. \ref{seds}). In H$\alpha$ or in [OII] (without accounting for dust extinction or Balmer absorption corrections), this value corresponded to a limit in the SFR of $\sim$0.008 $M_{\odot}$ yr$^{-1}$ at $z$=0.1 and of $\sim$0.5 $M_{\odot}$ yr$^{-1}$ for the most distant galaxy of our sample at $z$=1.50. A more detailed analysis of the observational limit to the SFR of our survey is discussed in section \ref{sfrevol}.

\begin{figure*}
\begin{center}
\includegraphics[width=0.32\textwidth]{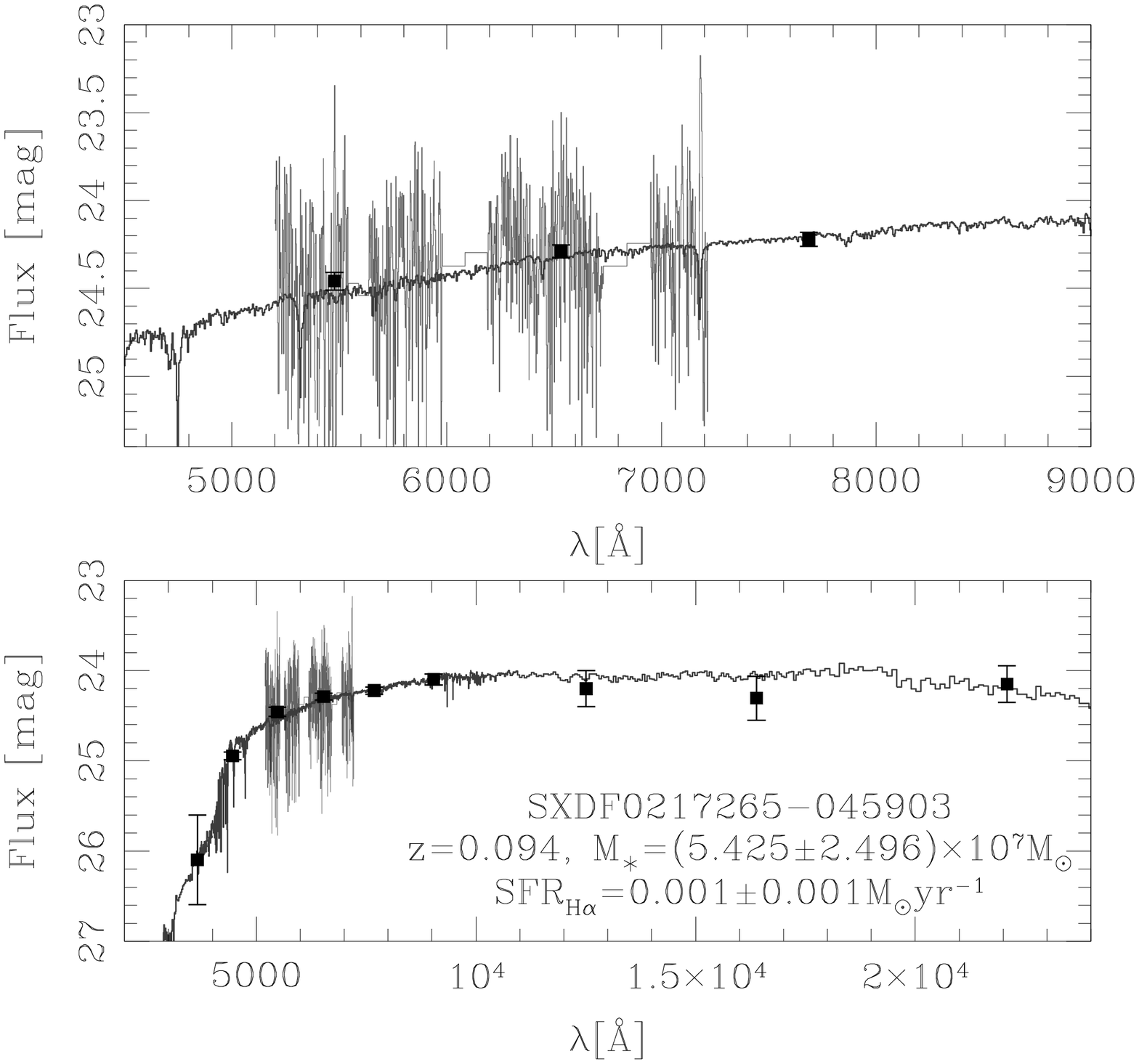}
\includegraphics[width=0.32\textwidth]{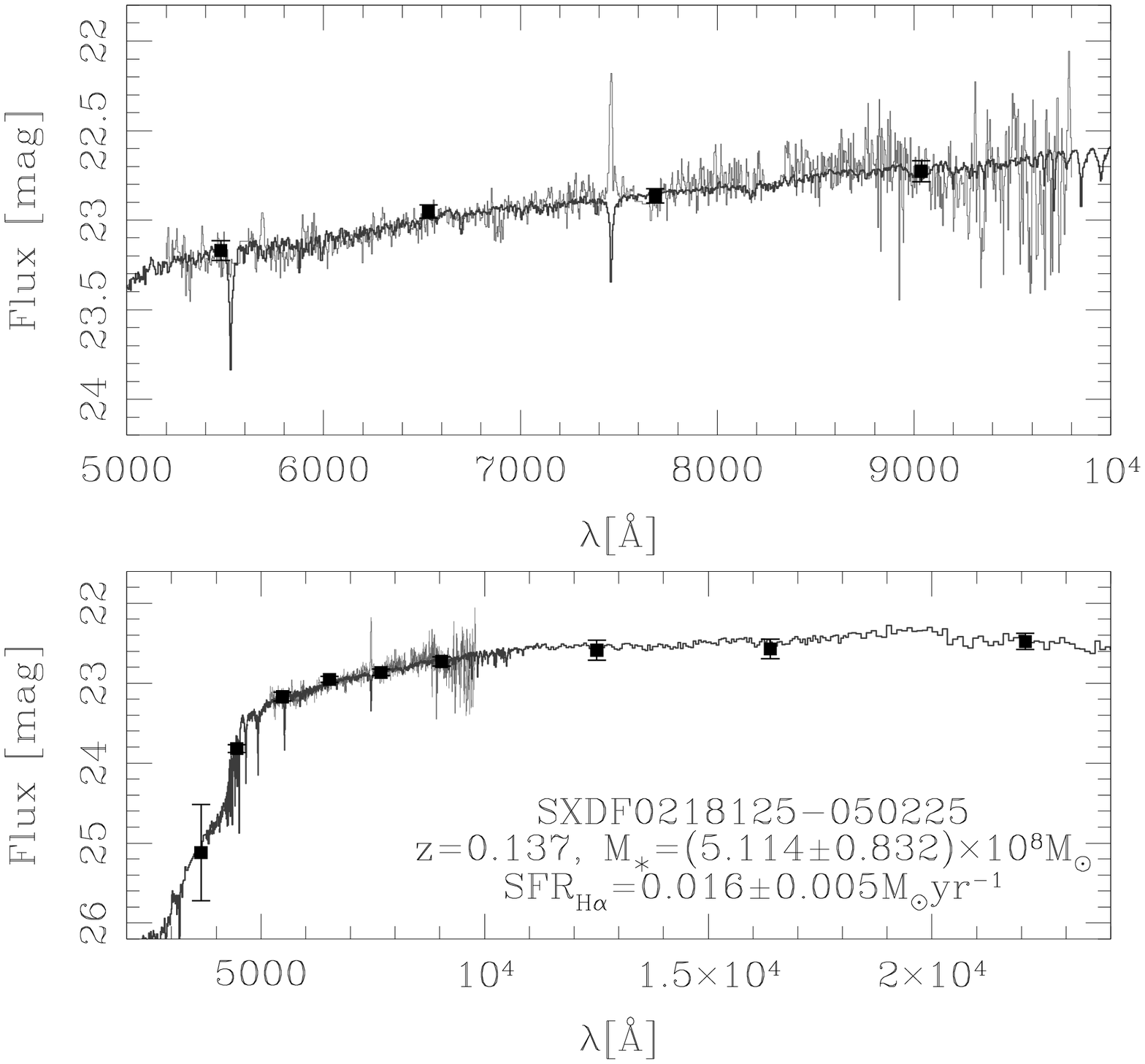}
\includegraphics[width=0.32\textwidth]{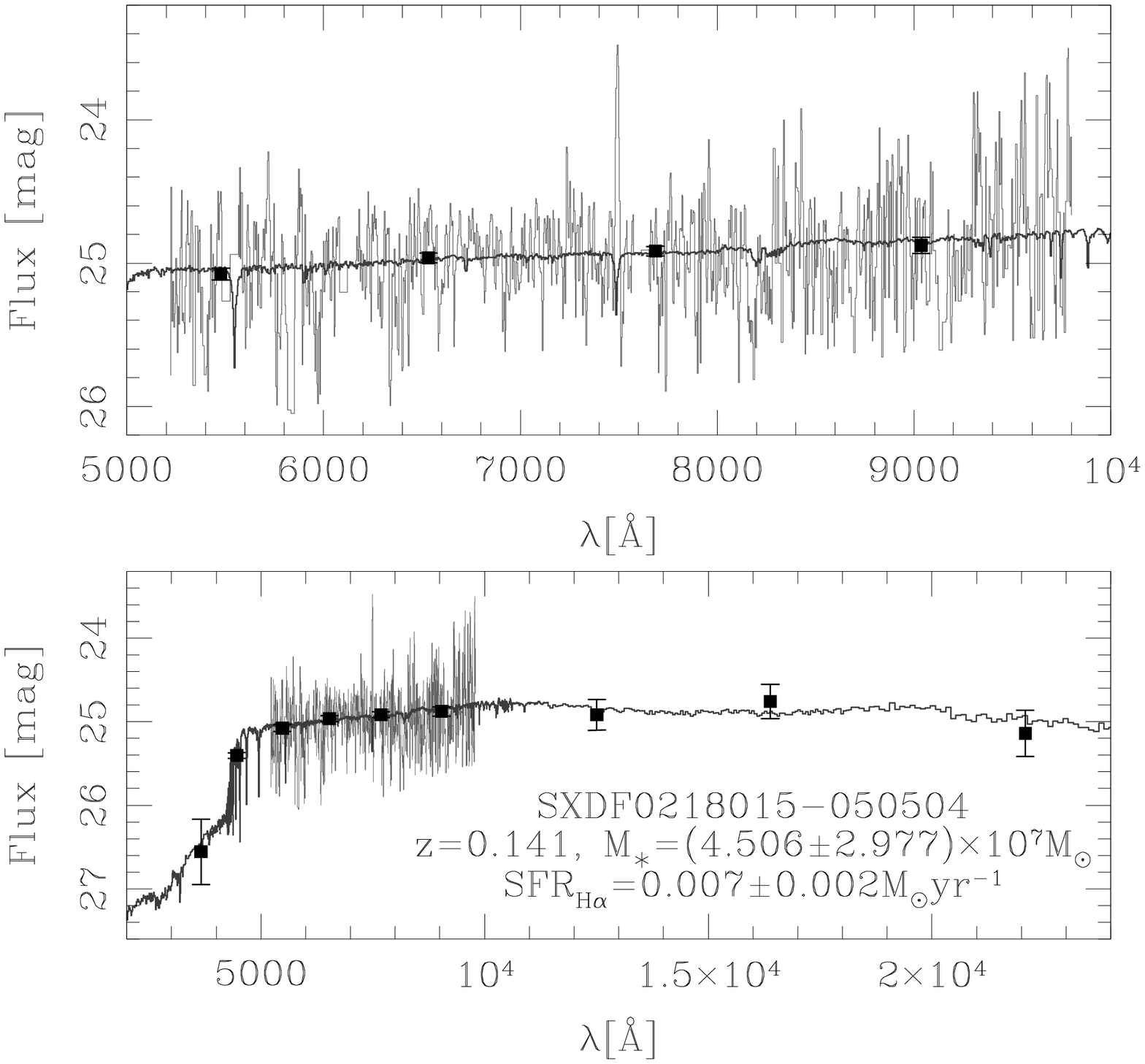}
\includegraphics[width=0.32\textwidth]{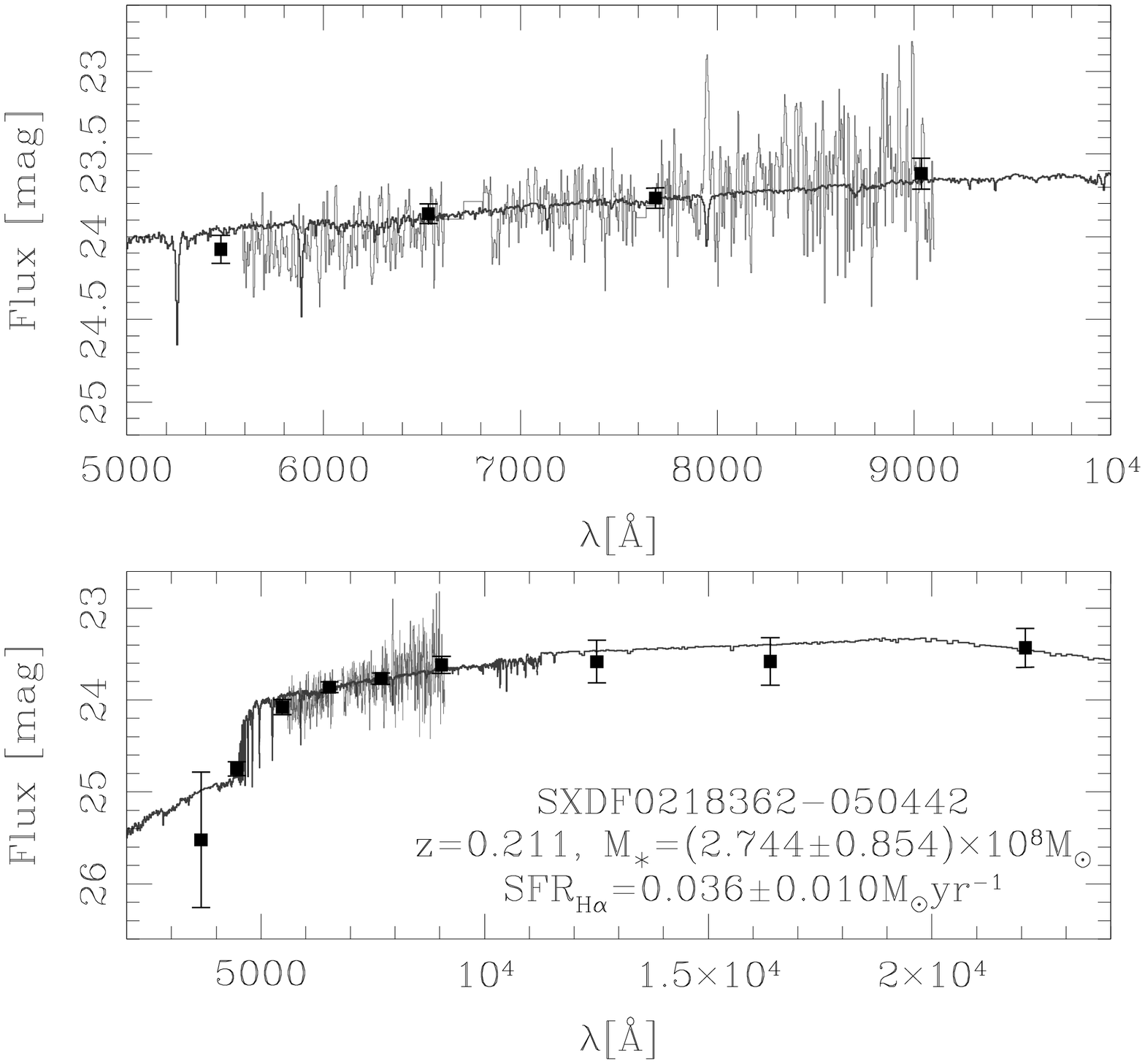}
\includegraphics[width=0.32\textwidth]{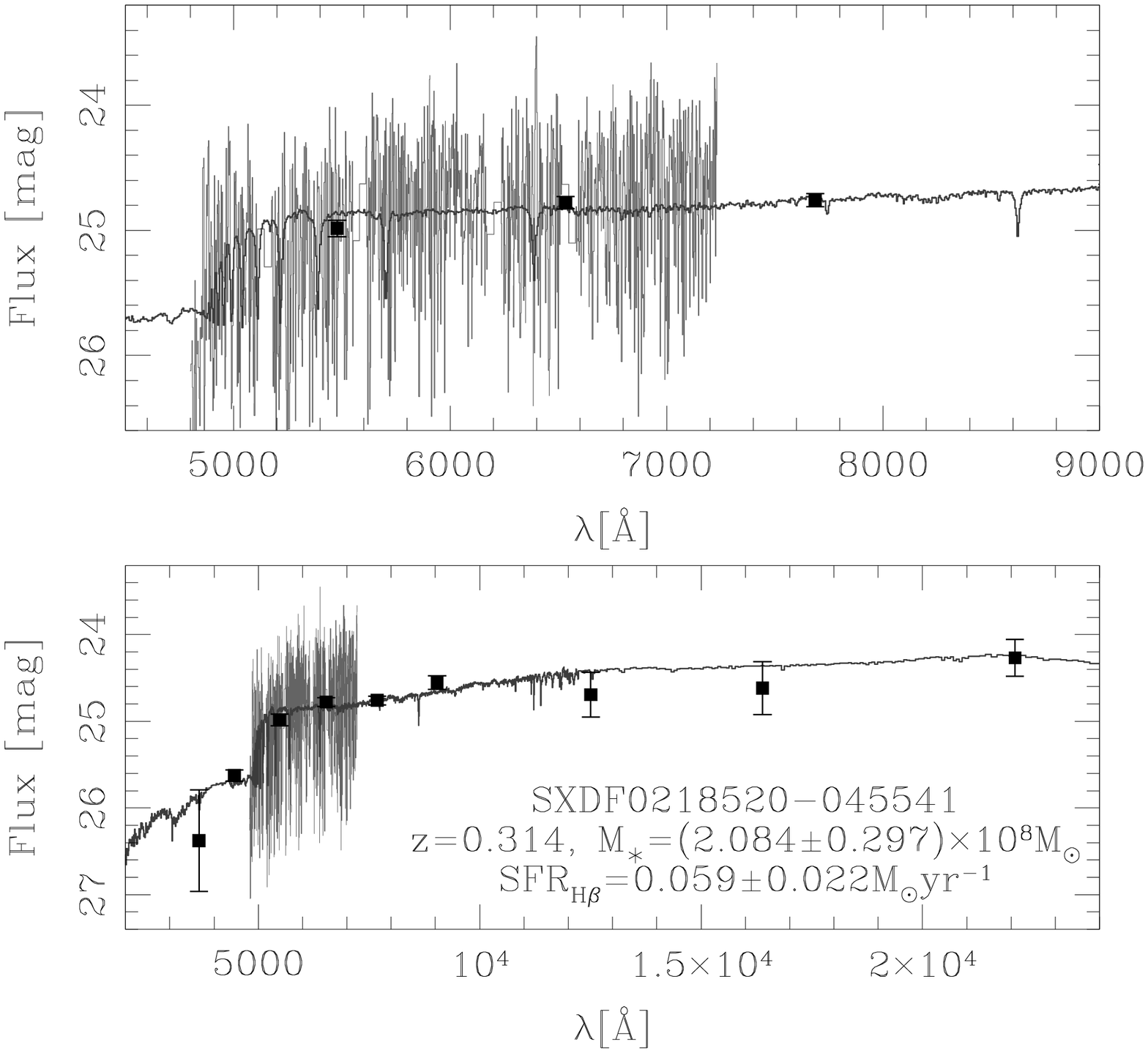}
\includegraphics[width=0.32\textwidth]{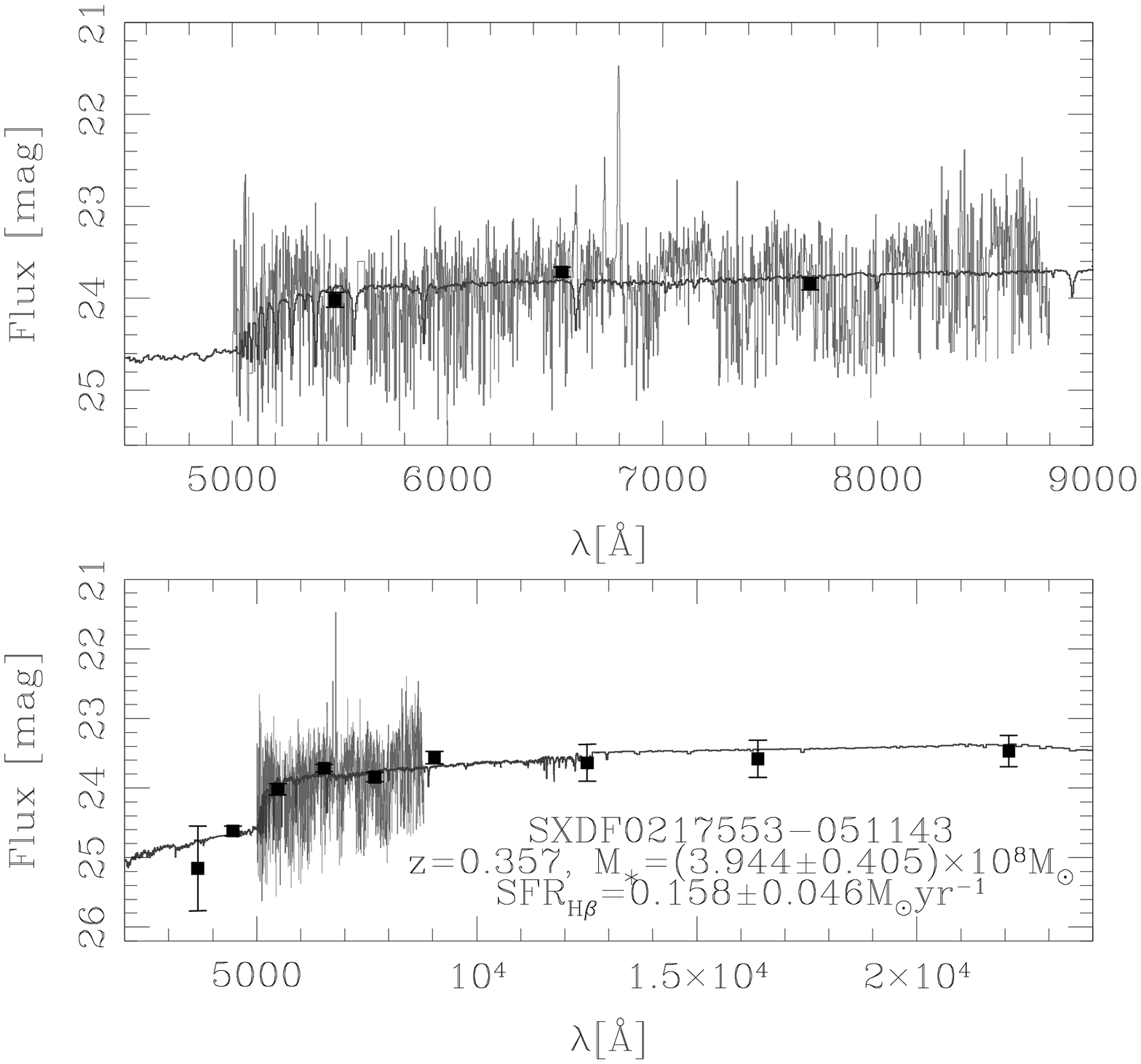}
\includegraphics[width=0.32\textwidth]{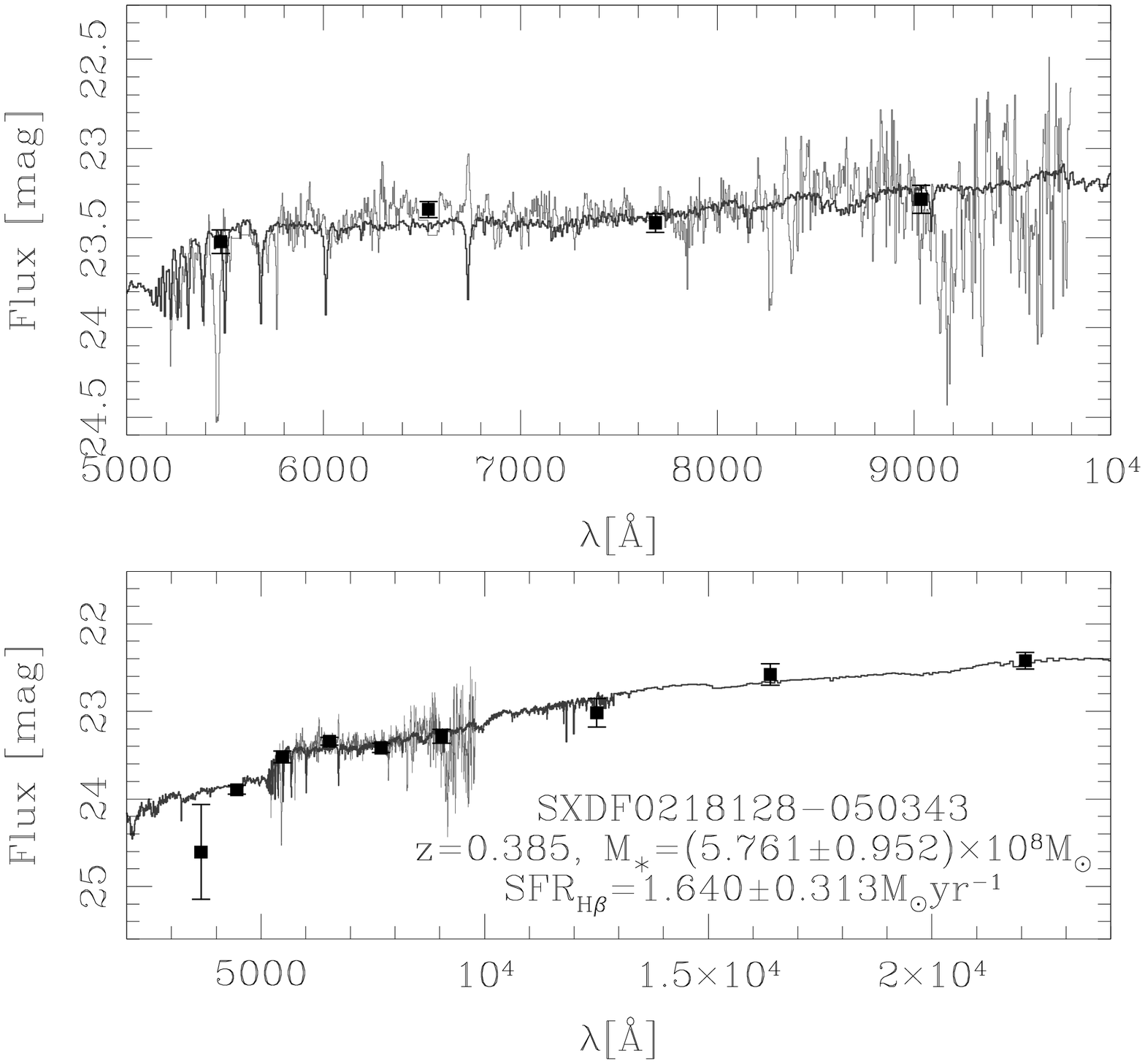}
\includegraphics[width=0.32\textwidth]{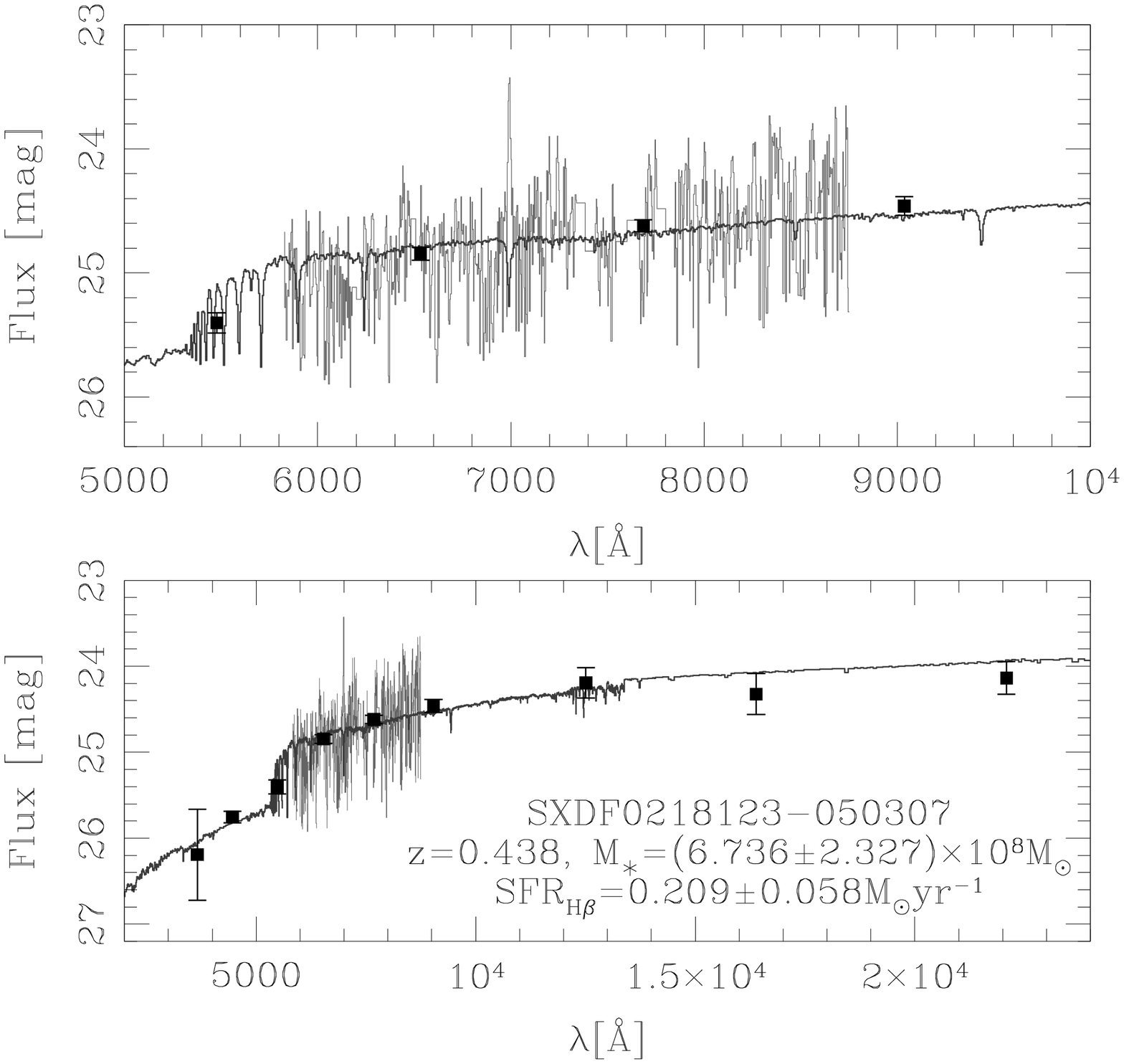}
\includegraphics[width=0.32\textwidth]{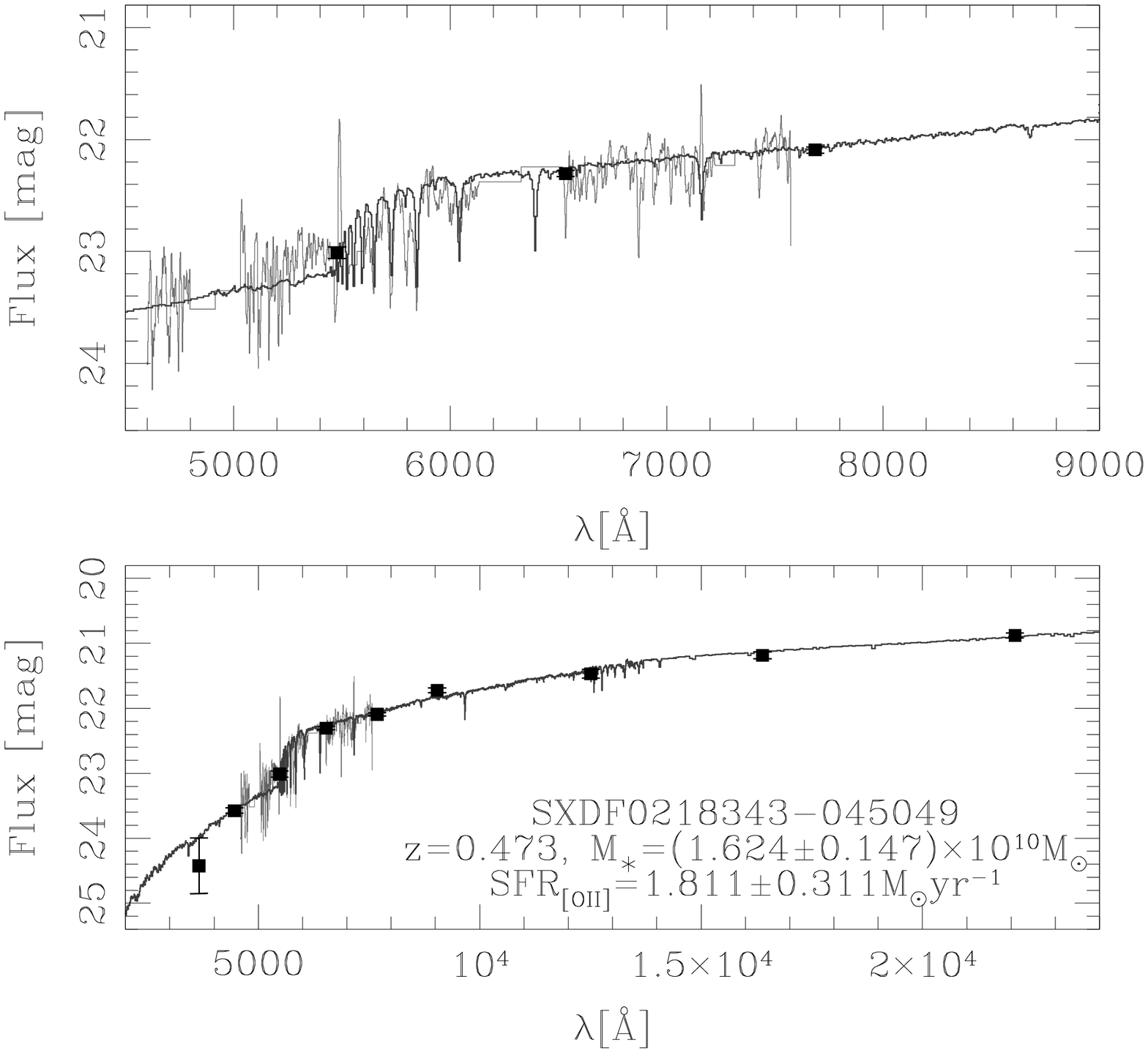}
\caption{Observed galaxy spectrum (light gray) together with aperture photometry (black squares) and SED models (dark gray). The upper panel shows the wavelength range of the observed spectrum, while the lower panel shows the full wavelength range of available photometric data.}\label{seds}
\end{center}
\end{figure*}

\begin{figure*}
\addtocounter{figure}{-1}
\begin{center}
\includegraphics[width=0.32\textwidth]{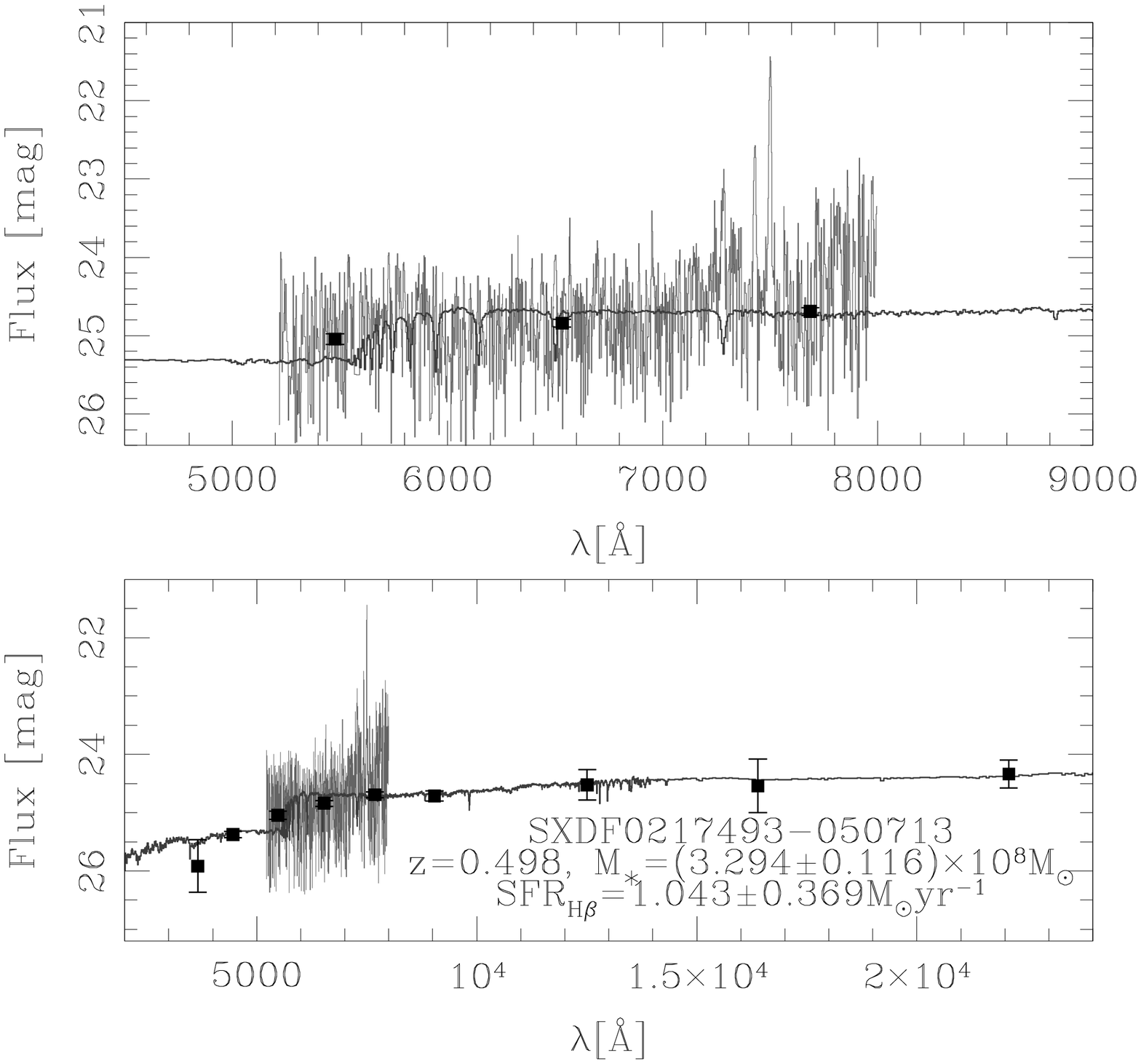}
\includegraphics[width=0.32\textwidth]{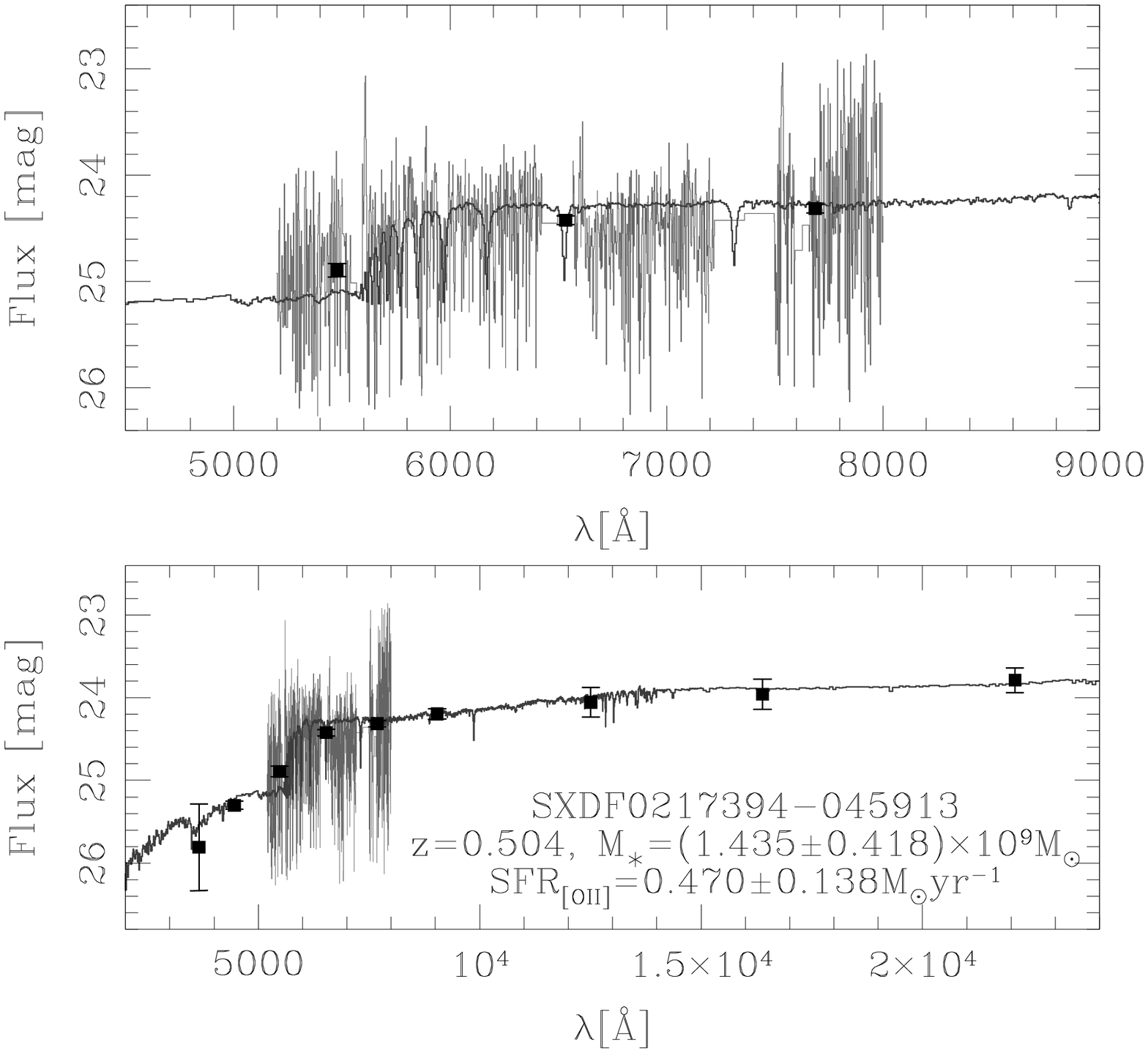}
\includegraphics[width=0.32\textwidth]{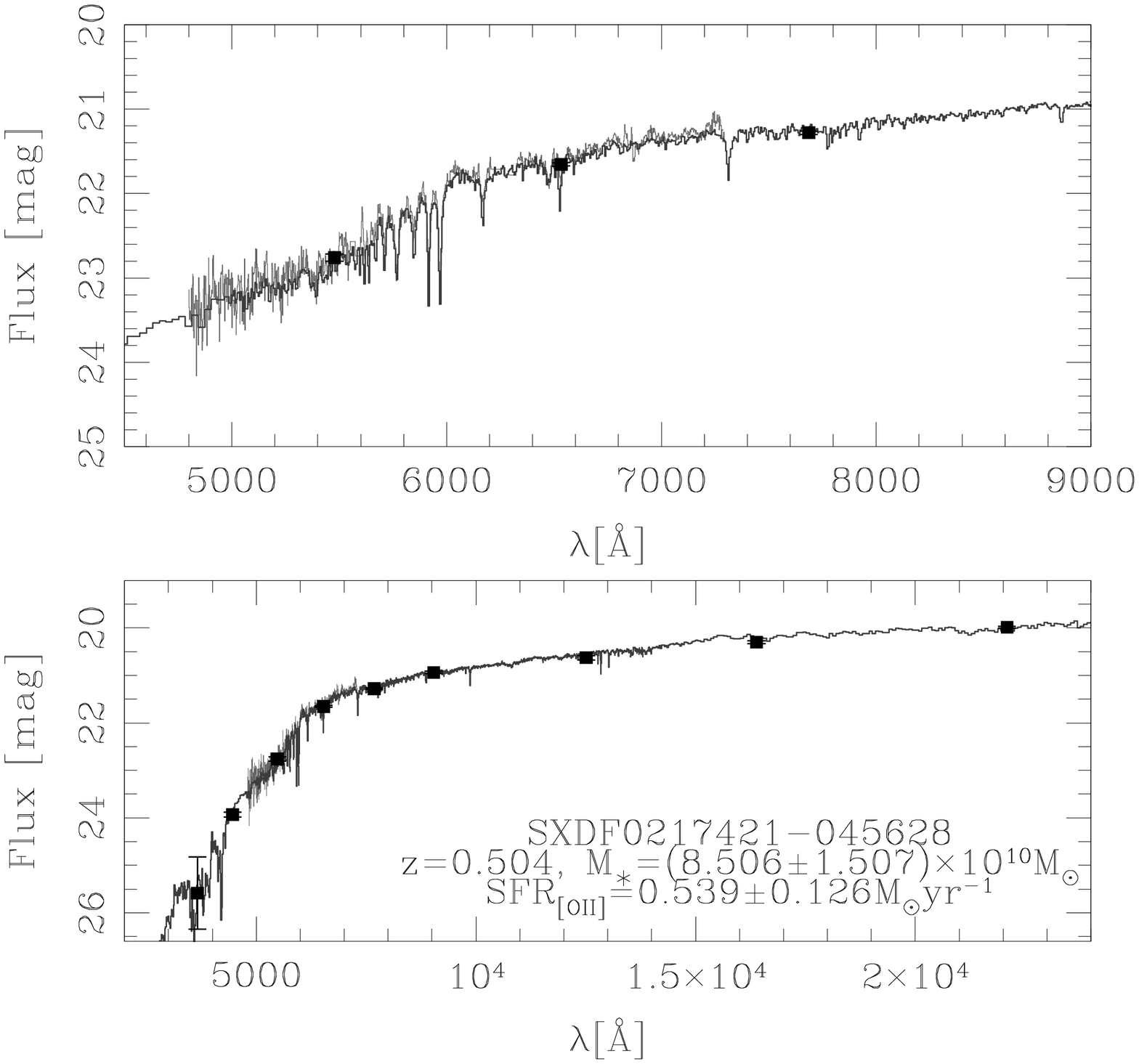}
\includegraphics[width=0.32\textwidth]{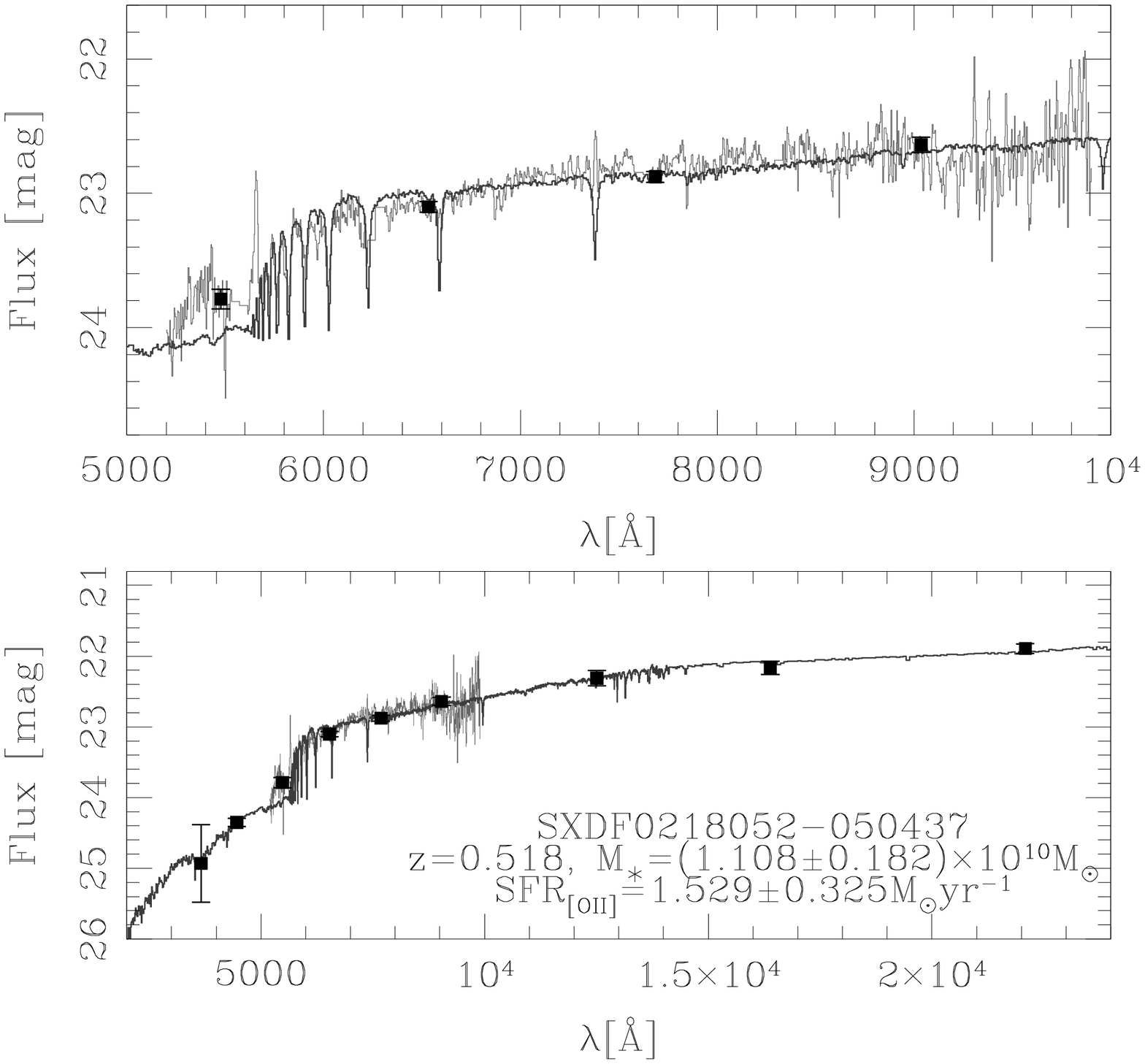}
\includegraphics[width=0.32\textwidth]{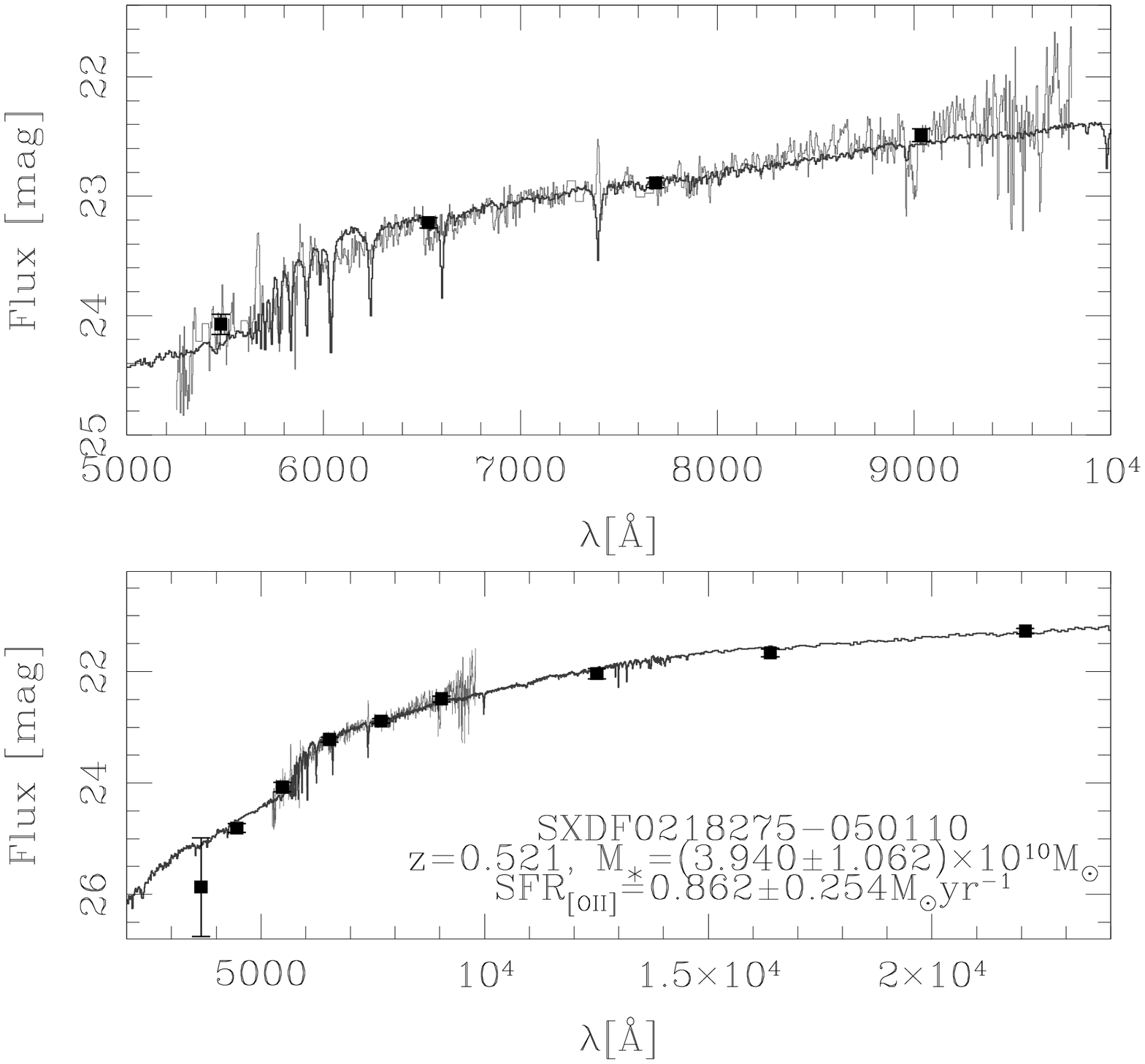}
\includegraphics[width=0.32\textwidth]{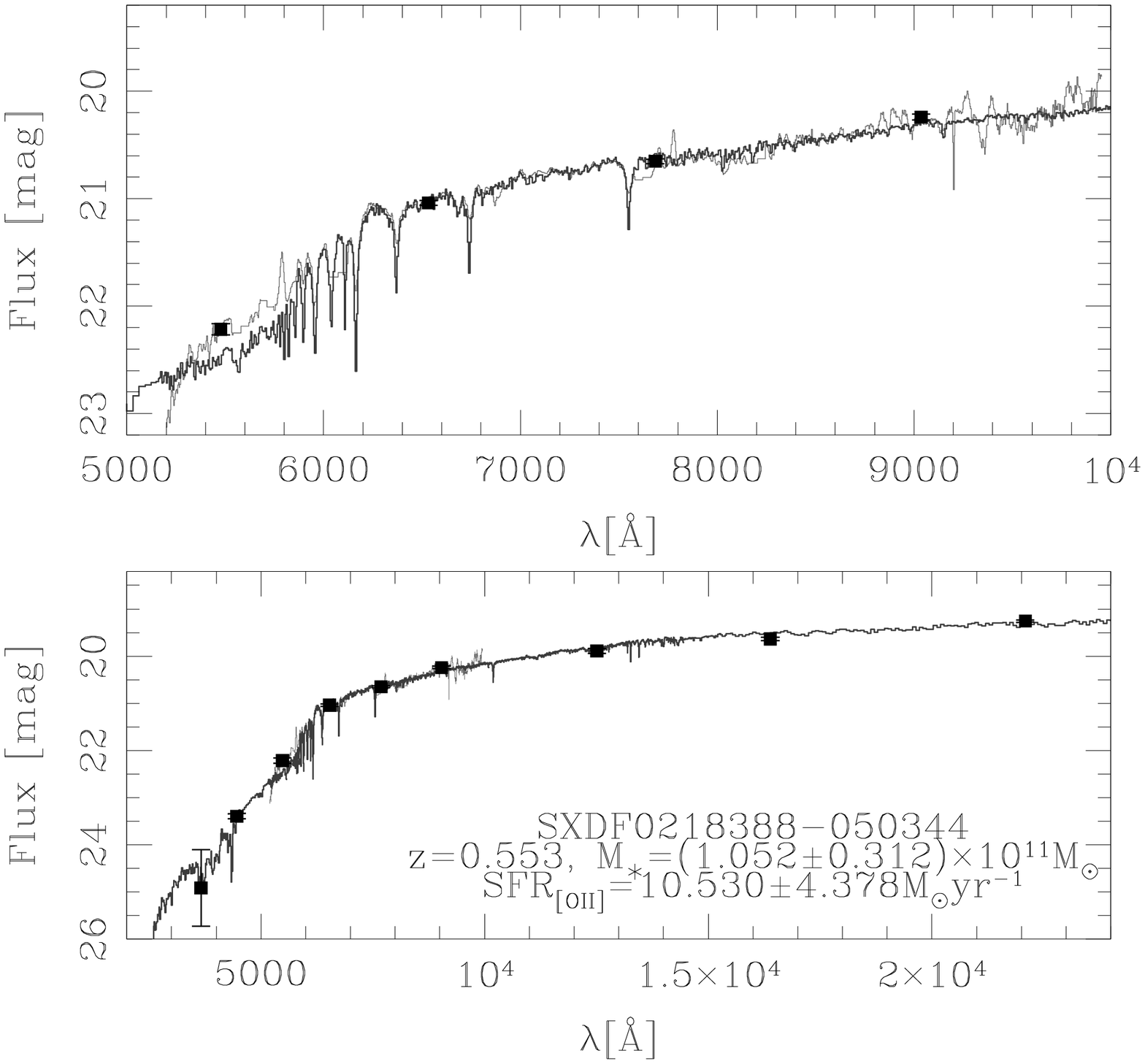}
\includegraphics[width=0.32\textwidth]{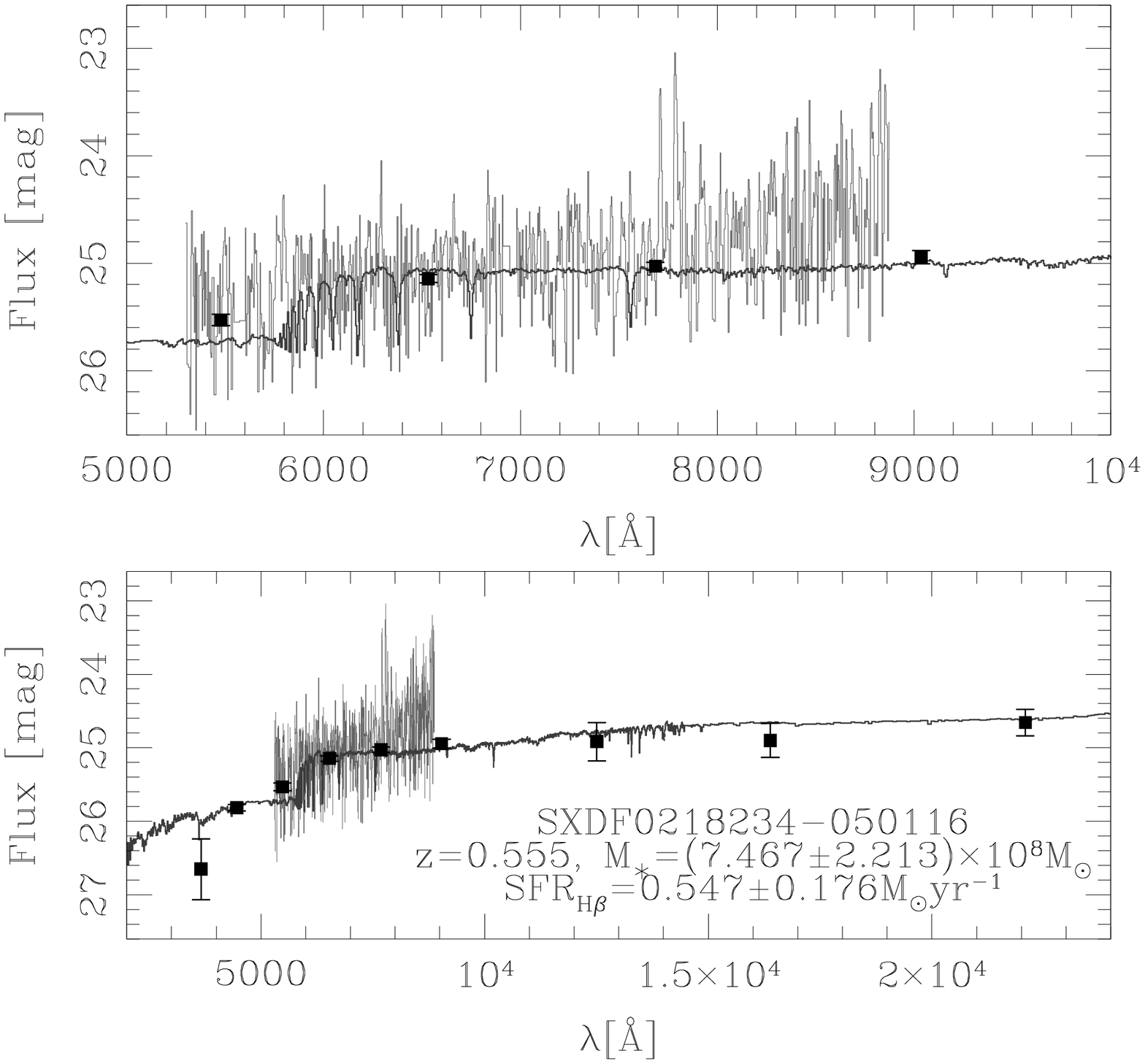}
\includegraphics[width=0.32\textwidth]{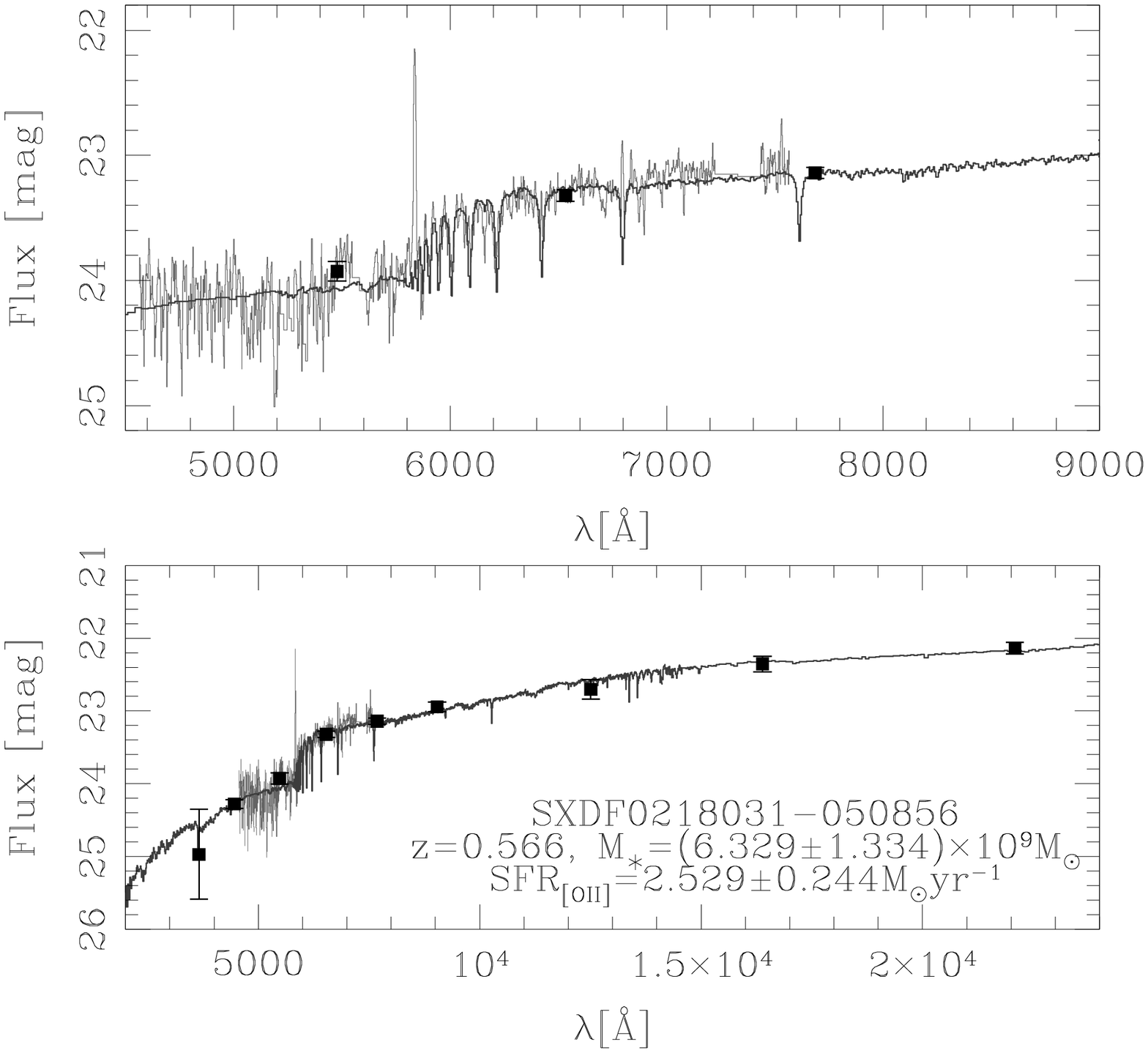}
\includegraphics[width=0.32\textwidth]{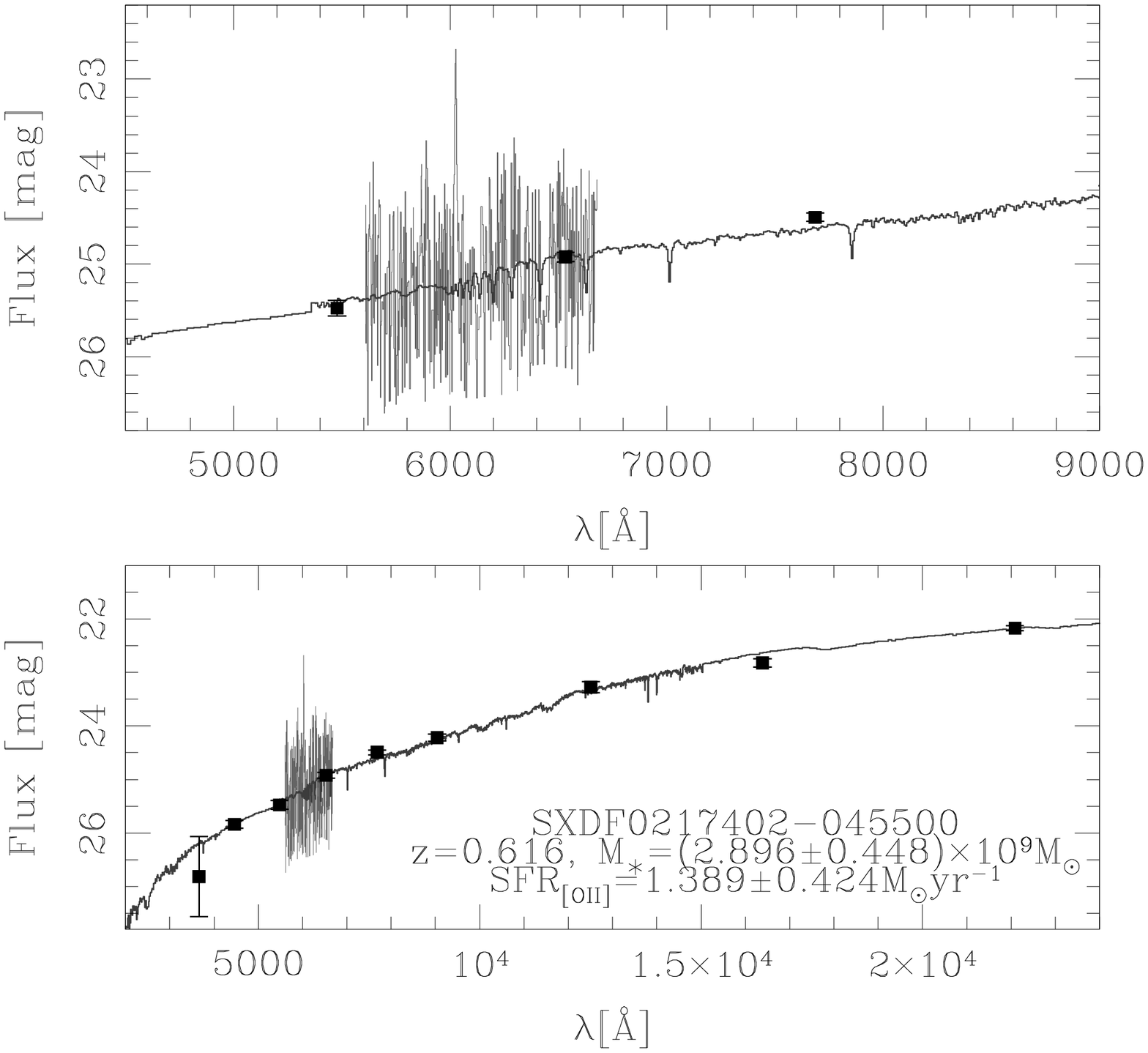}
\includegraphics[width=0.32\textwidth]{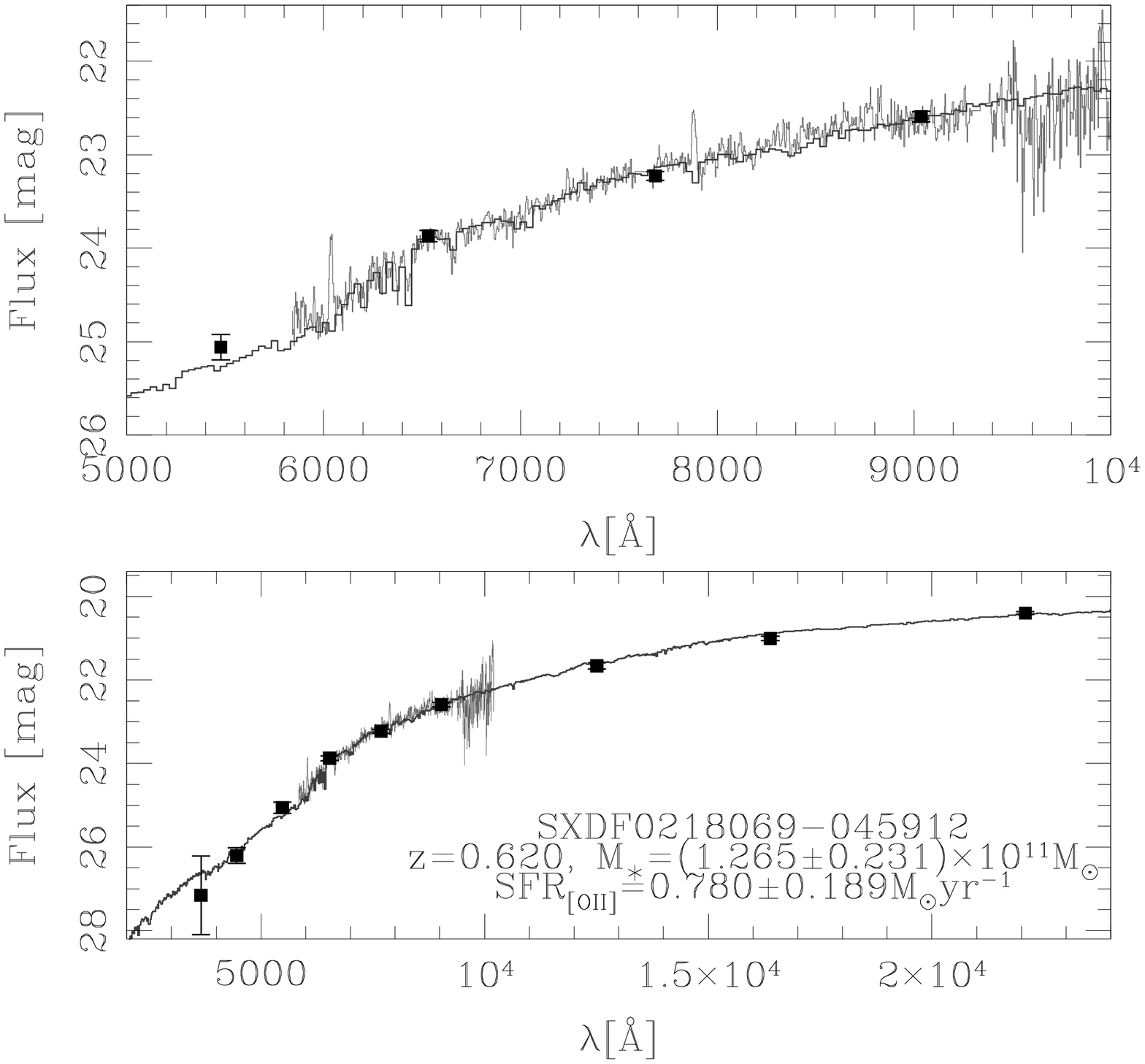}
\includegraphics[width=0.32\textwidth]{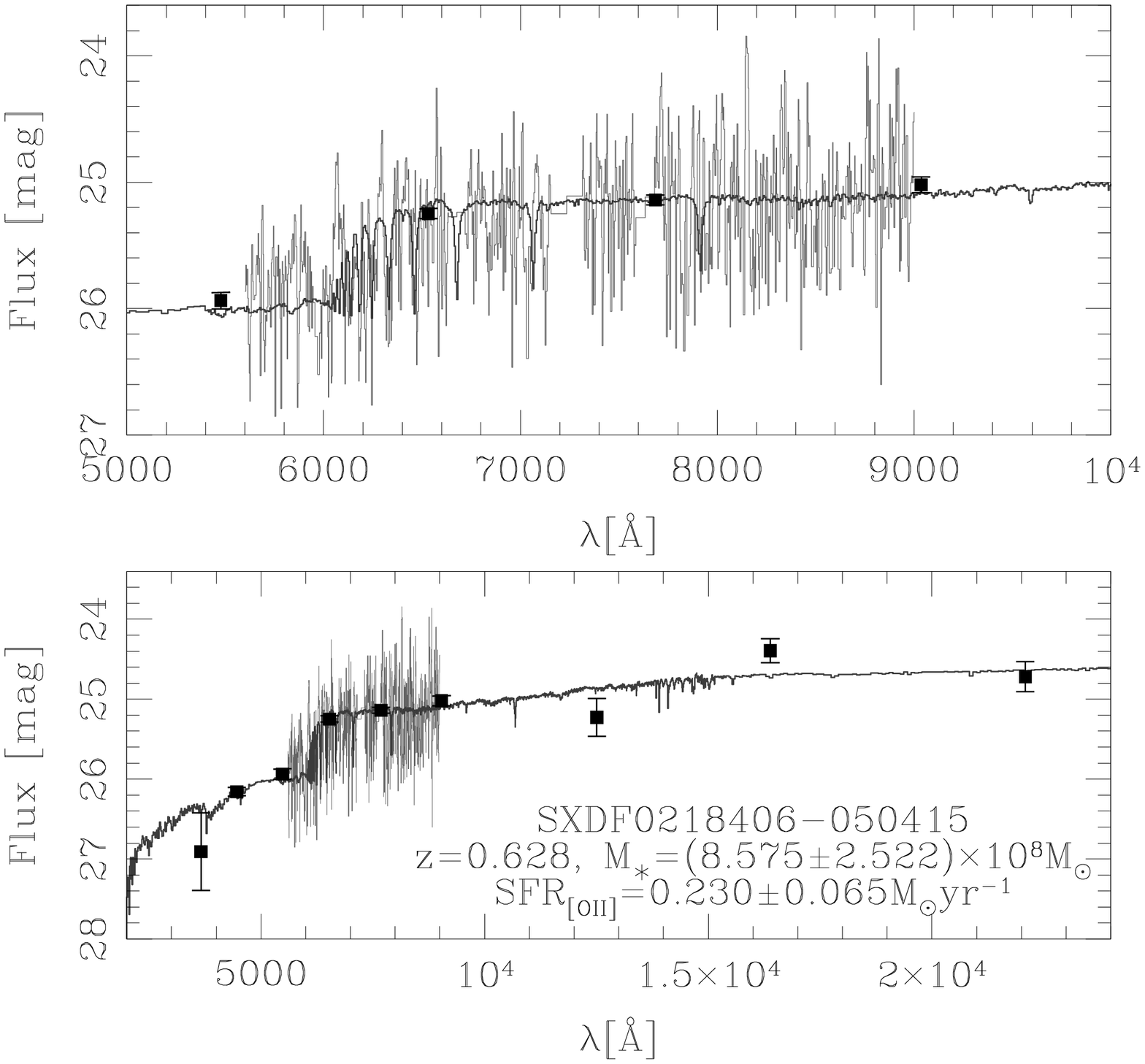}
\includegraphics[width=0.32\textwidth]{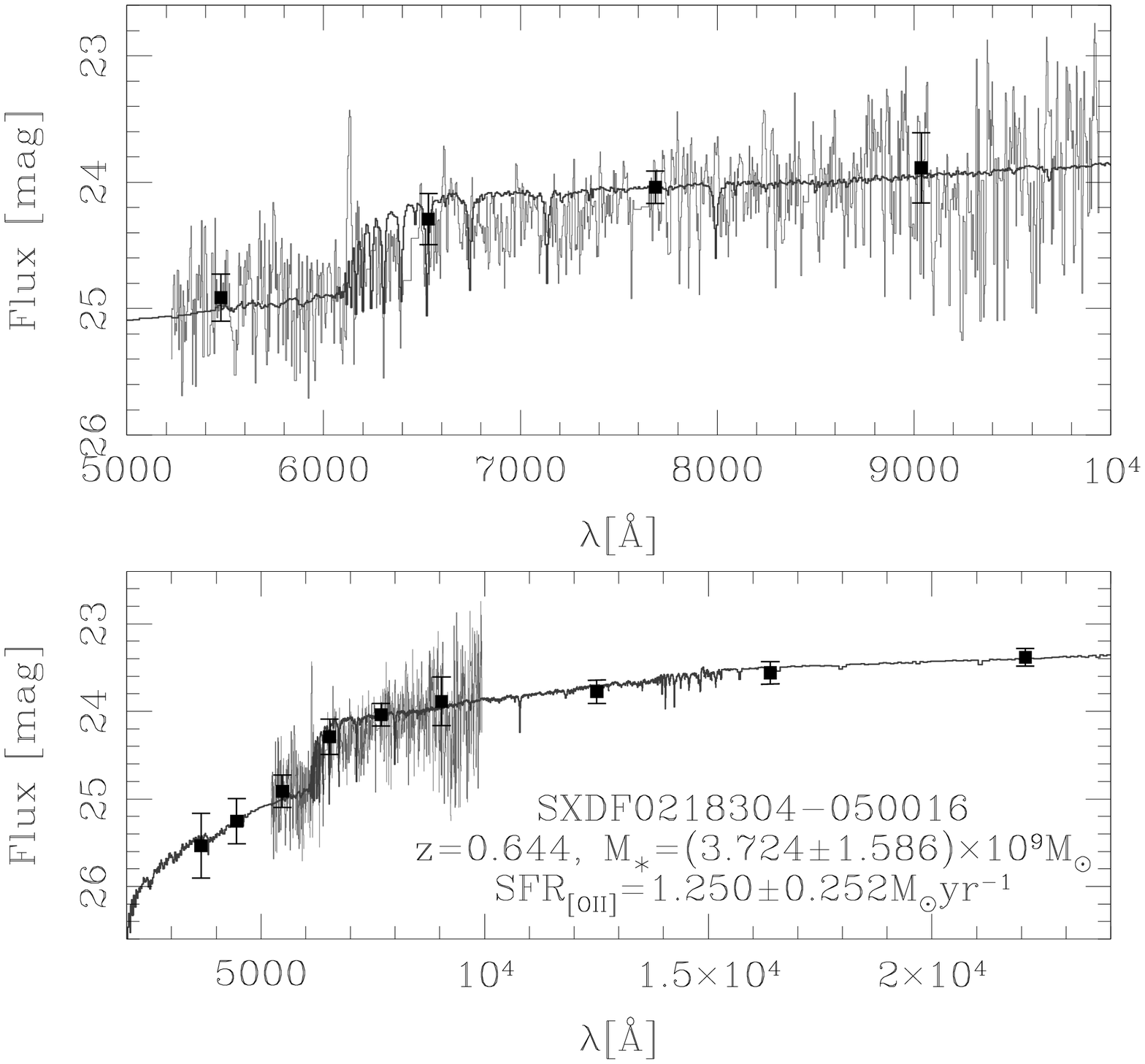}
\caption{\textit{- continued}}
\end{center}
\end{figure*}

\begin{figure*}
\addtocounter{figure}{-1}
\begin{center}
\includegraphics[width=0.32\textwidth]{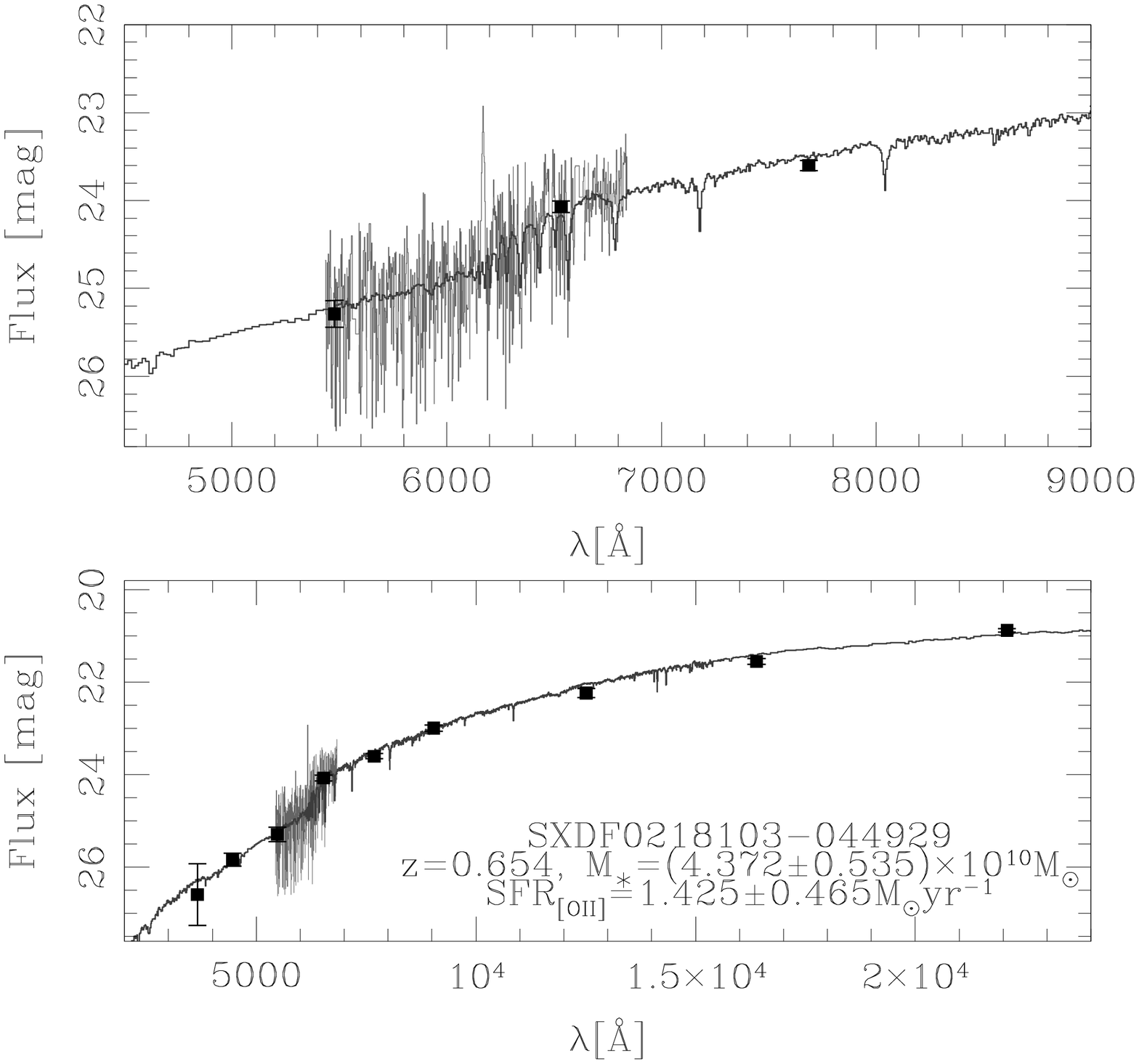}
\includegraphics[width=0.32\textwidth]{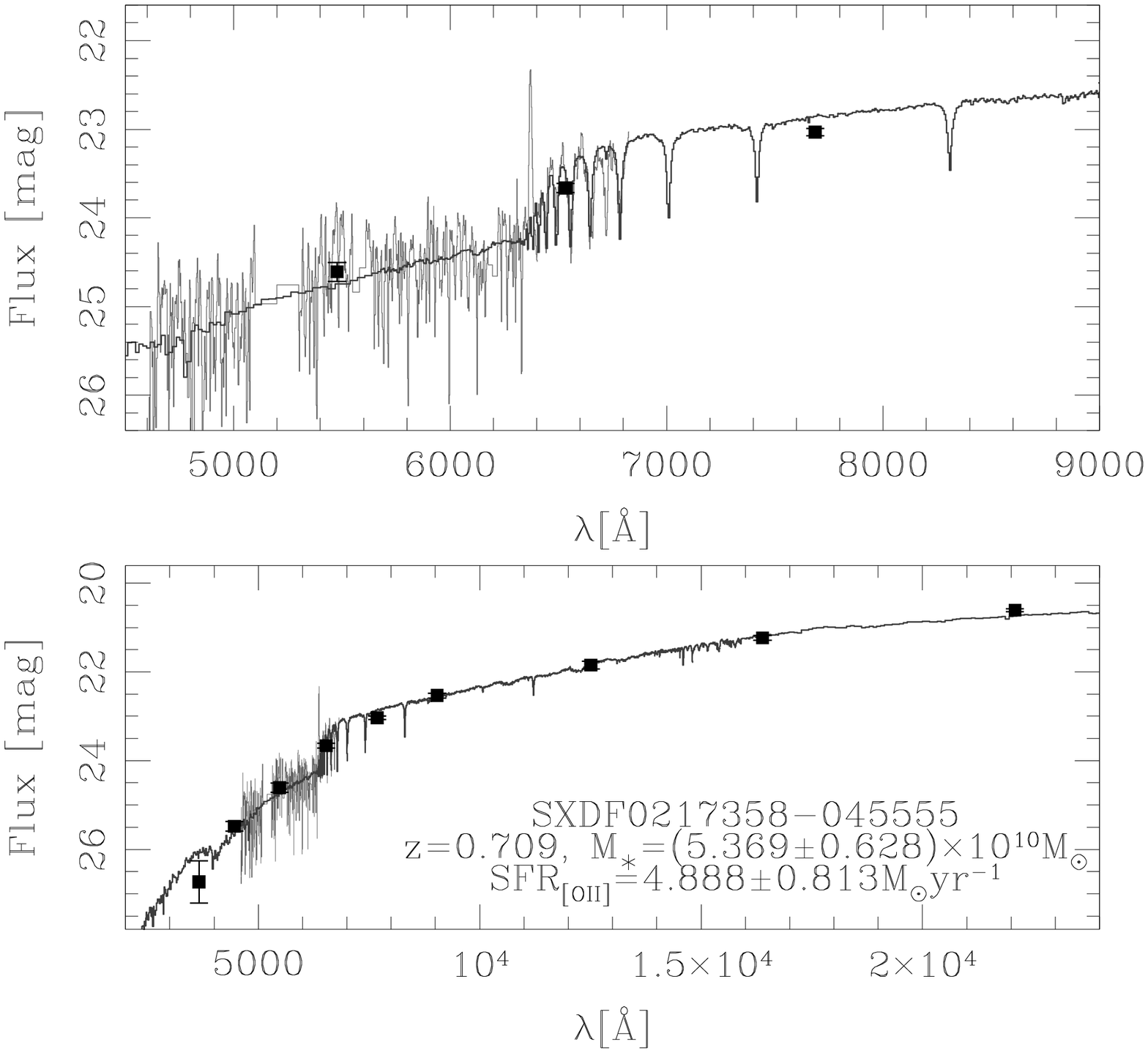}
\includegraphics[width=0.32\textwidth]{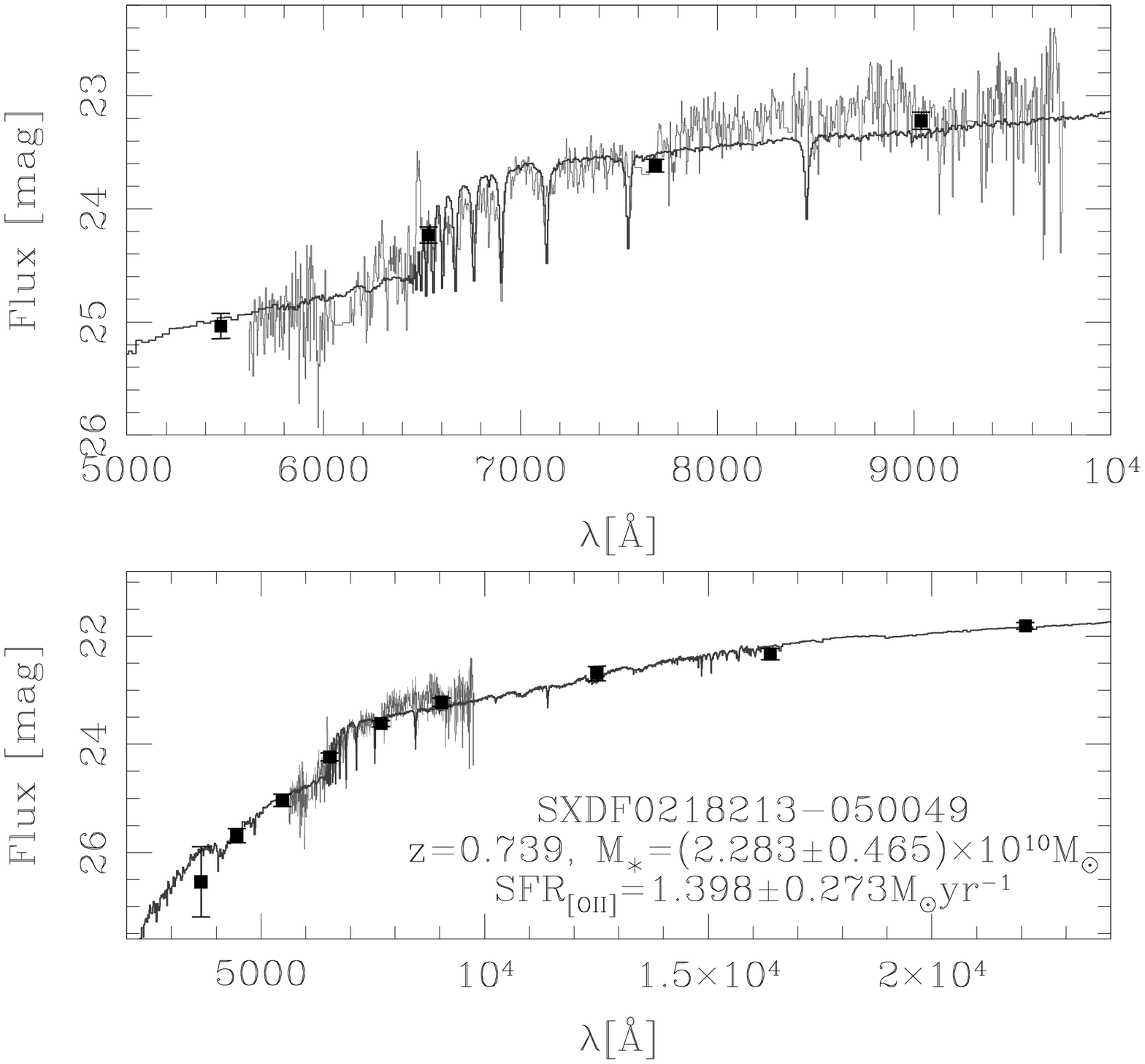}
\includegraphics[width=0.32\textwidth]{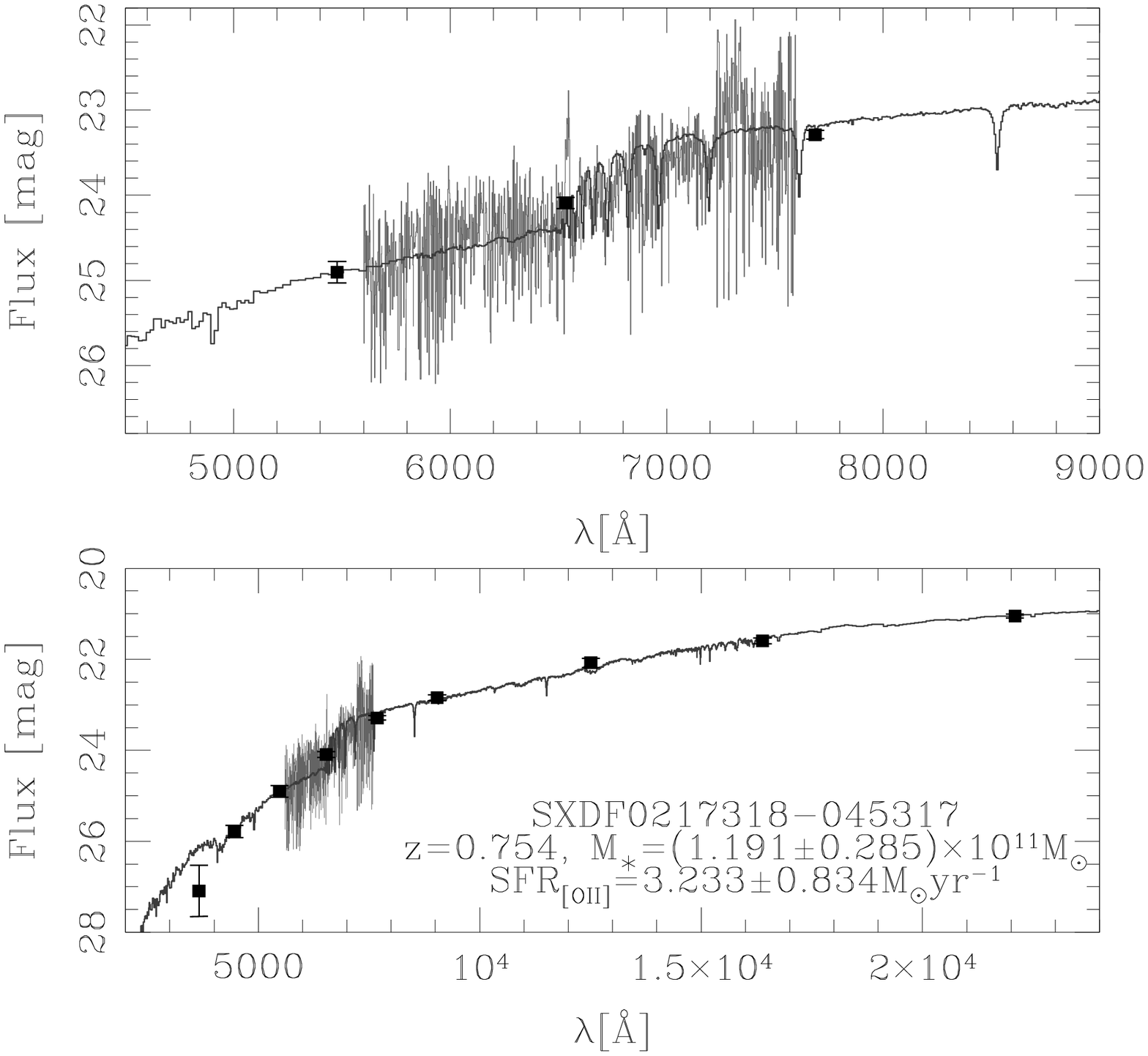}
\includegraphics[width=0.32\textwidth]{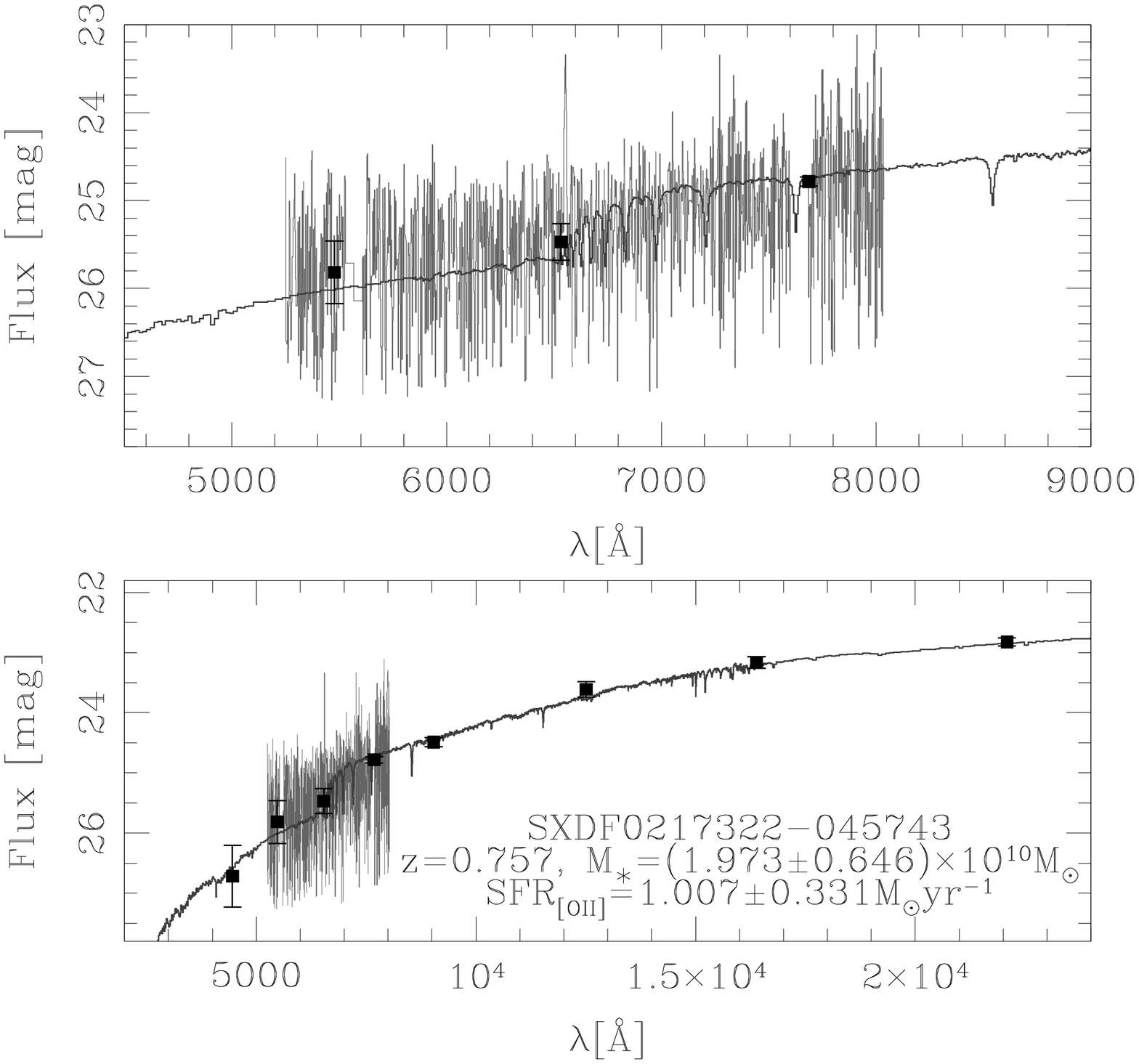}
\includegraphics[width=0.32\textwidth]{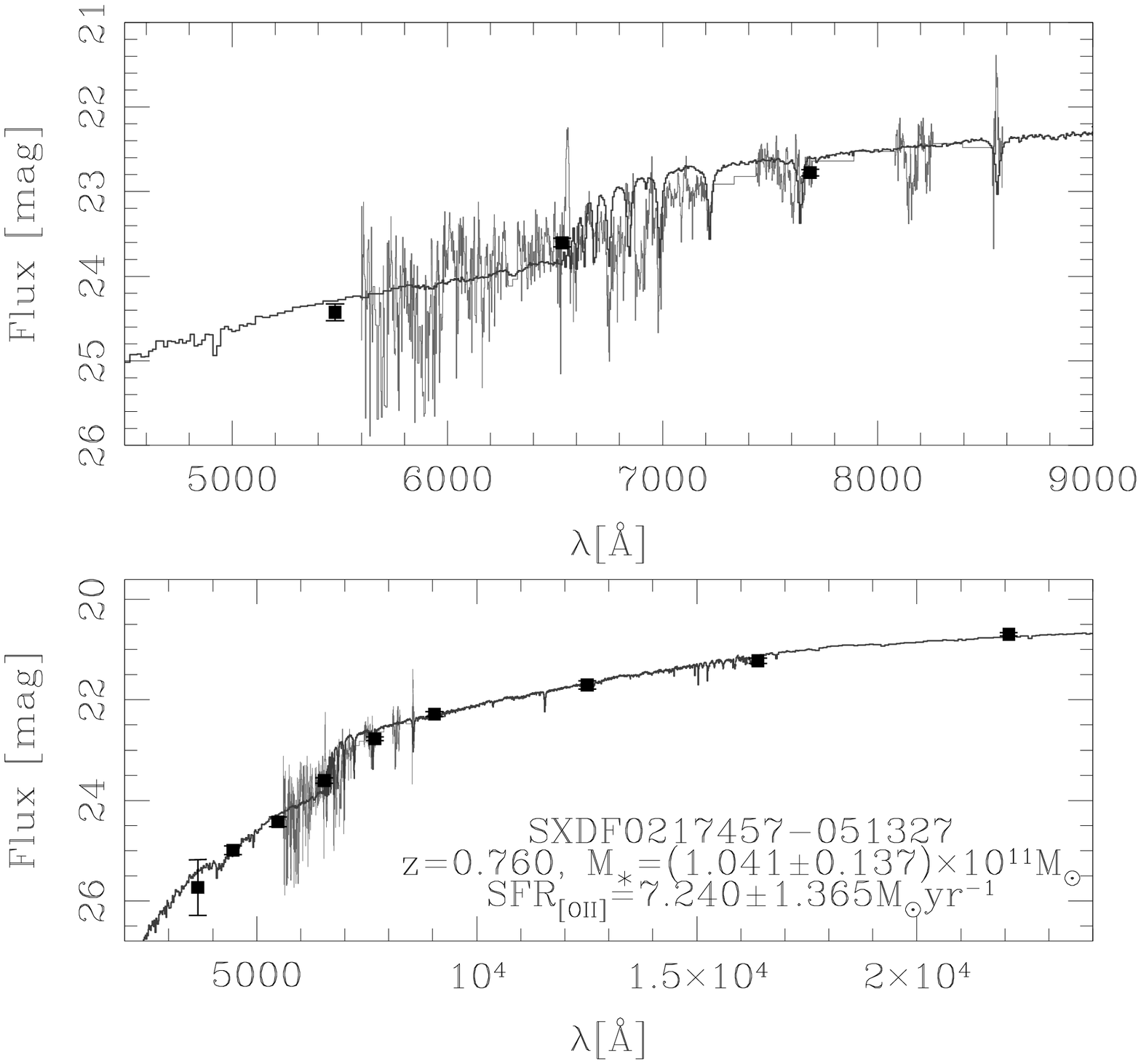}
\includegraphics[width=0.32\textwidth]{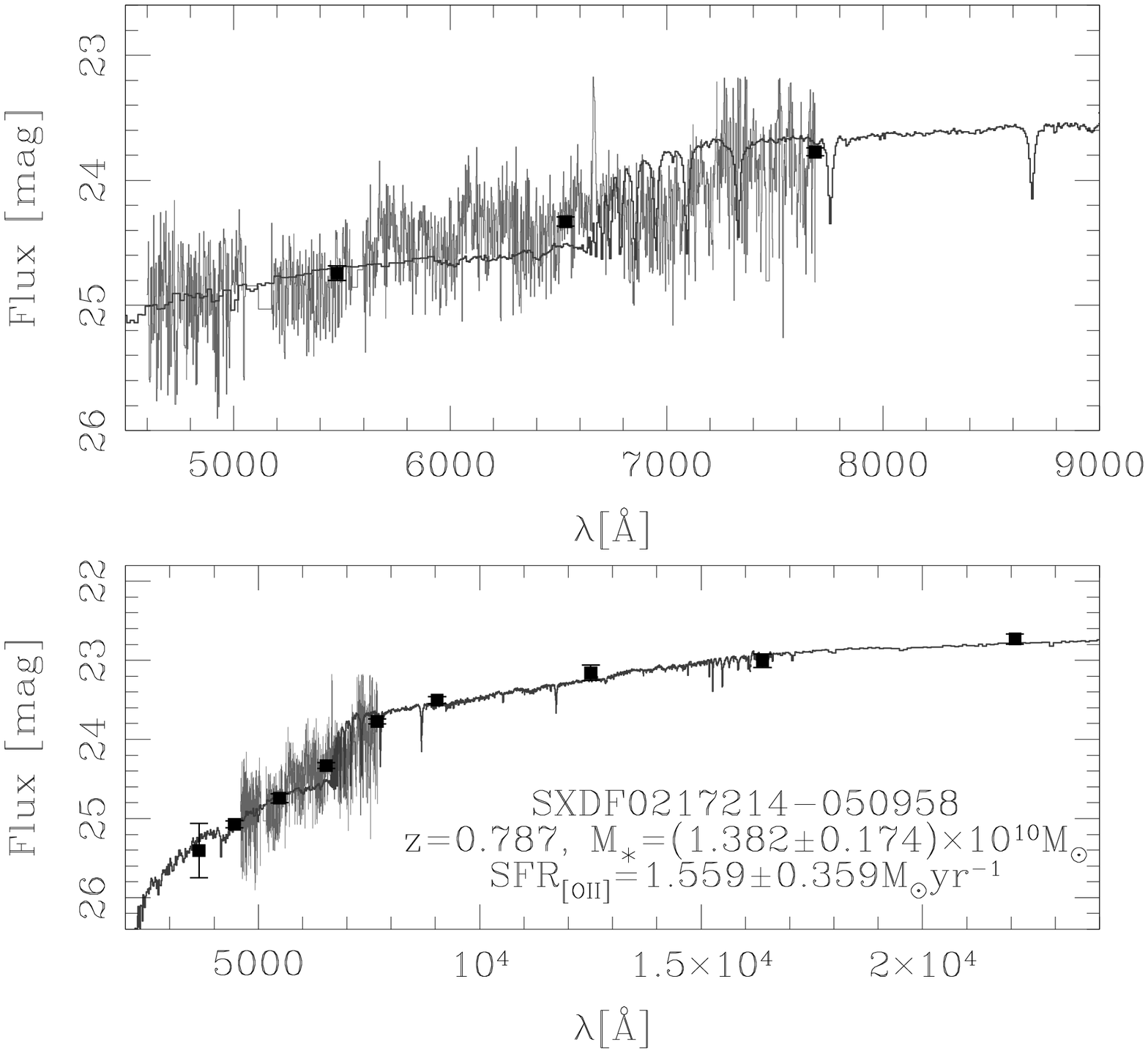}
\includegraphics[width=0.32\textwidth]{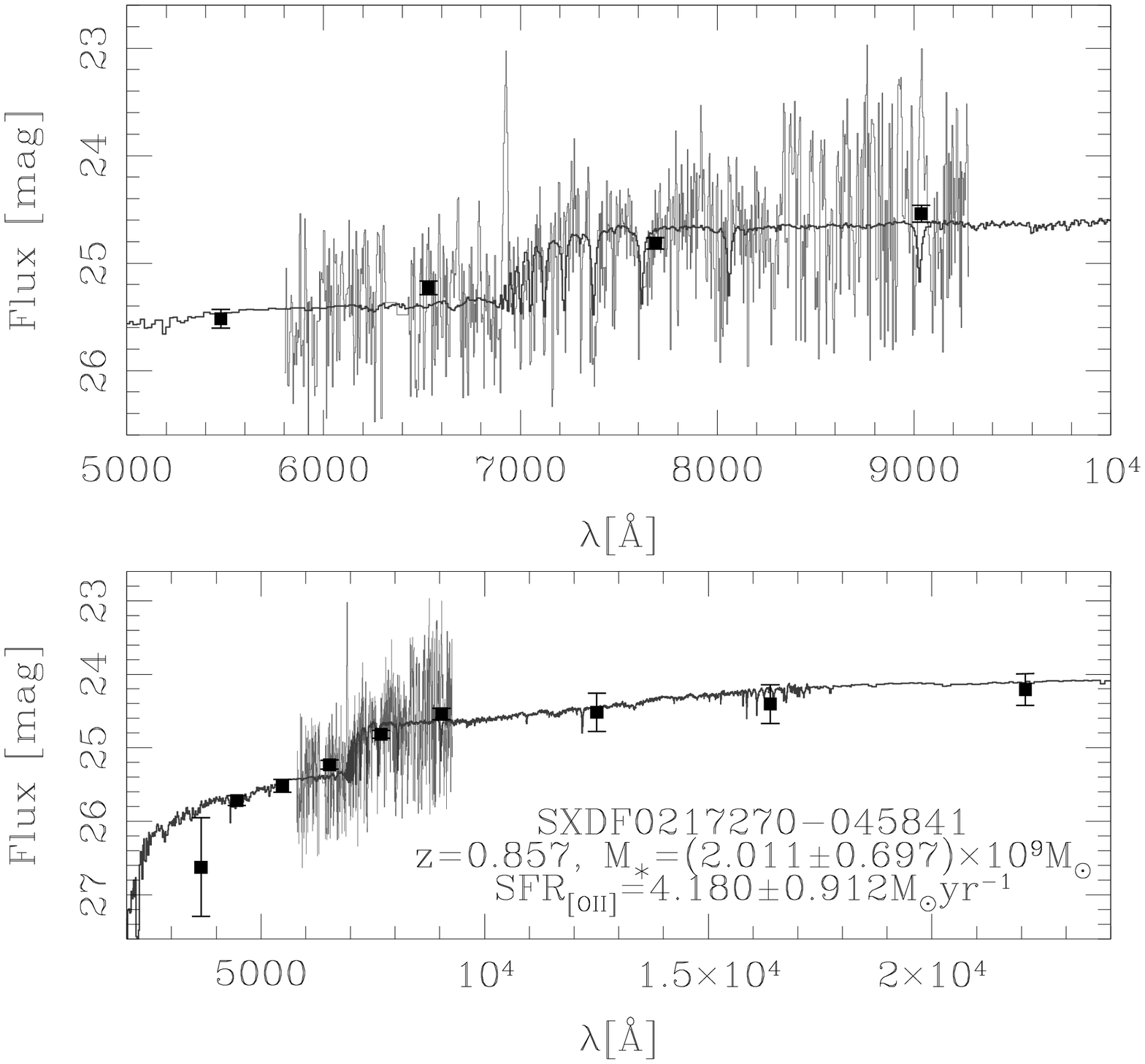}
\includegraphics[width=0.32\textwidth]{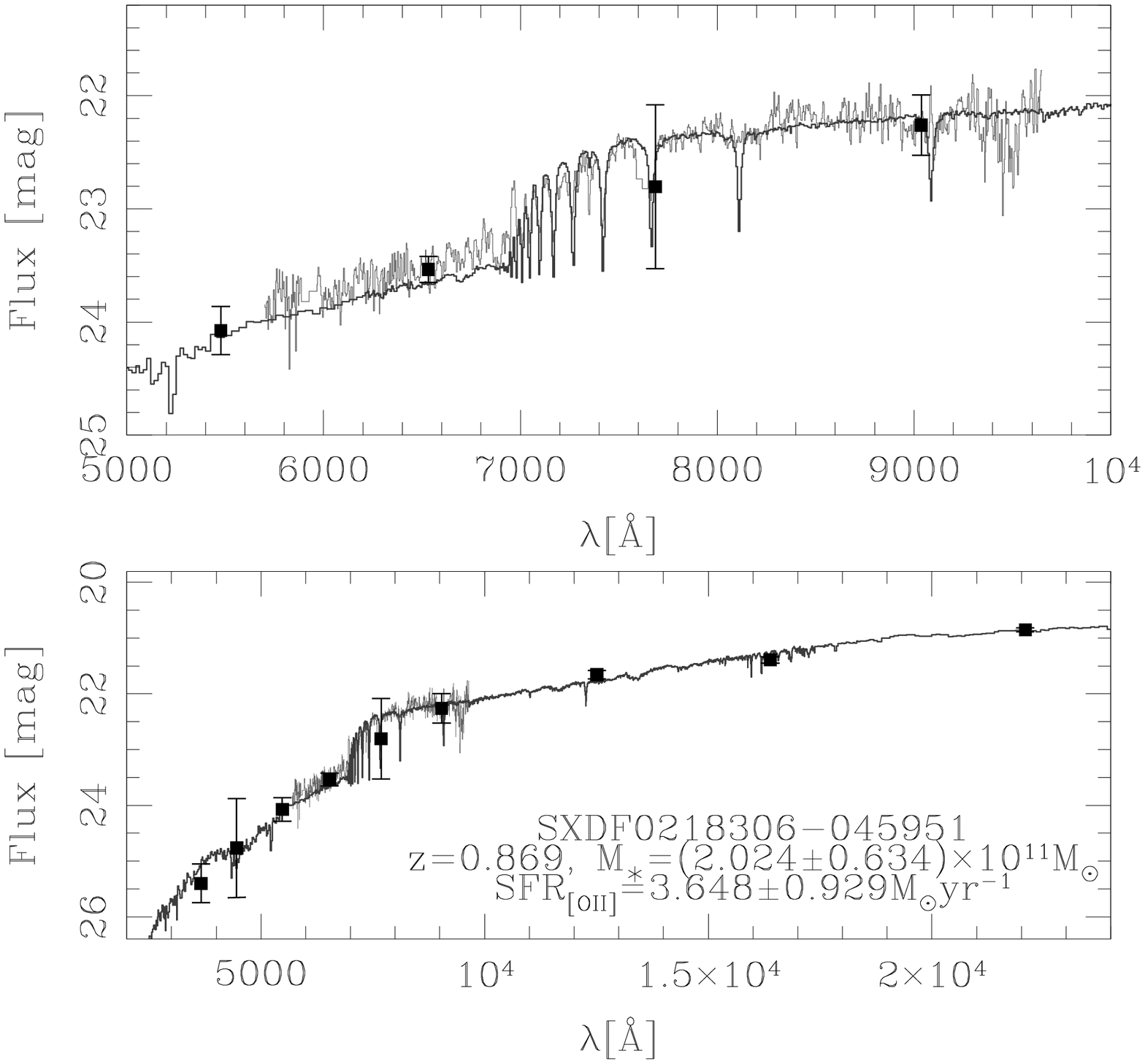}
\includegraphics[width=0.32\textwidth]{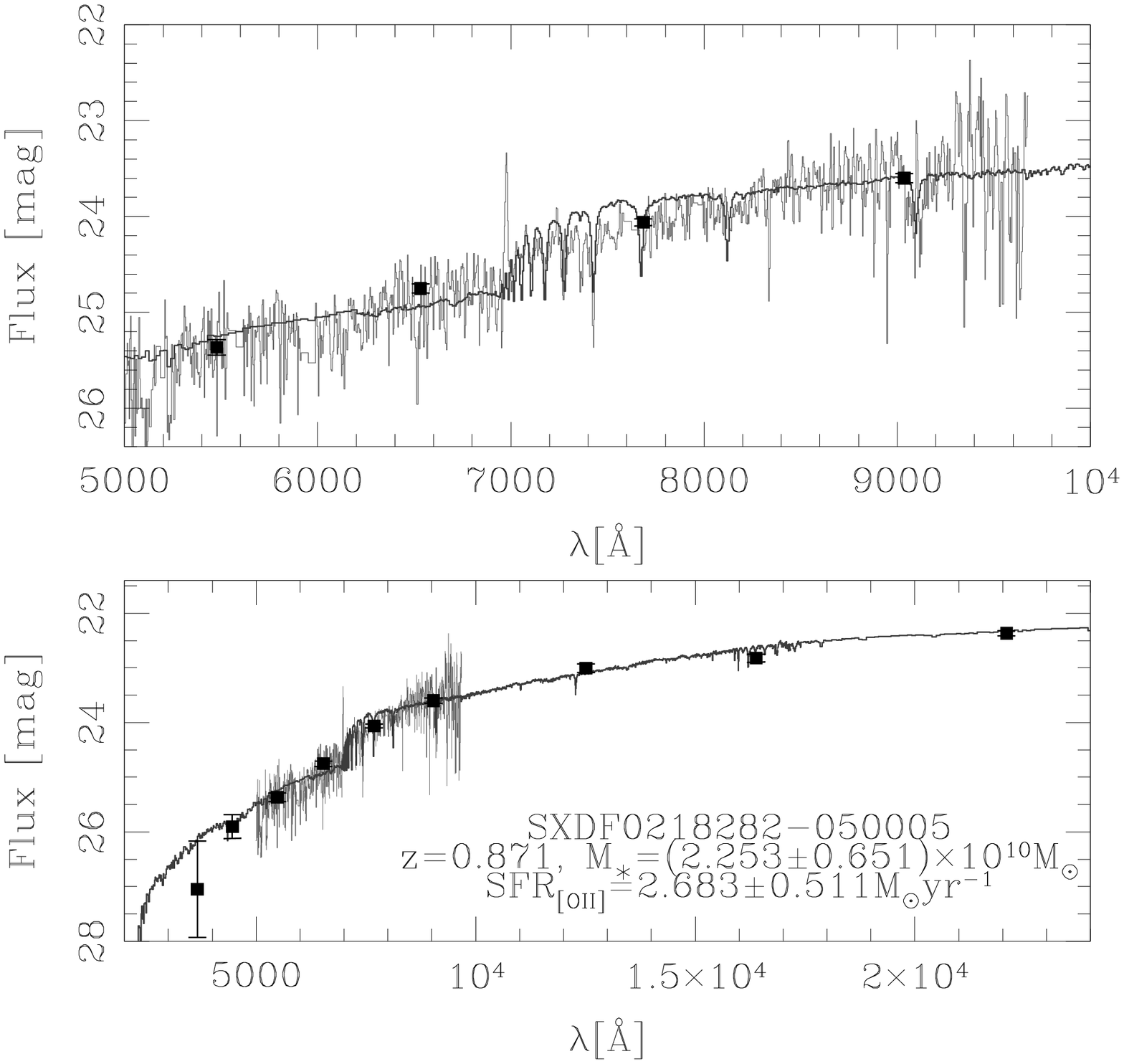}
\includegraphics[width=0.32\textwidth]{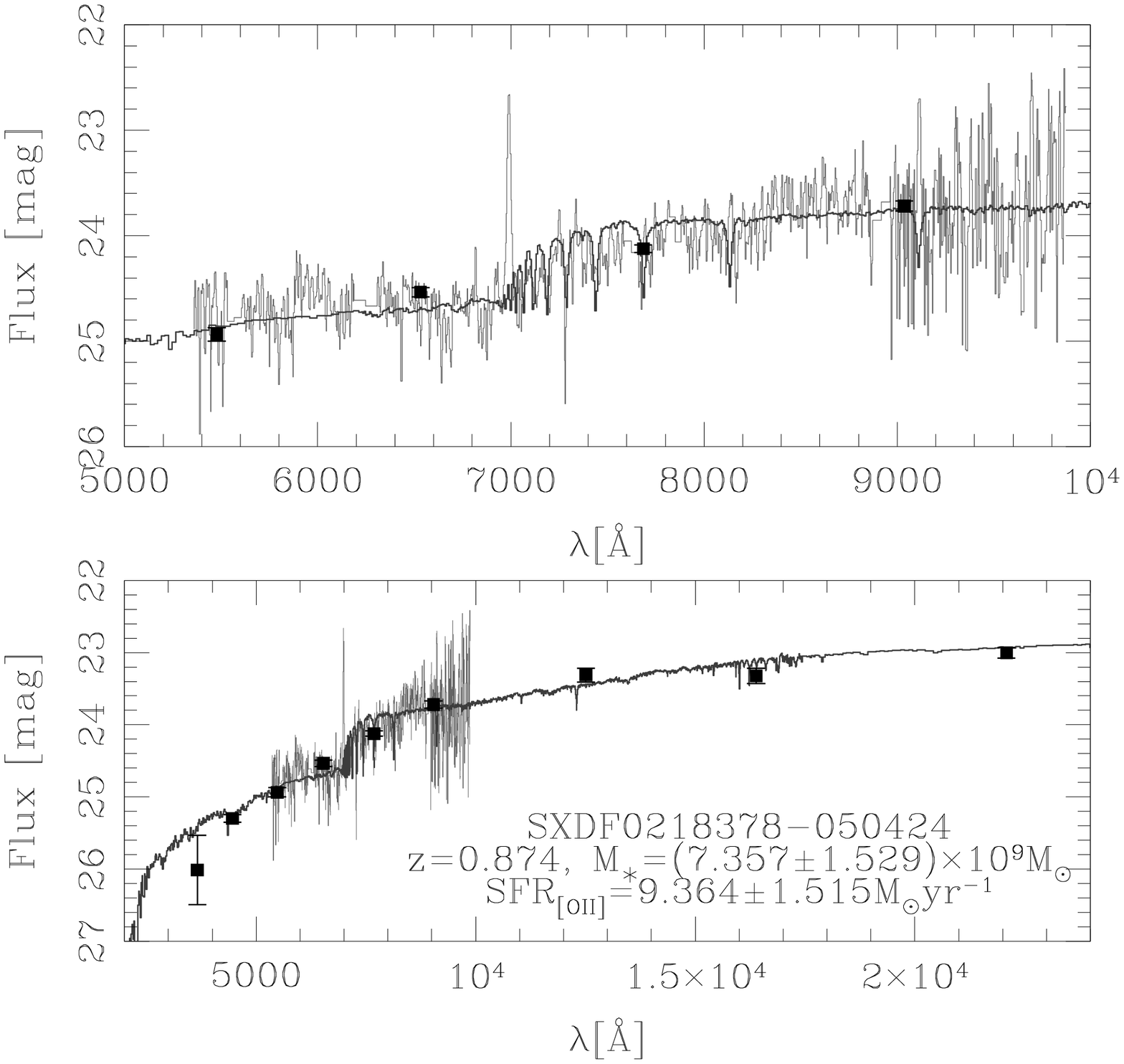}
\includegraphics[width=0.32\textwidth]{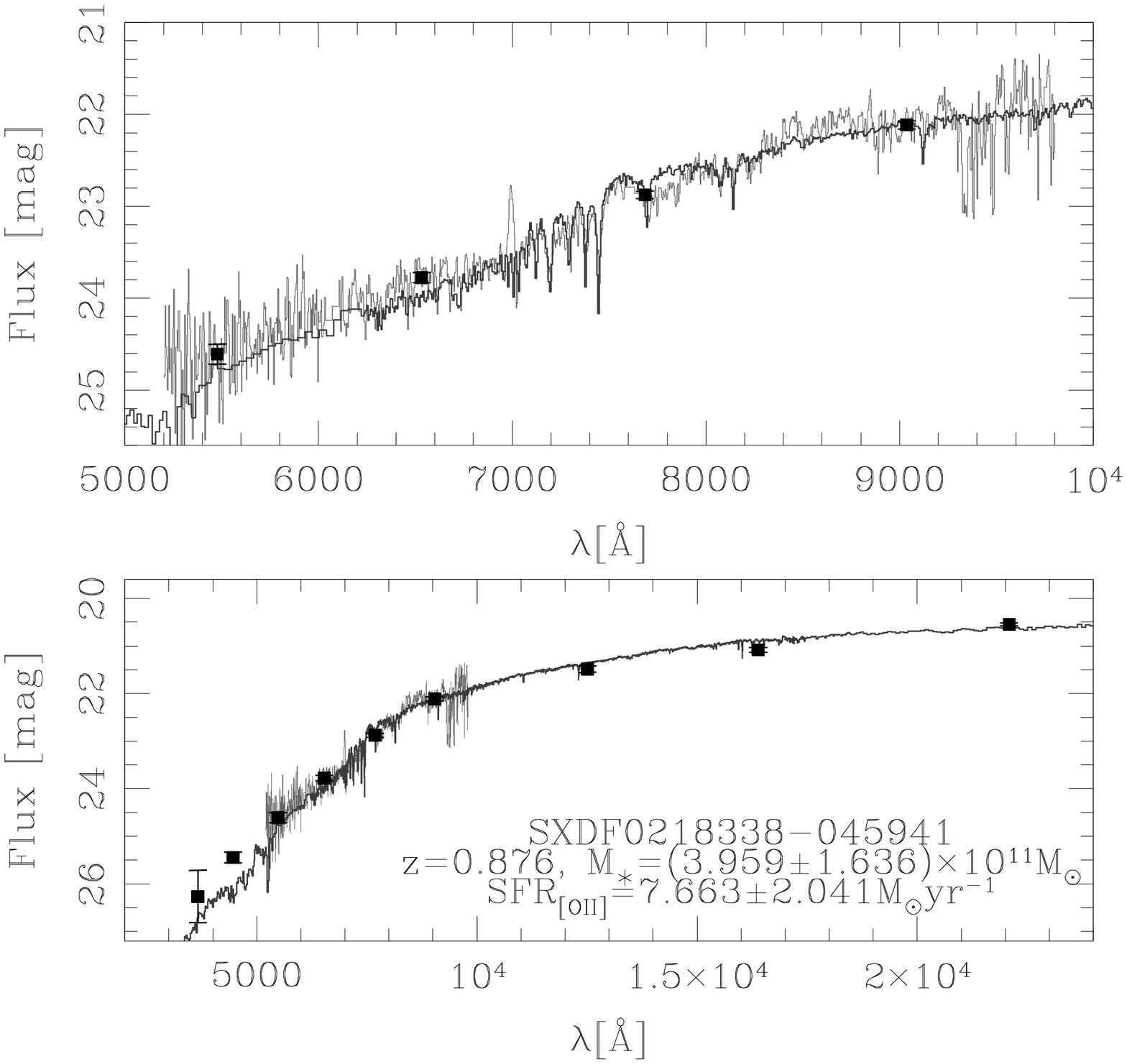}
\caption{\textit{- continued}}
\end{center}
\end{figure*}

\begin{figure*}
\addtocounter{figure}{-1}
\begin{center}
\includegraphics[width=0.32\textwidth]{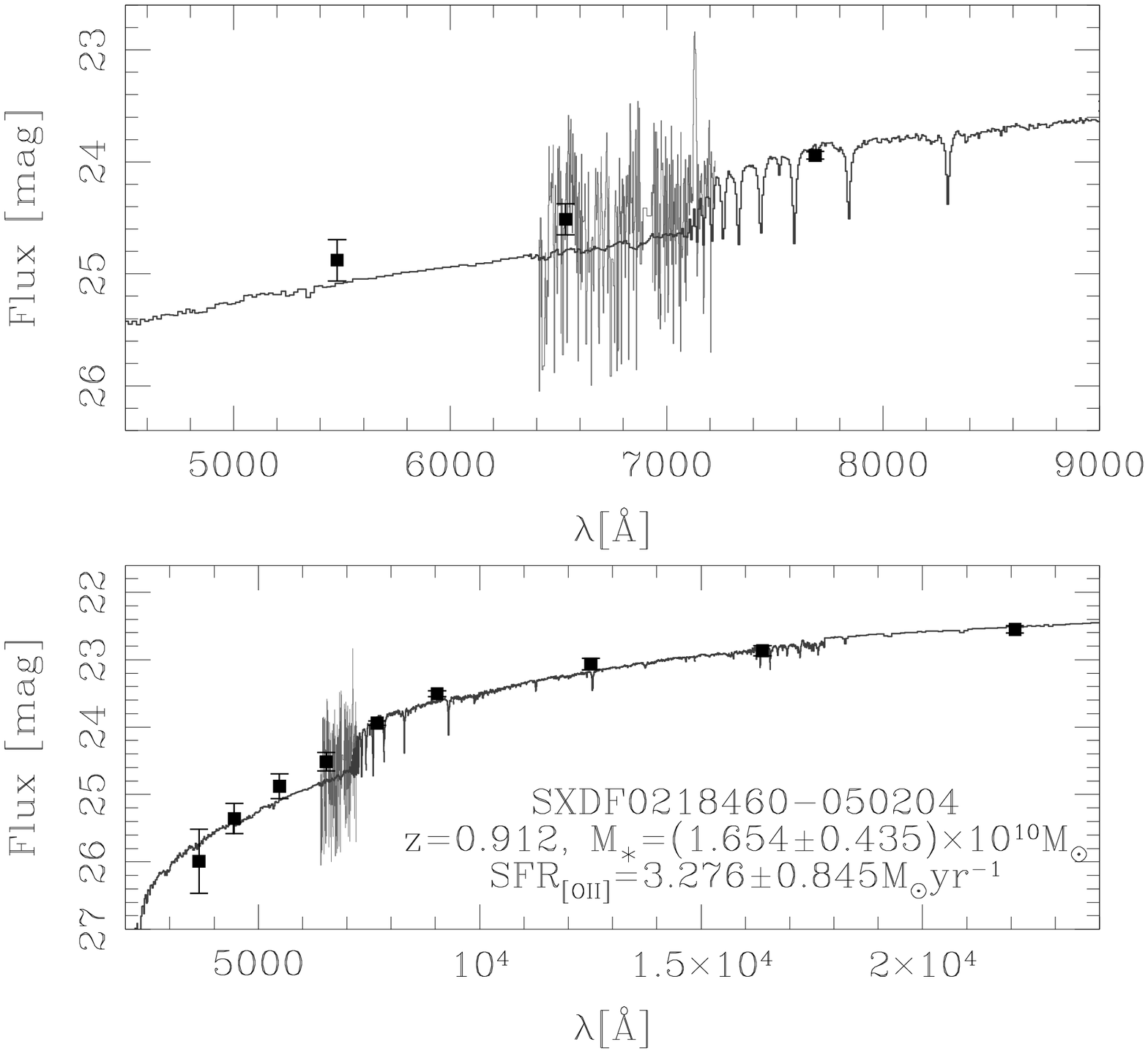}
\includegraphics[width=0.32\textwidth]{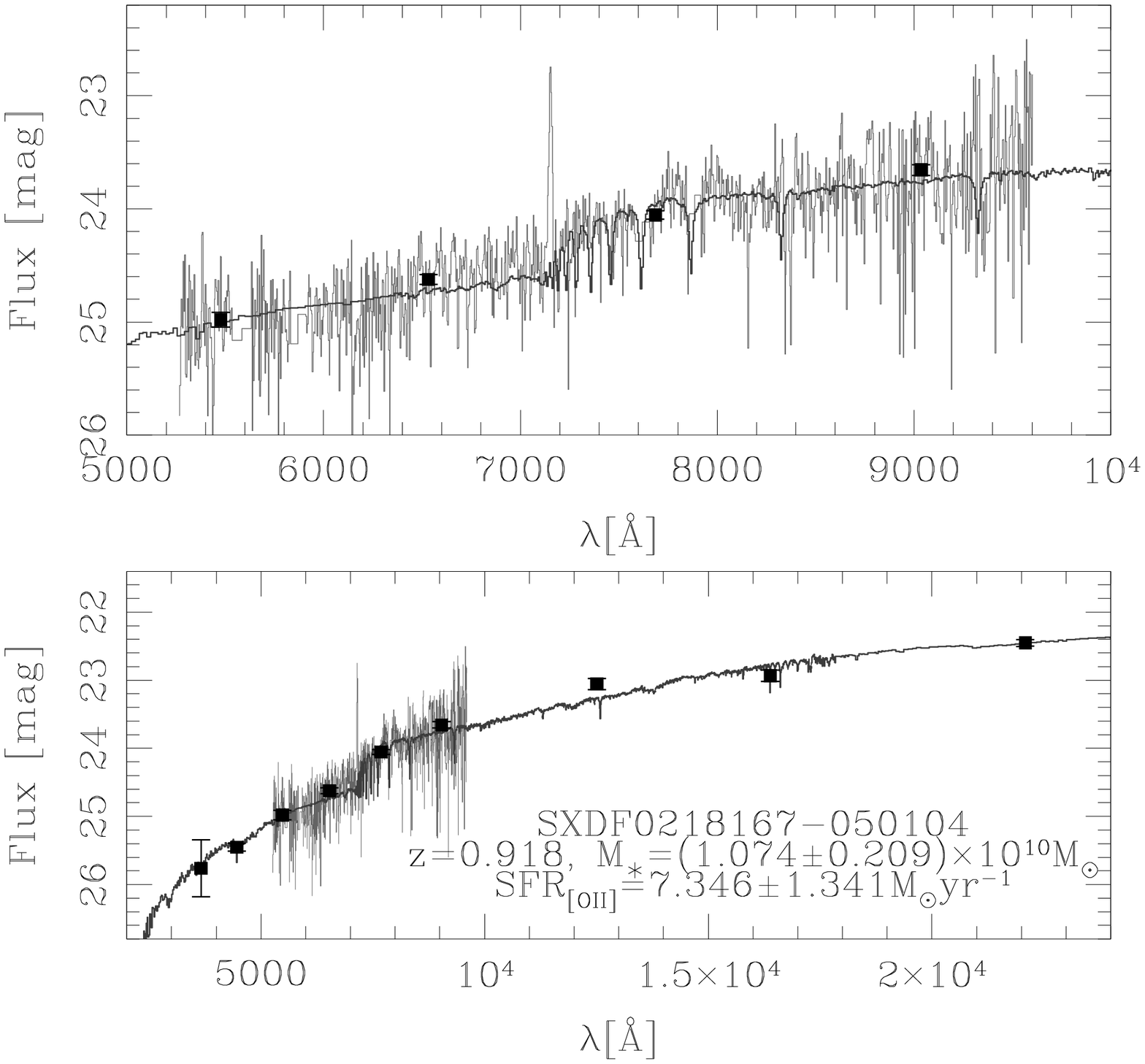}
\includegraphics[width=0.32\textwidth]{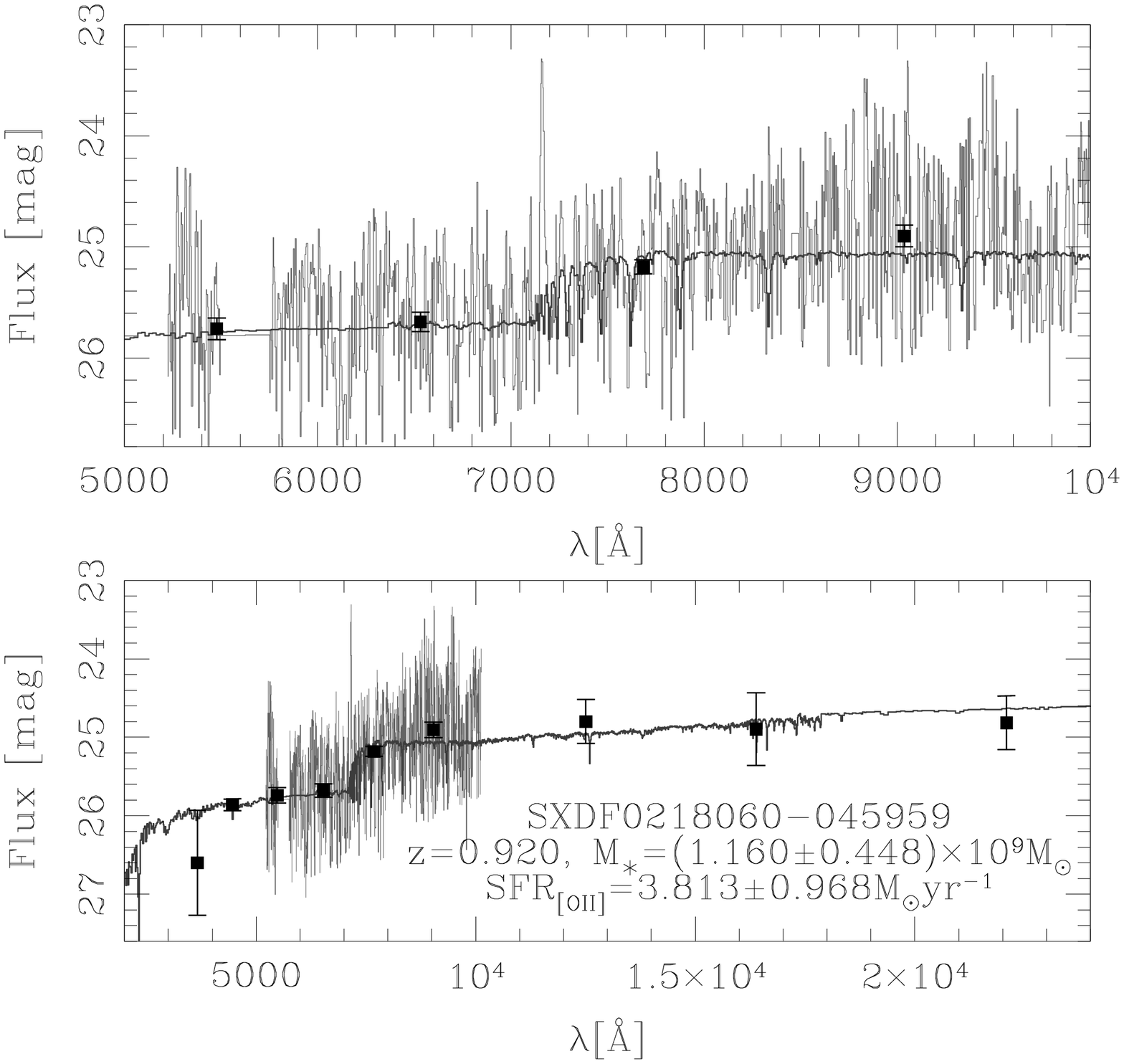}
\includegraphics[width=0.32\textwidth]{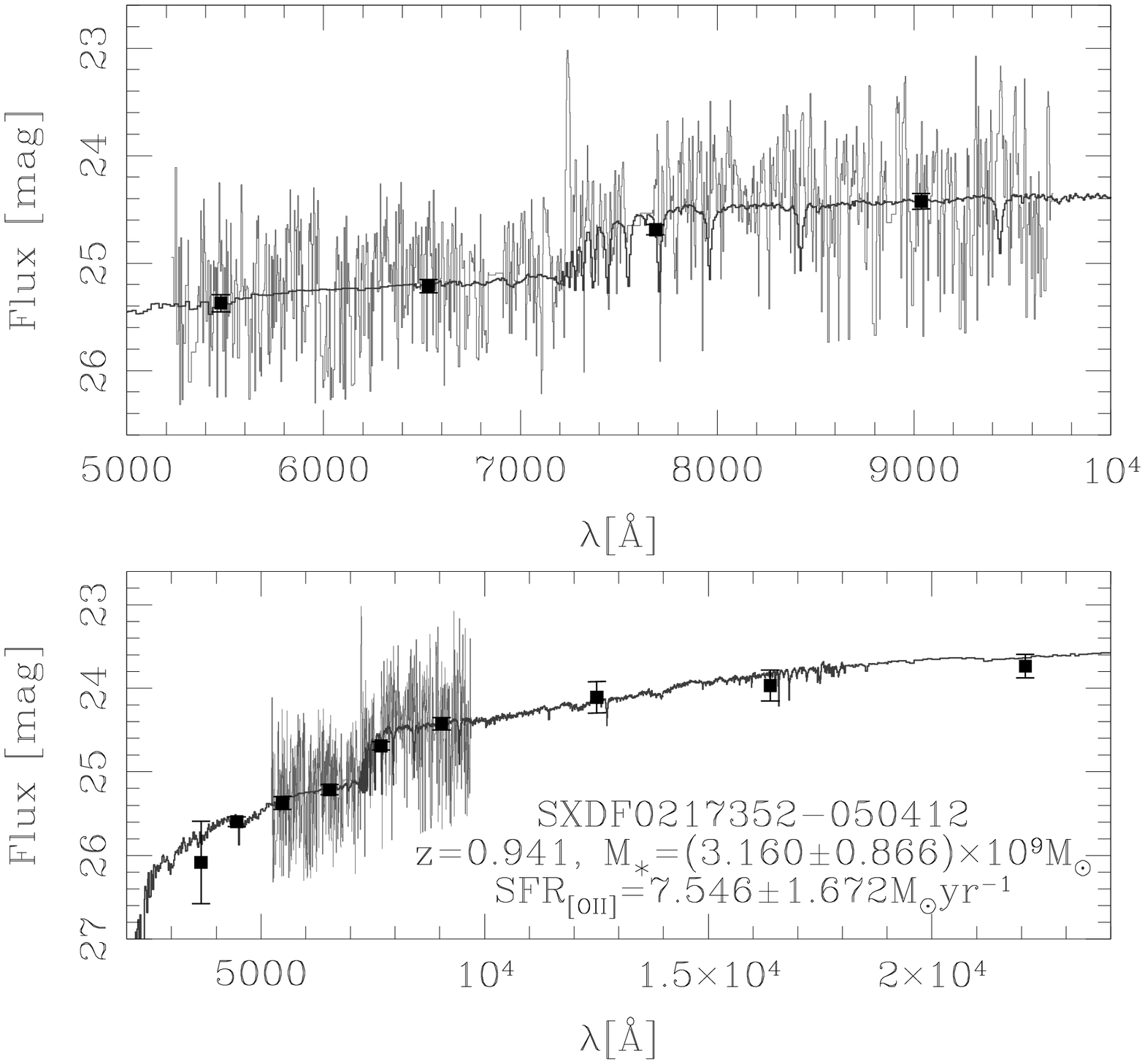}
\includegraphics[width=0.32\textwidth]{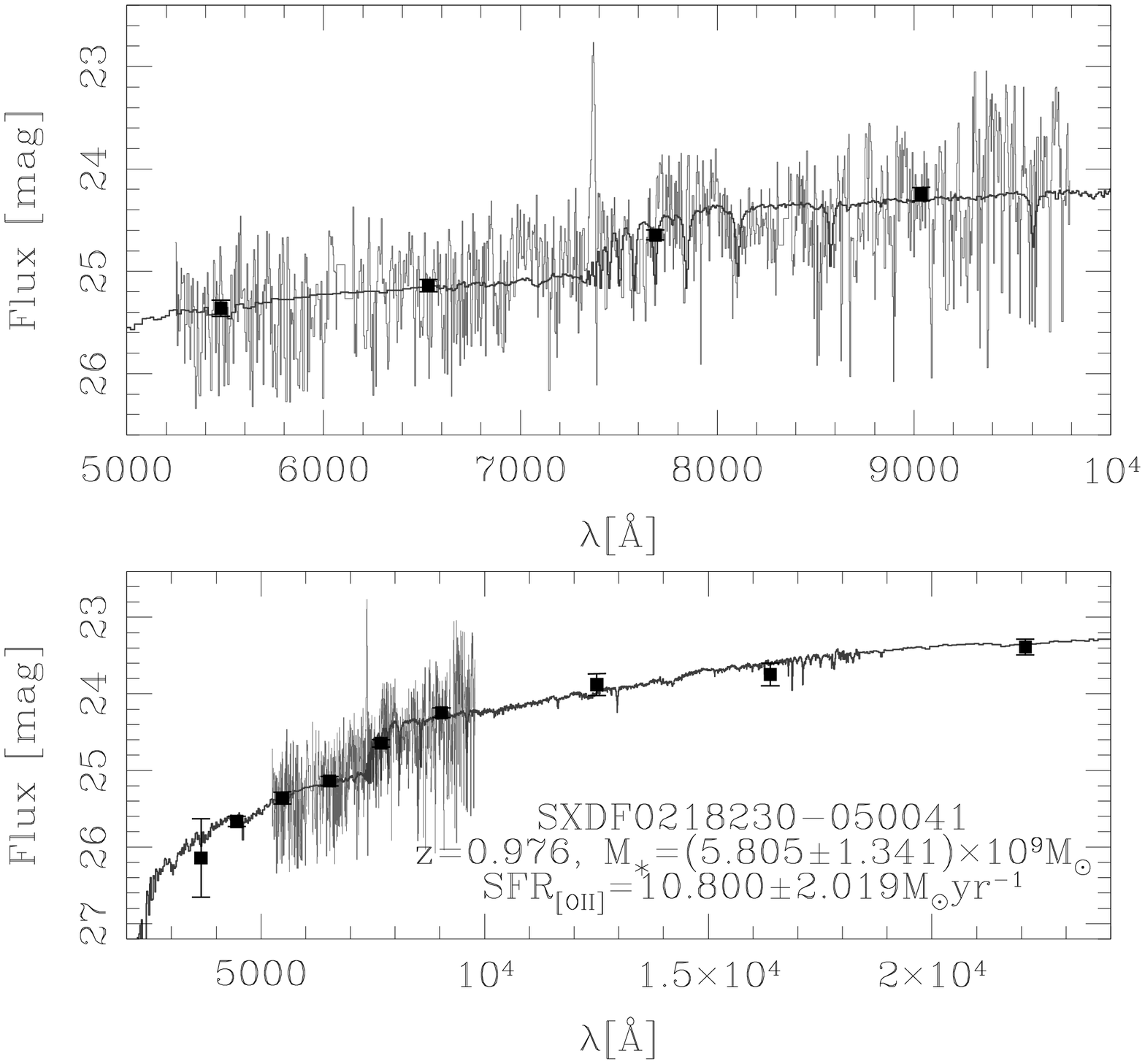}
\includegraphics[width=0.32\textwidth]{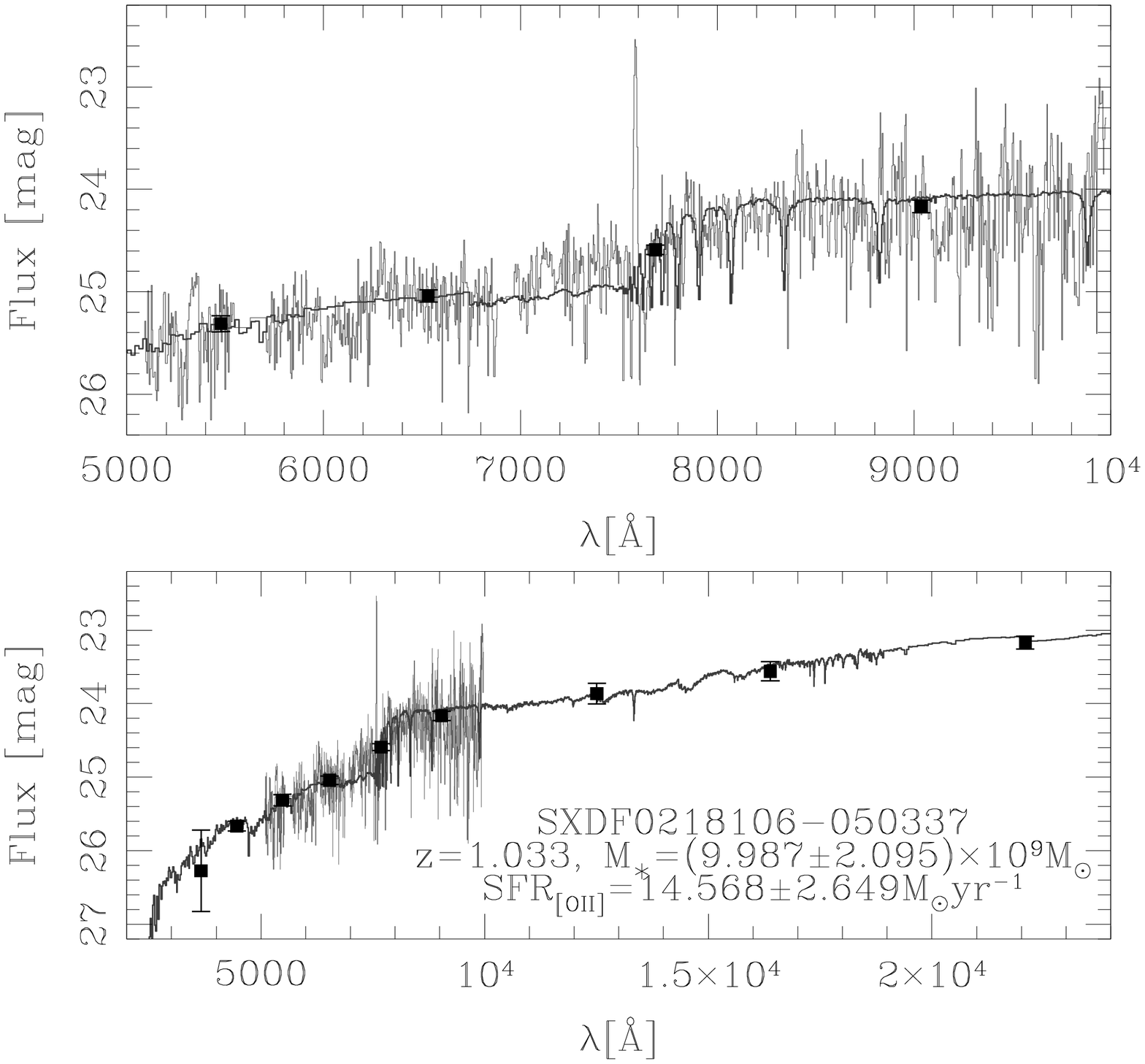}
\includegraphics[width=0.32\textwidth]{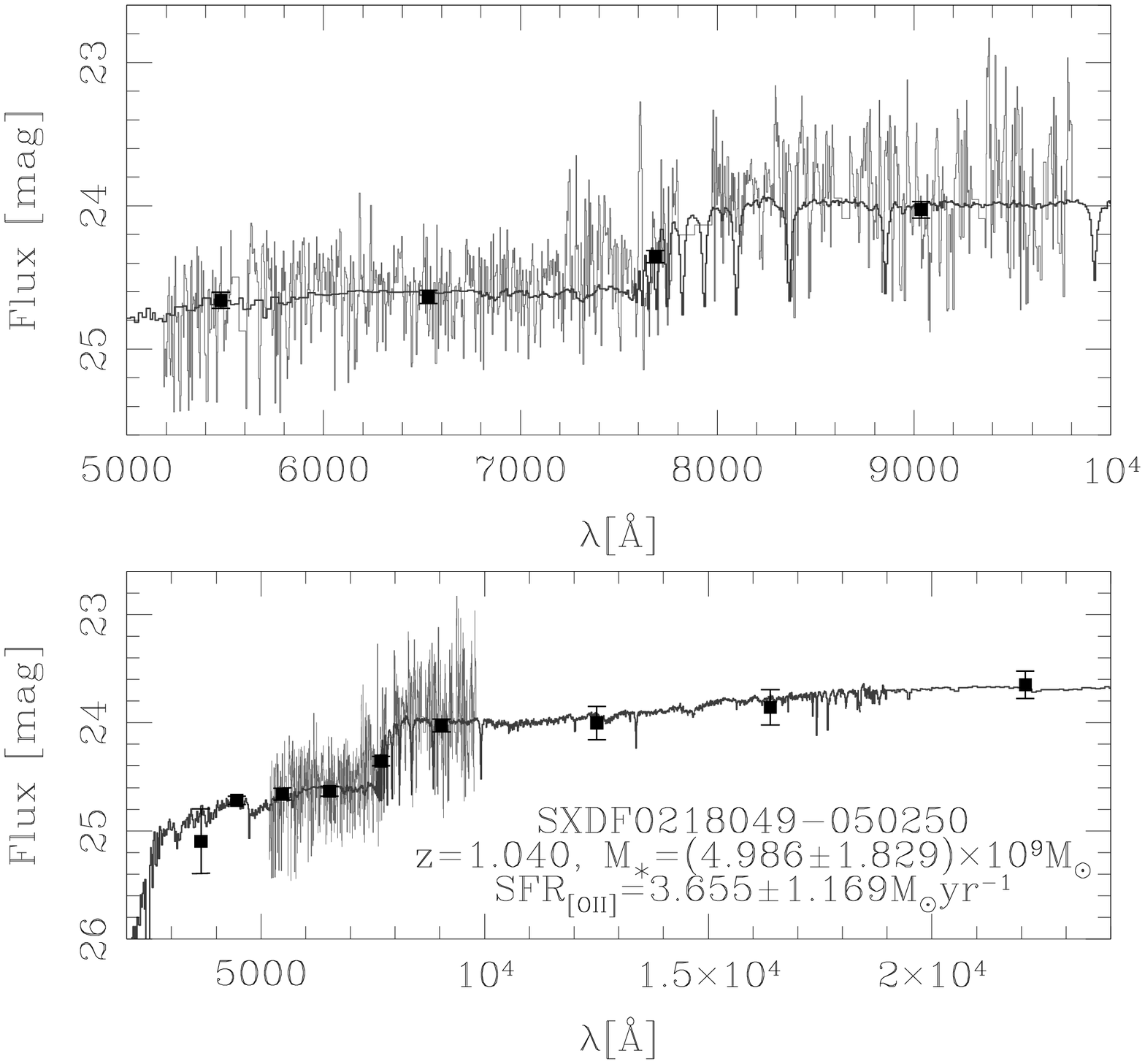}
\includegraphics[width=0.32\textwidth]{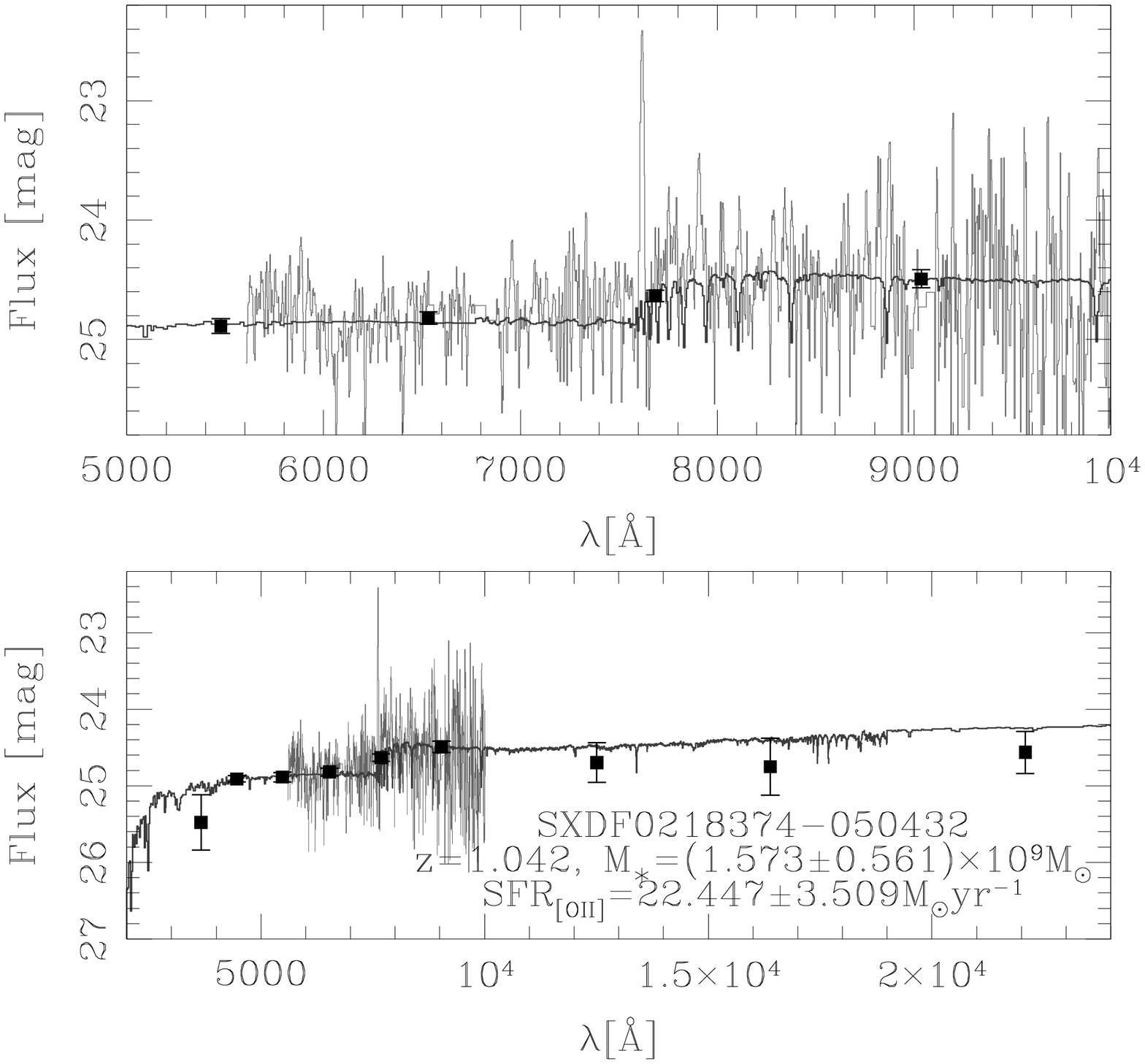}
\includegraphics[width=0.32\textwidth]{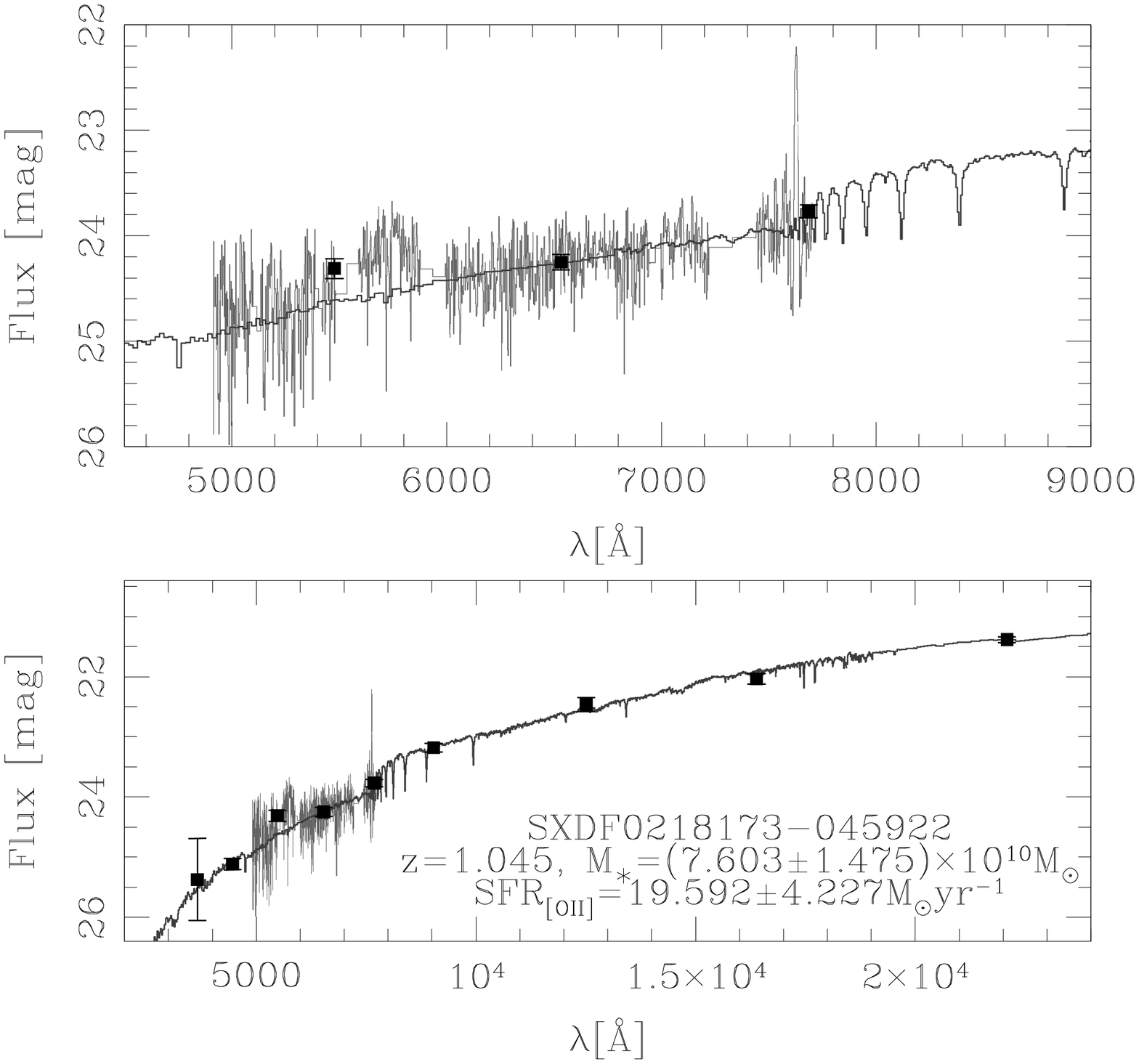}
\includegraphics[width=0.32\textwidth]{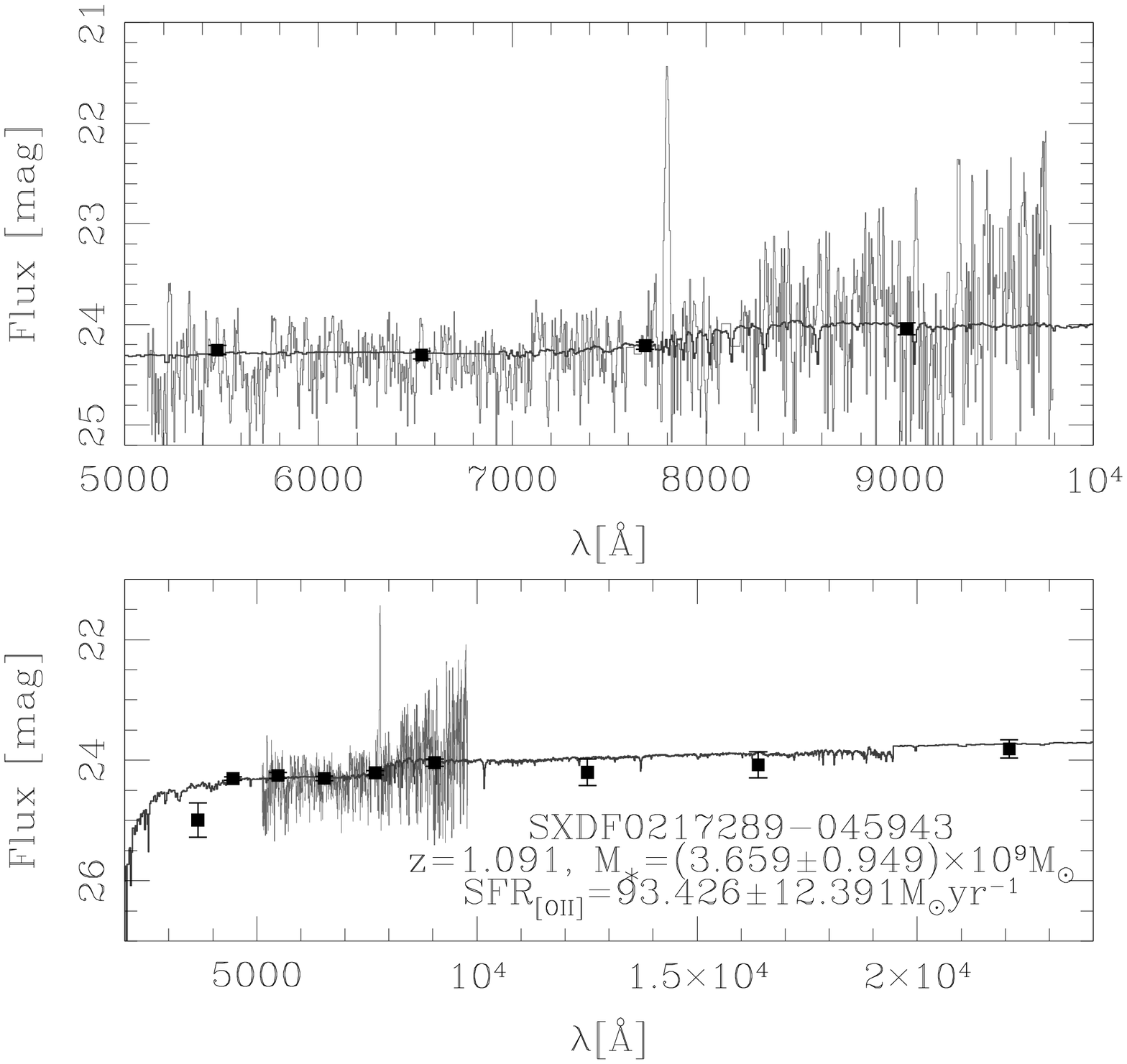}
\includegraphics[width=0.32\textwidth]{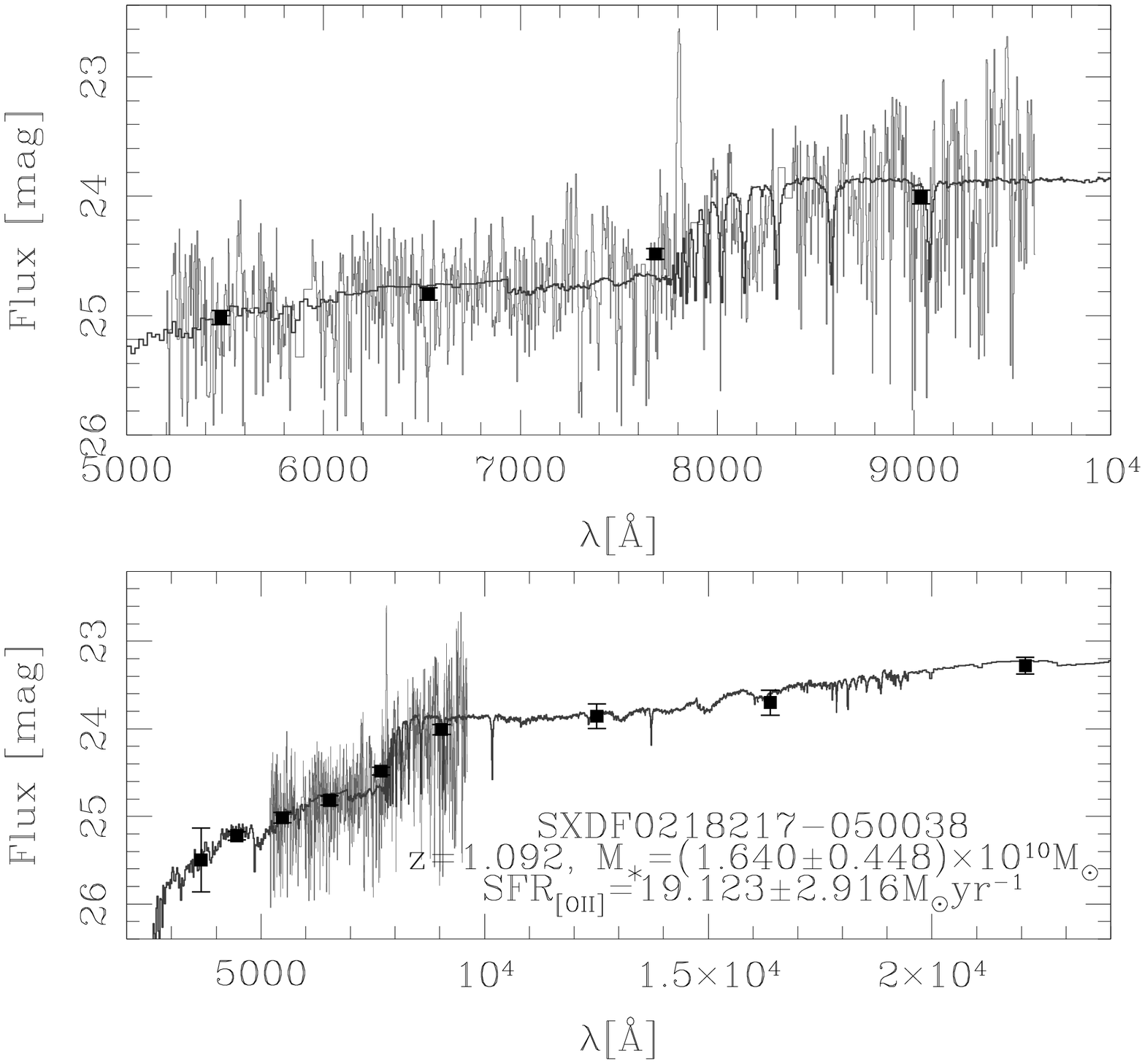}
\includegraphics[width=0.32\textwidth]{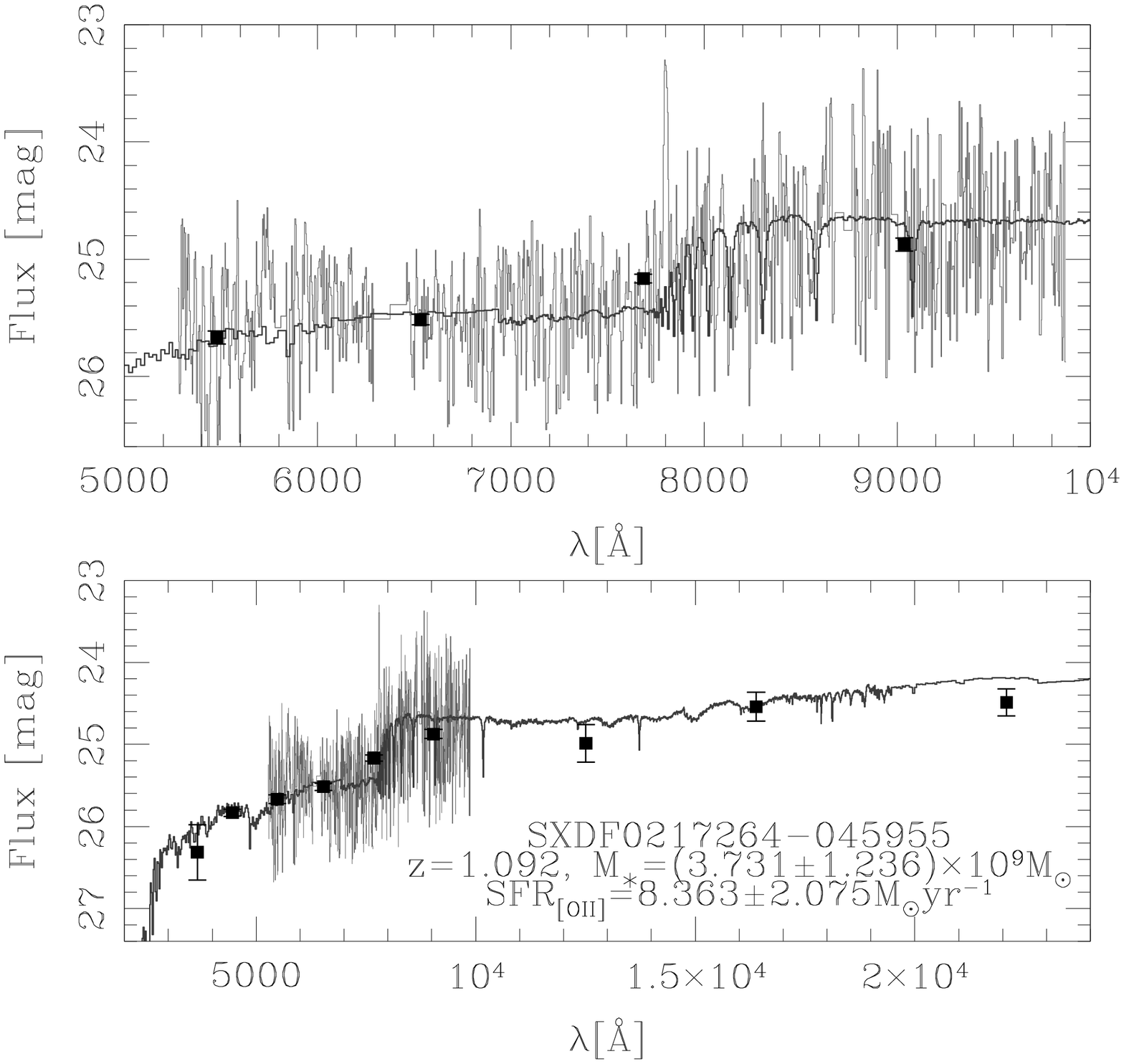}
\caption{\textit{- continued}}
\end{center}
\end{figure*}

\begin{figure*}
\addtocounter{figure}{-1}
\begin{center}
\includegraphics[width=0.32\textwidth]{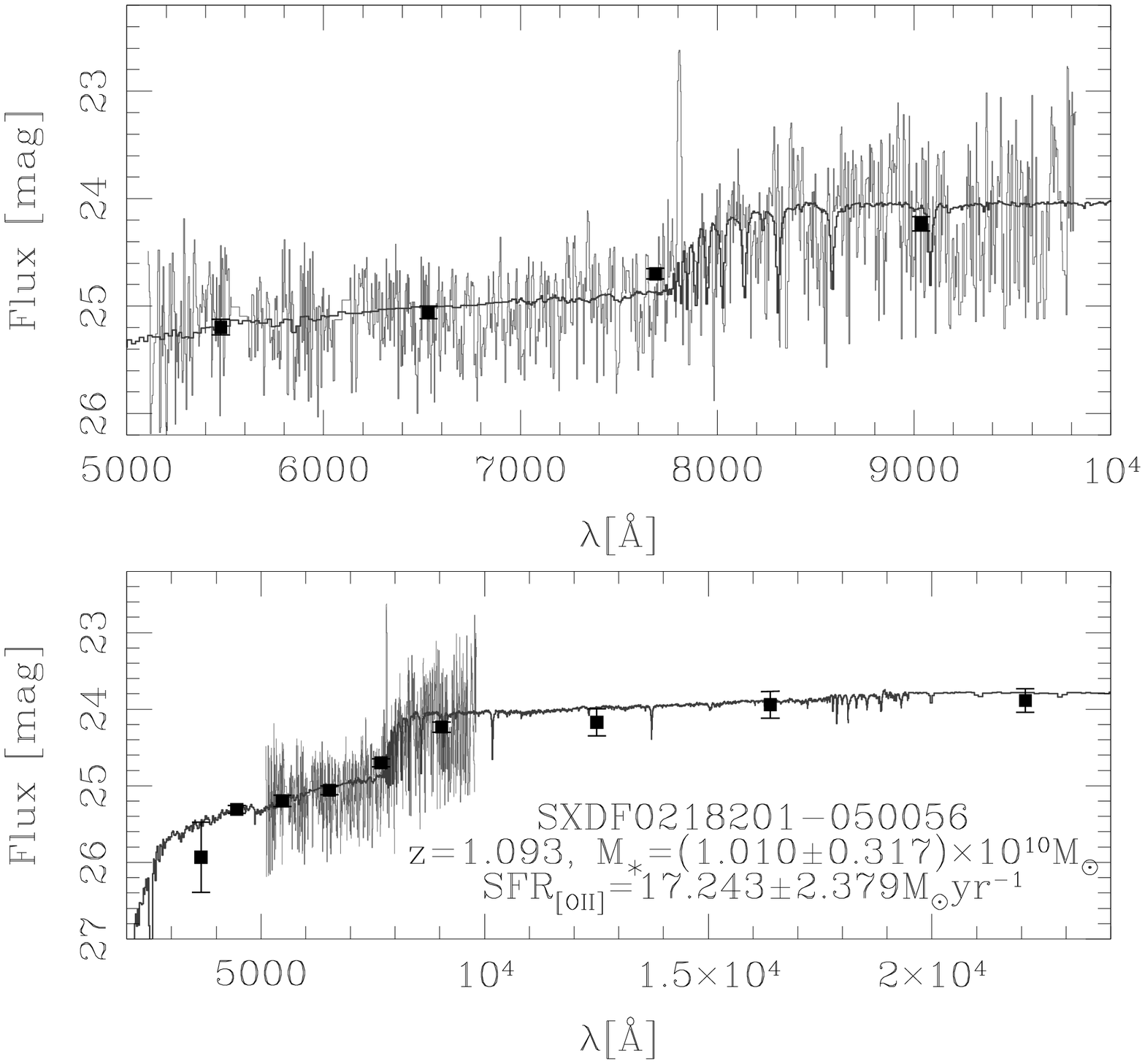}
\includegraphics[width=0.32\textwidth]{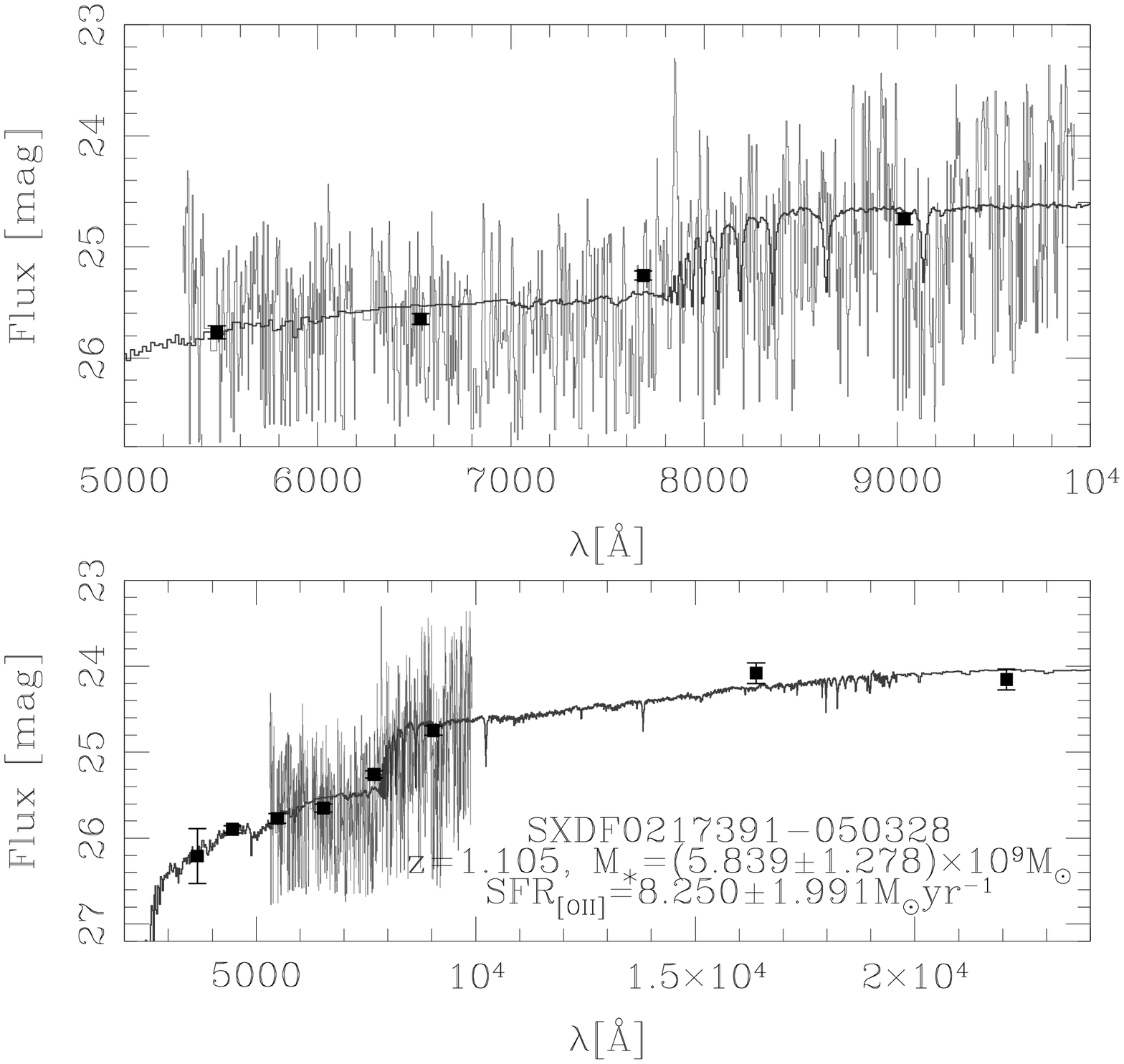}
\includegraphics[width=0.32\textwidth]{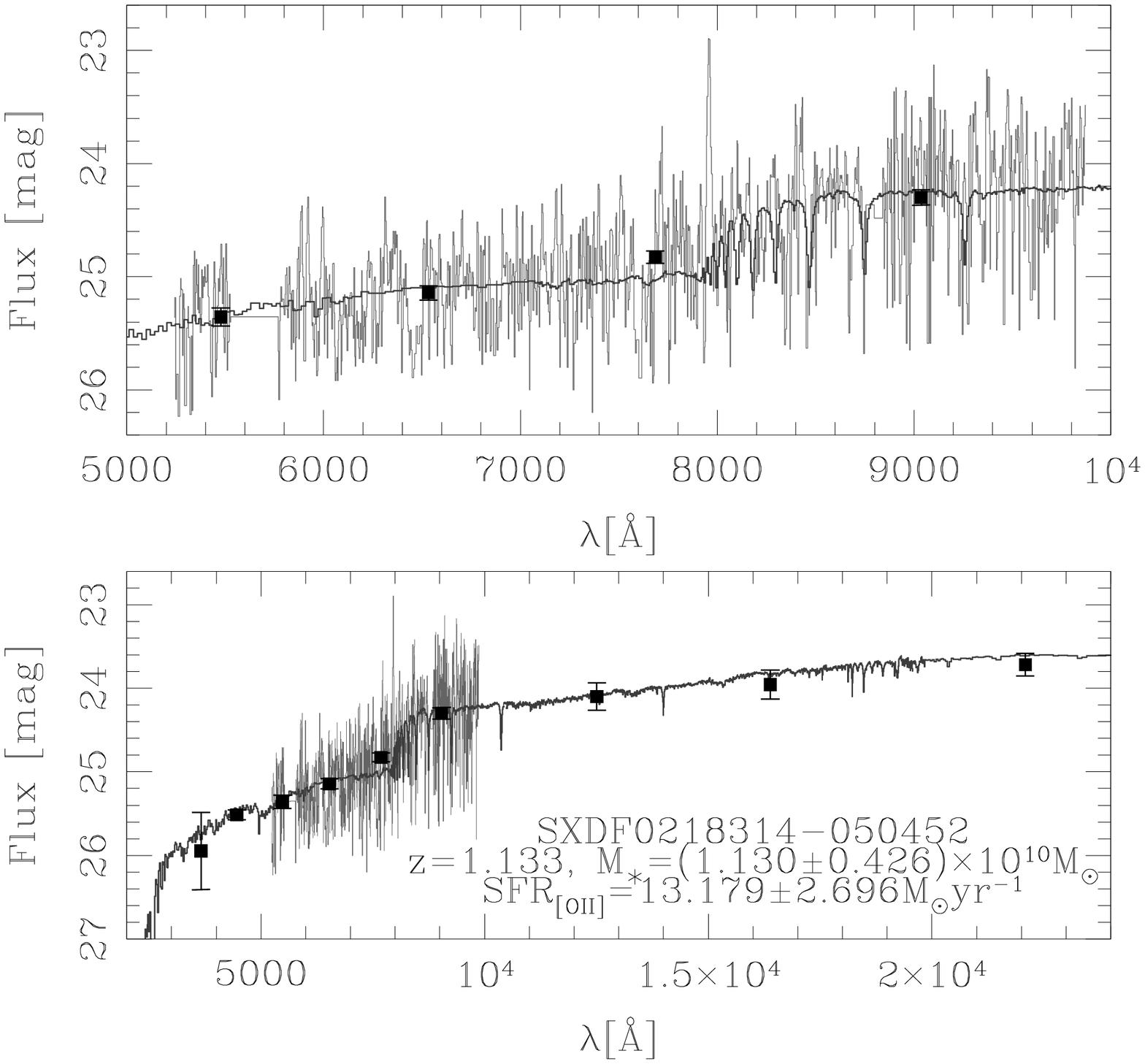}
\includegraphics[width=0.32\textwidth]{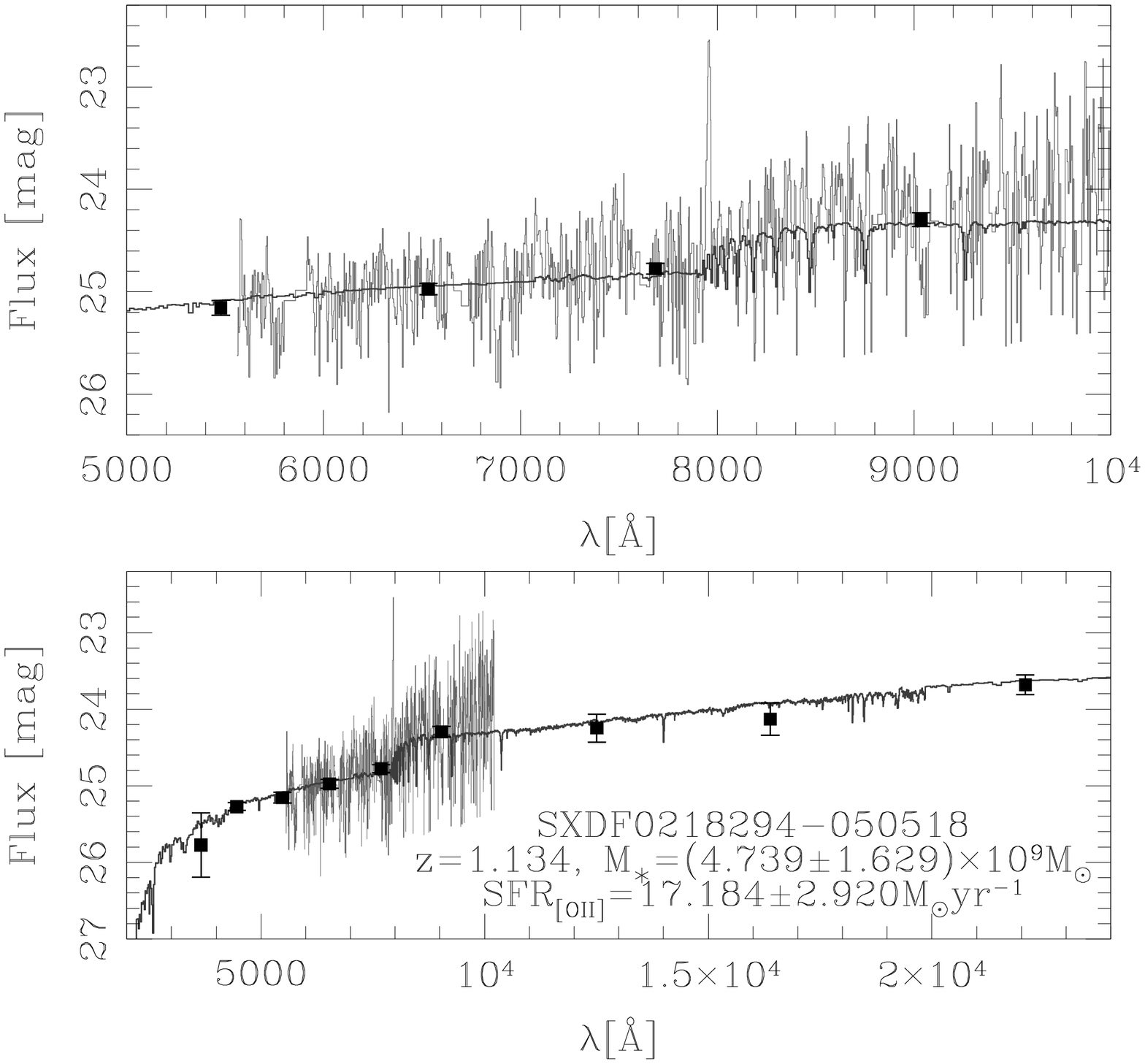}
\includegraphics[width=0.32\textwidth]{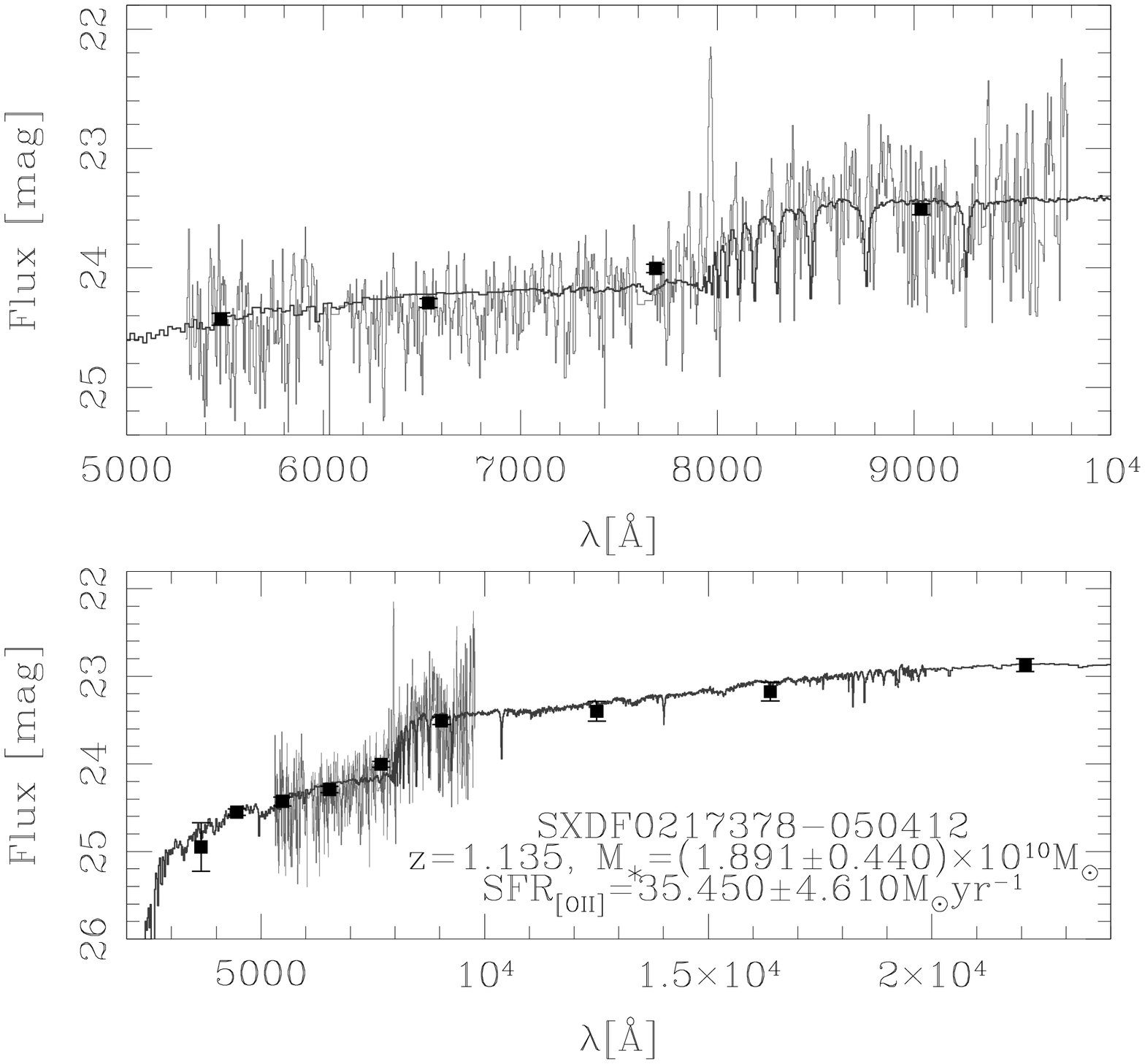}
\includegraphics[width=0.32\textwidth]{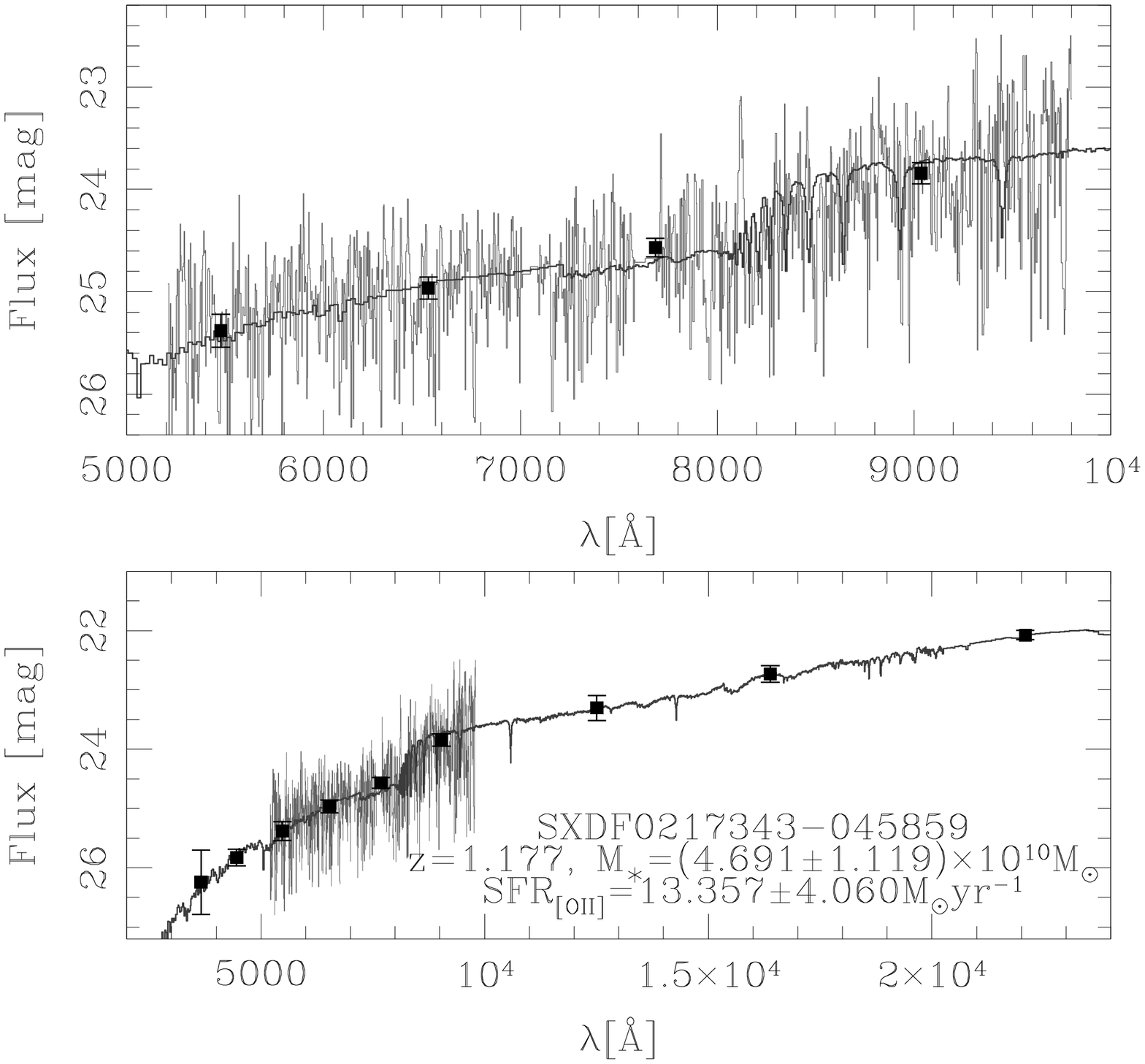}
\includegraphics[width=0.32\textwidth]{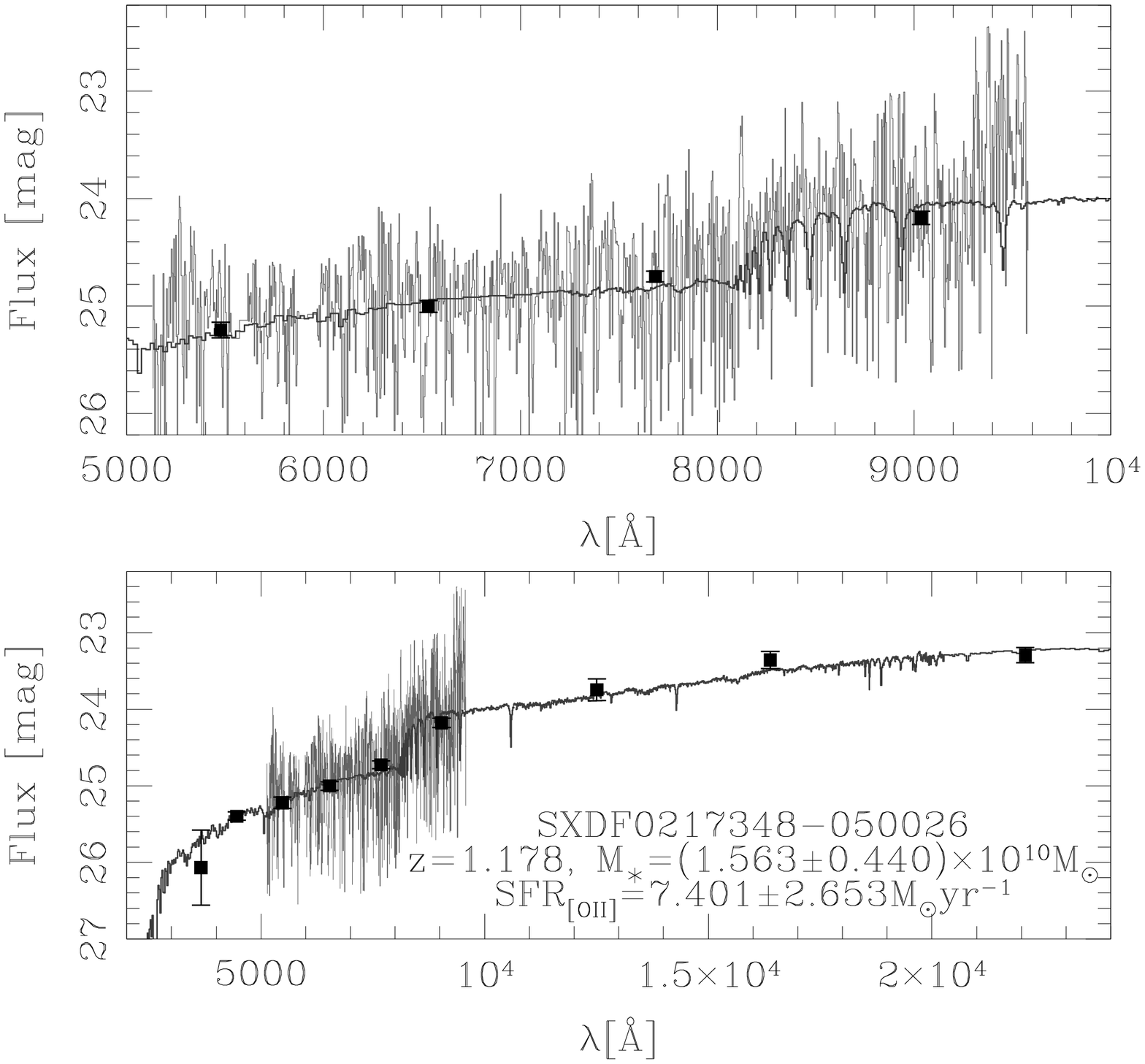}
\includegraphics[width=0.32\textwidth]{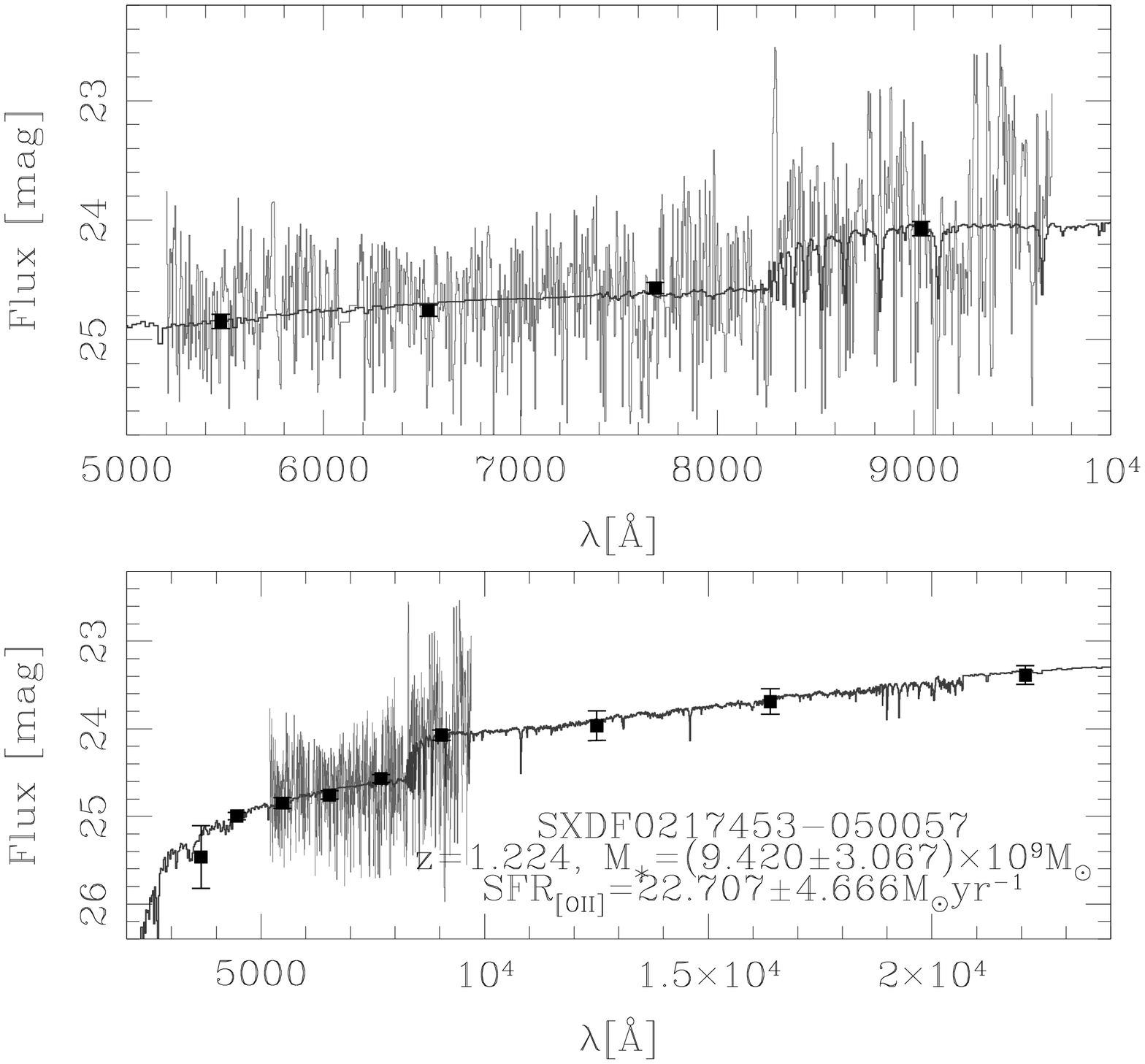}
\includegraphics[width=0.32\textwidth]{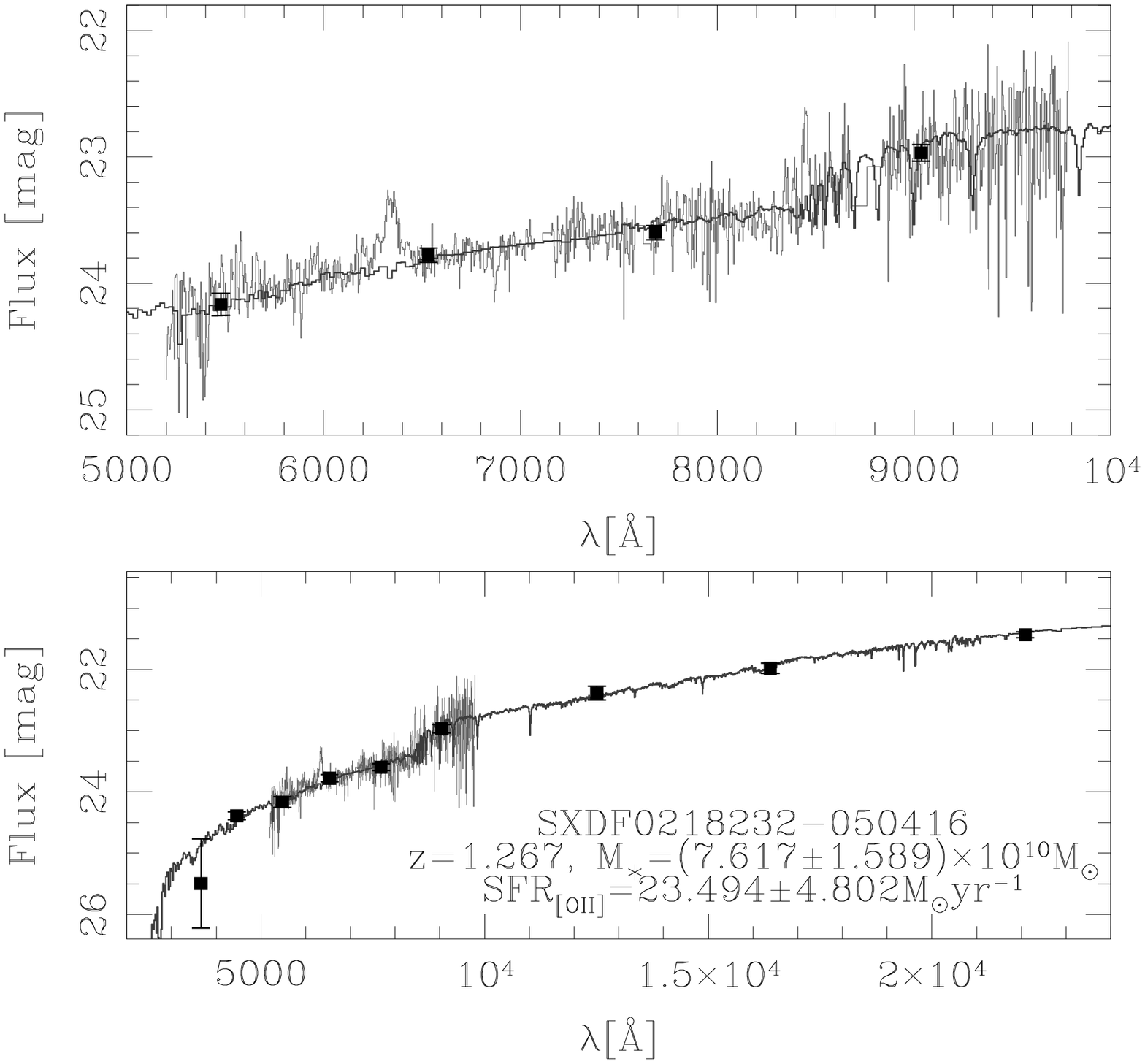}
\includegraphics[width=0.32\textwidth]{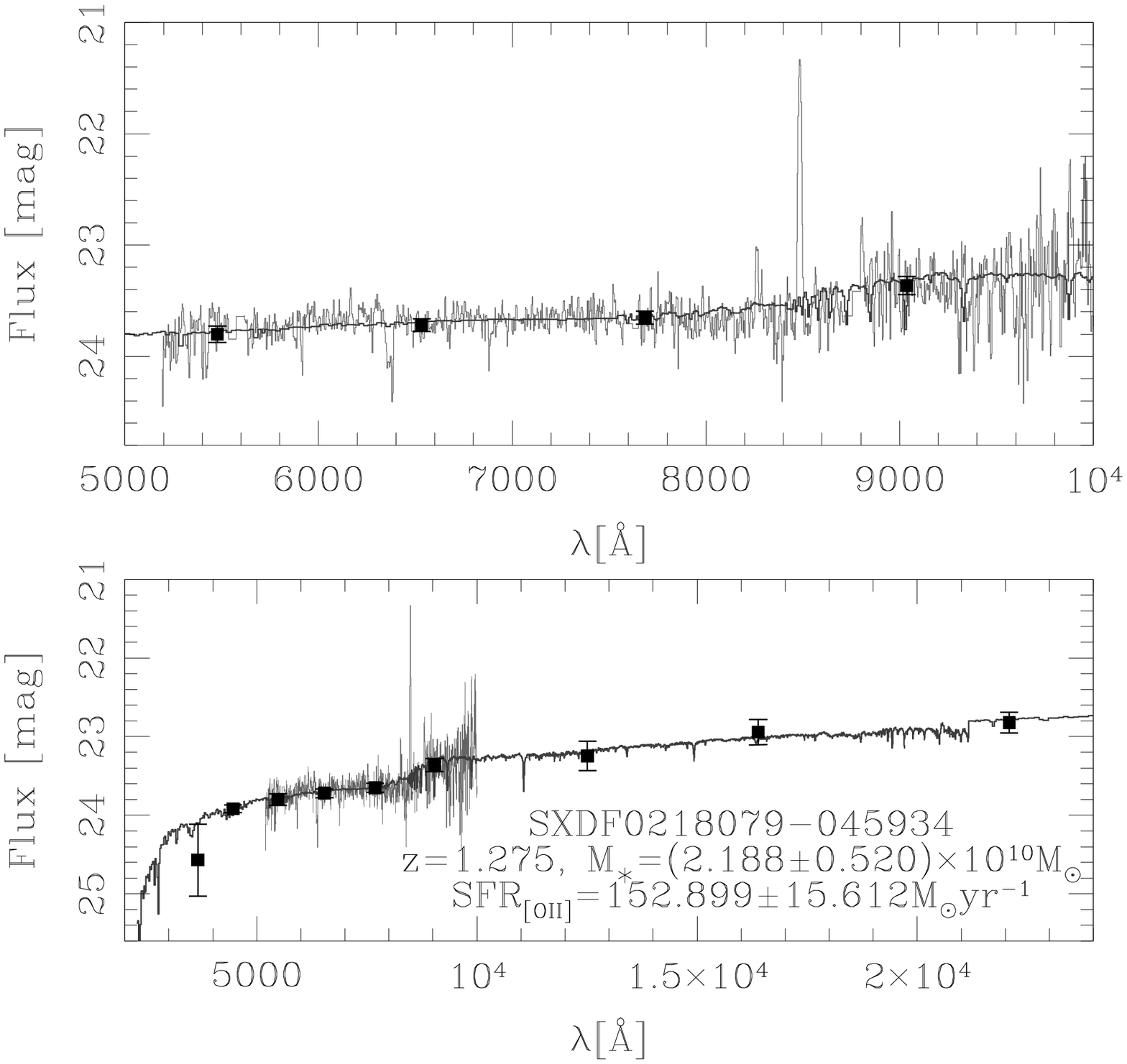}
\includegraphics[width=0.32\textwidth]{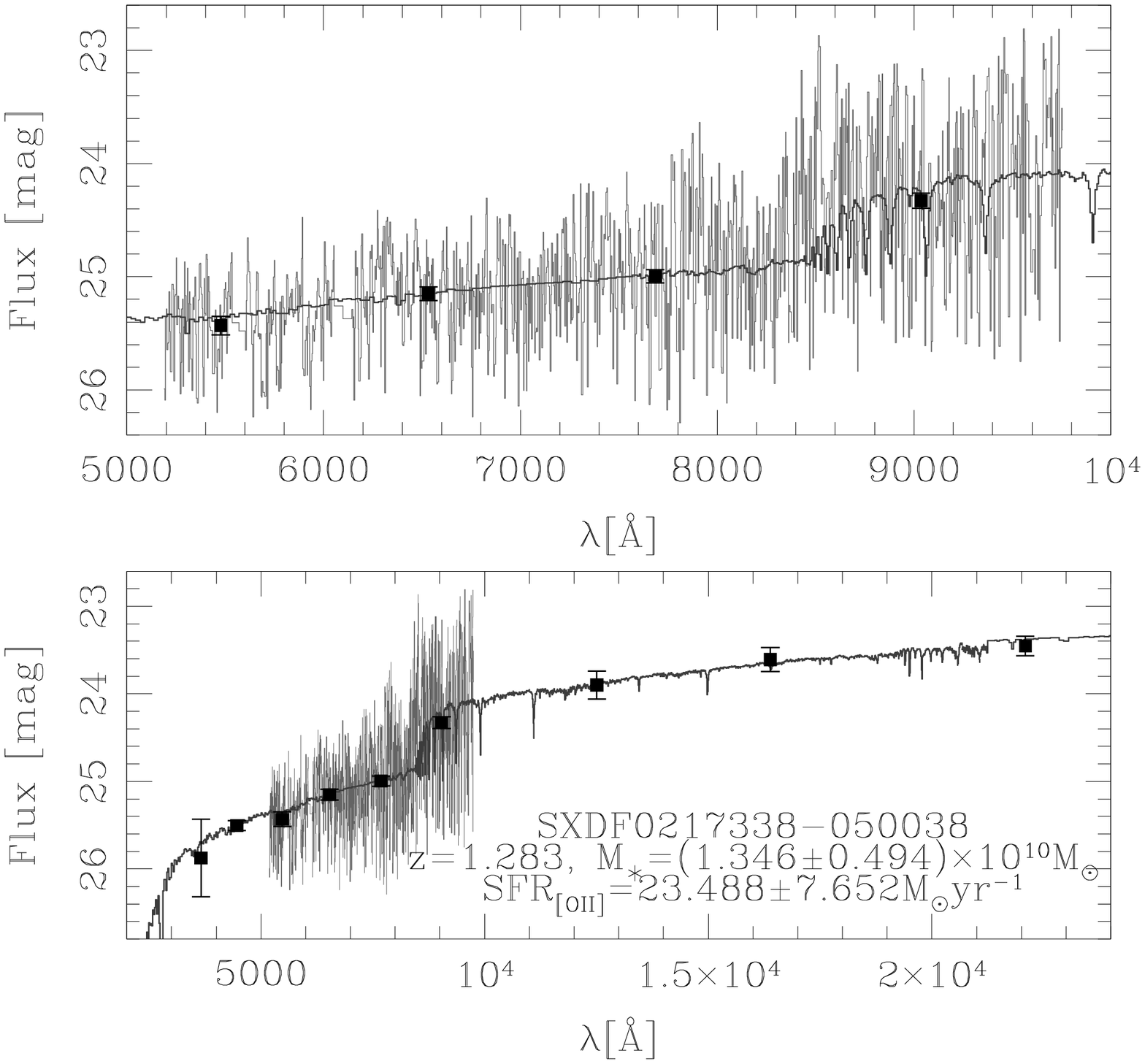}
\includegraphics[width=0.32\textwidth]{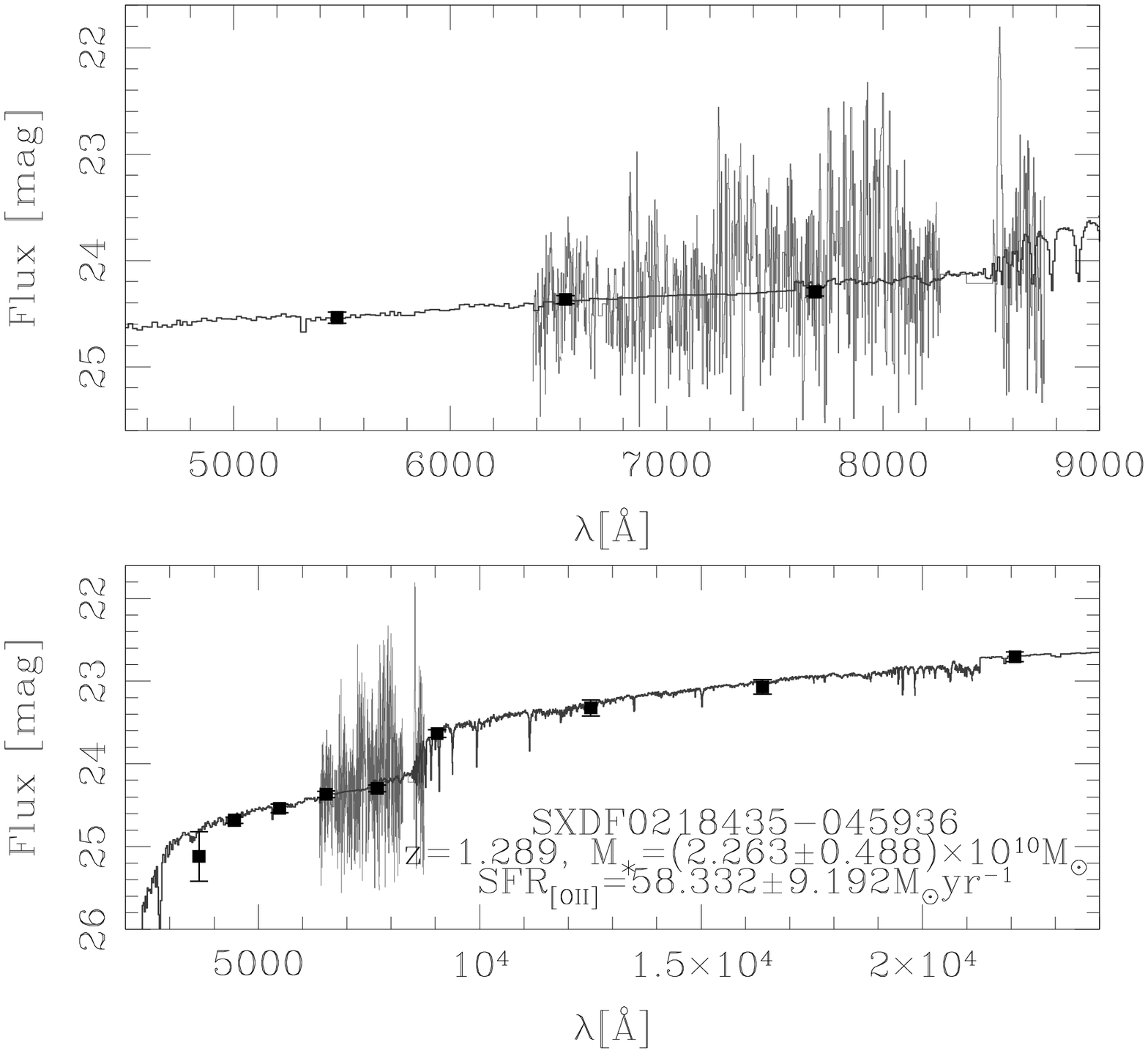}
\caption{\textit{- continued}}
\end{center}
\end{figure*}

\begin{figure*}
\addtocounter{figure}{-1}
\begin{center}
\includegraphics[width=0.32\textwidth]{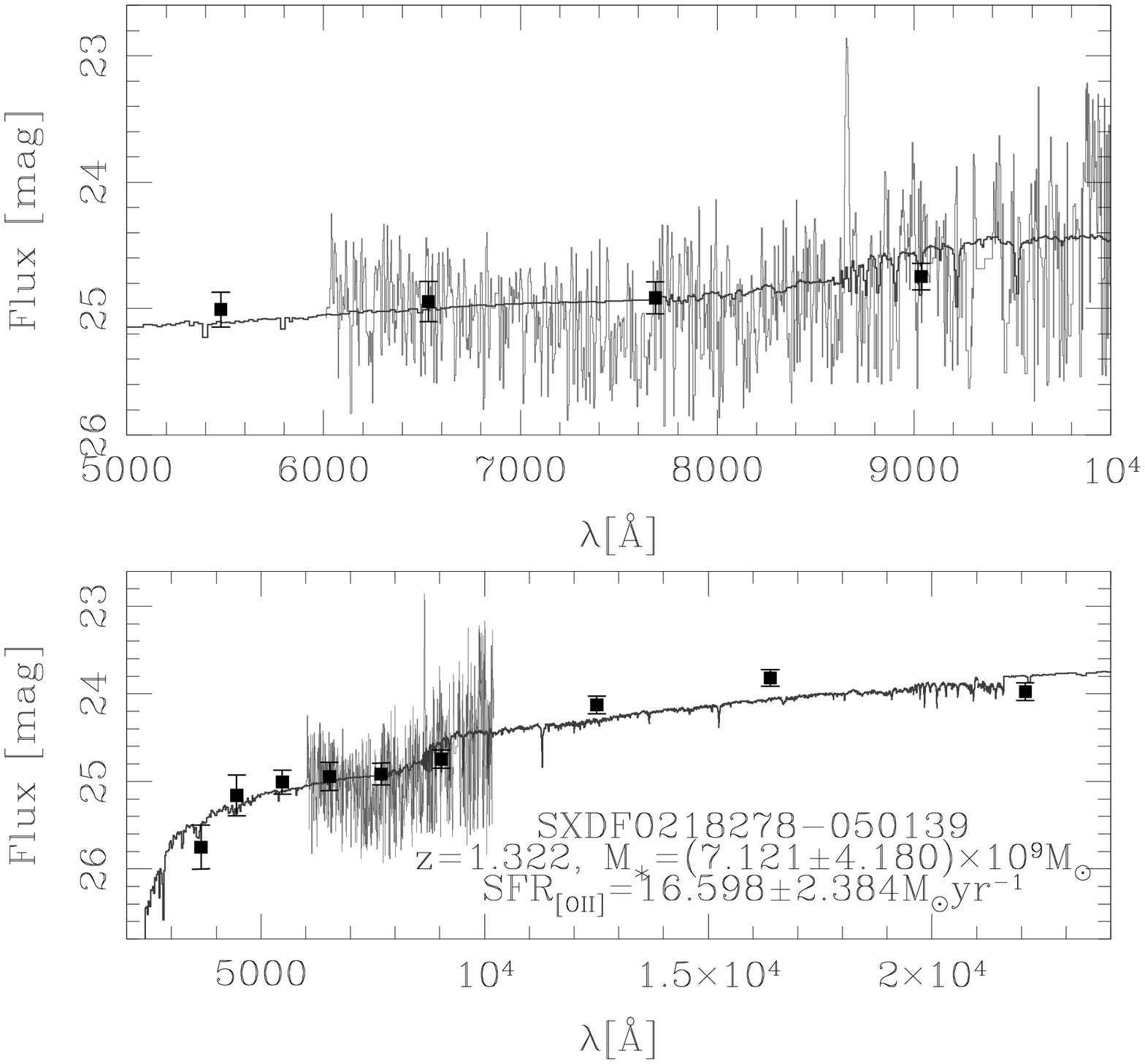}
\includegraphics[width=0.32\textwidth]{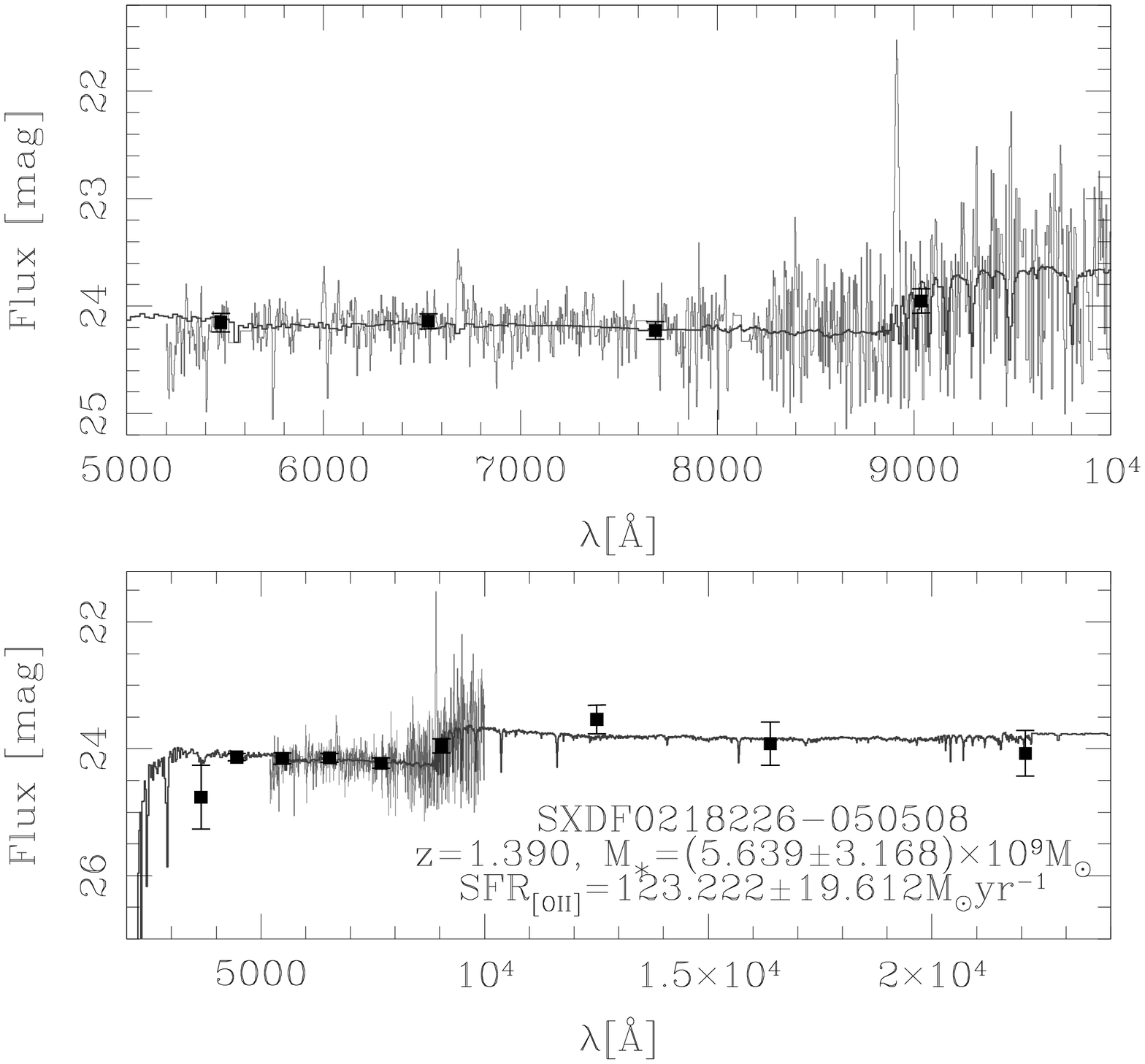}
\includegraphics[width=0.32\textwidth]{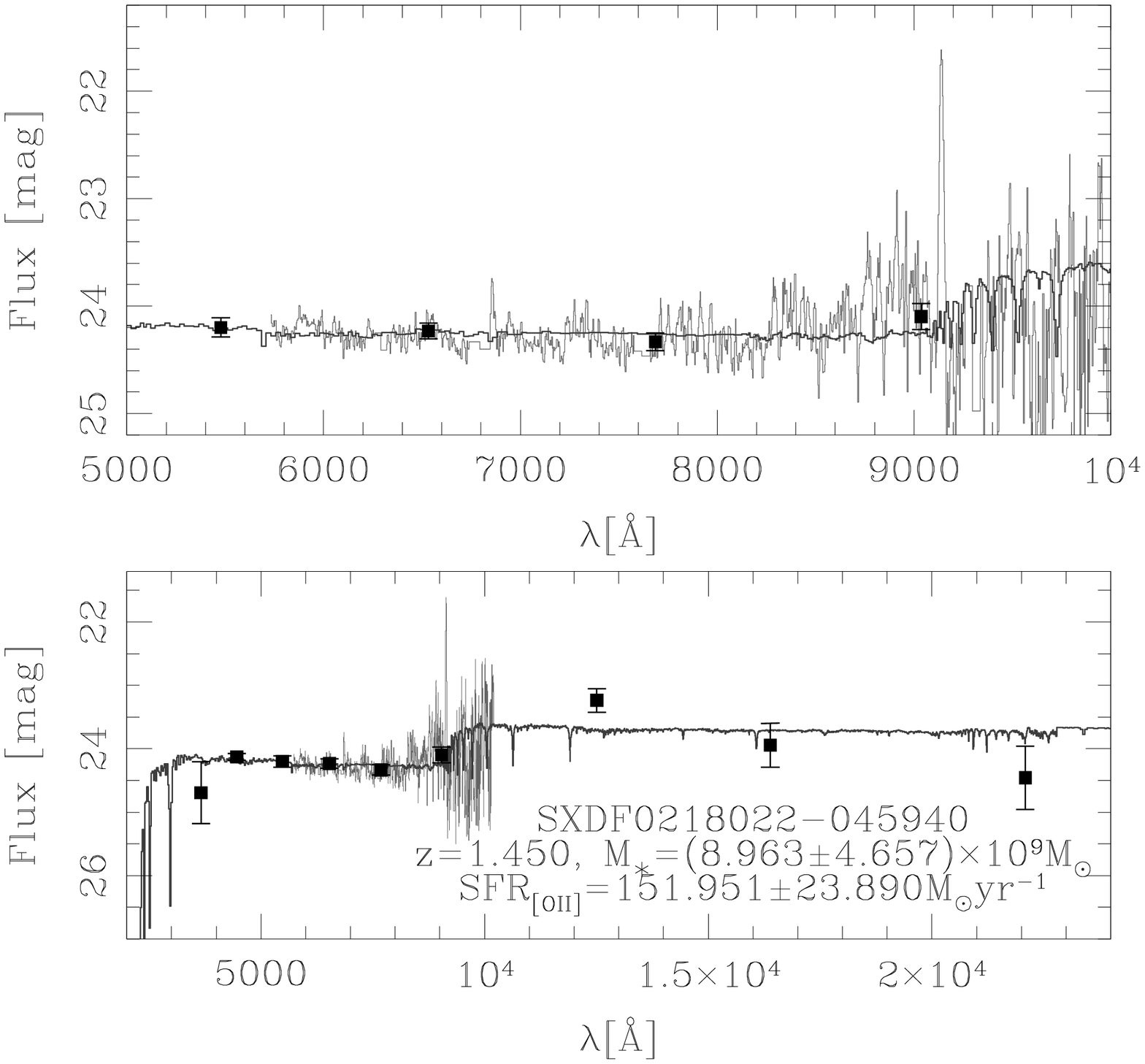}
\includegraphics[width=0.32\textwidth]{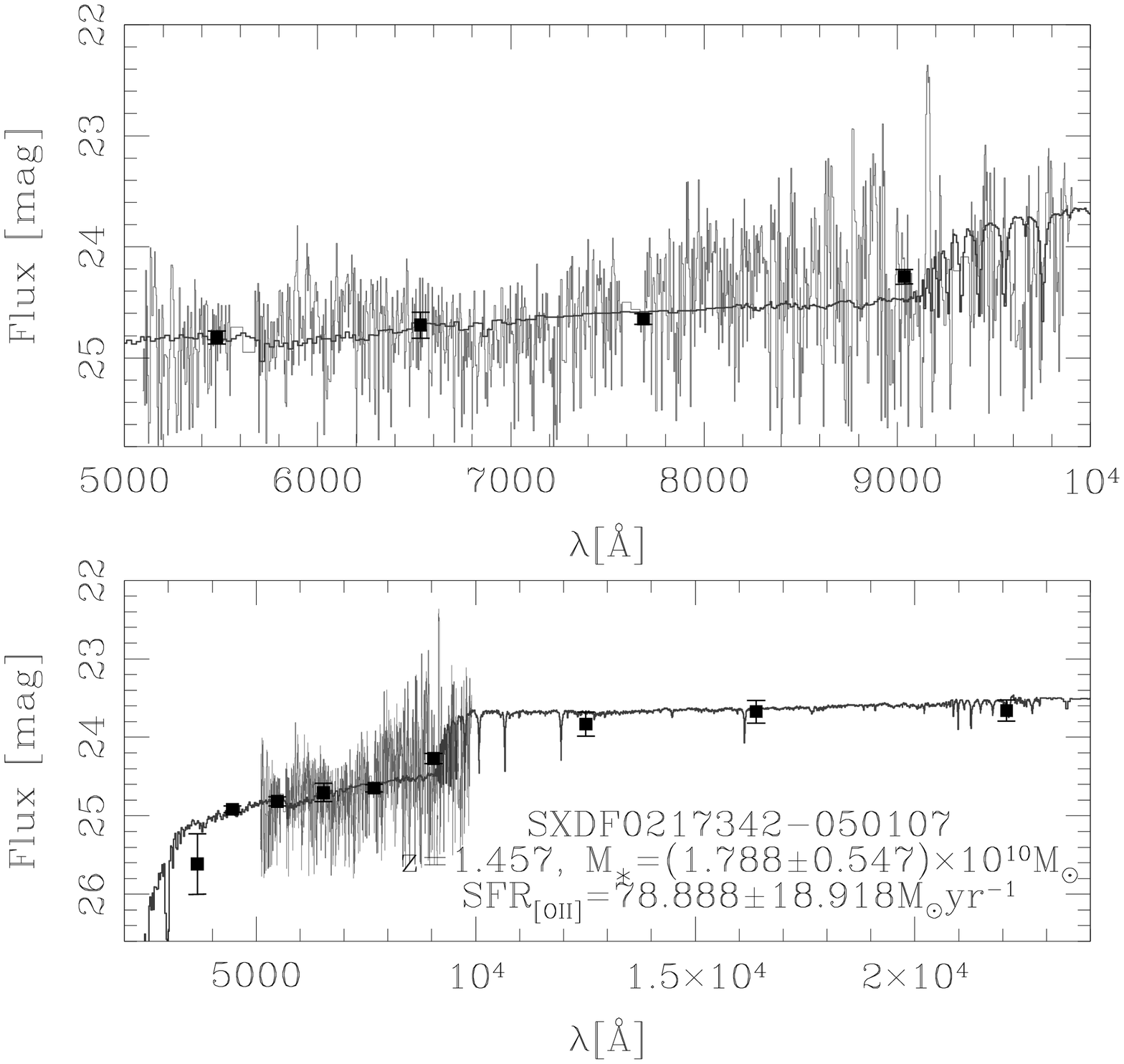}
\includegraphics[width=0.32\textwidth]{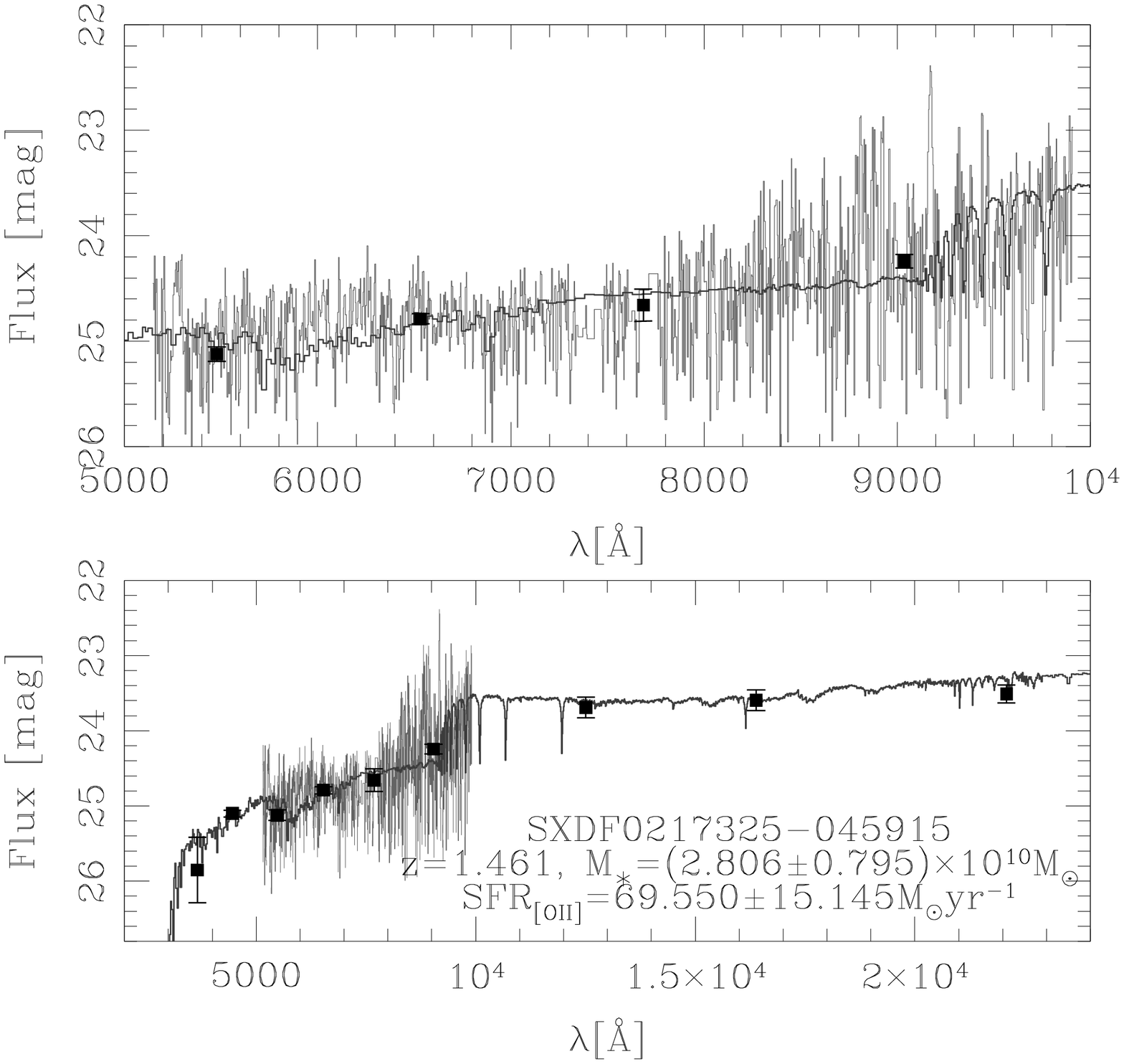}
\includegraphics[width=0.32\textwidth]{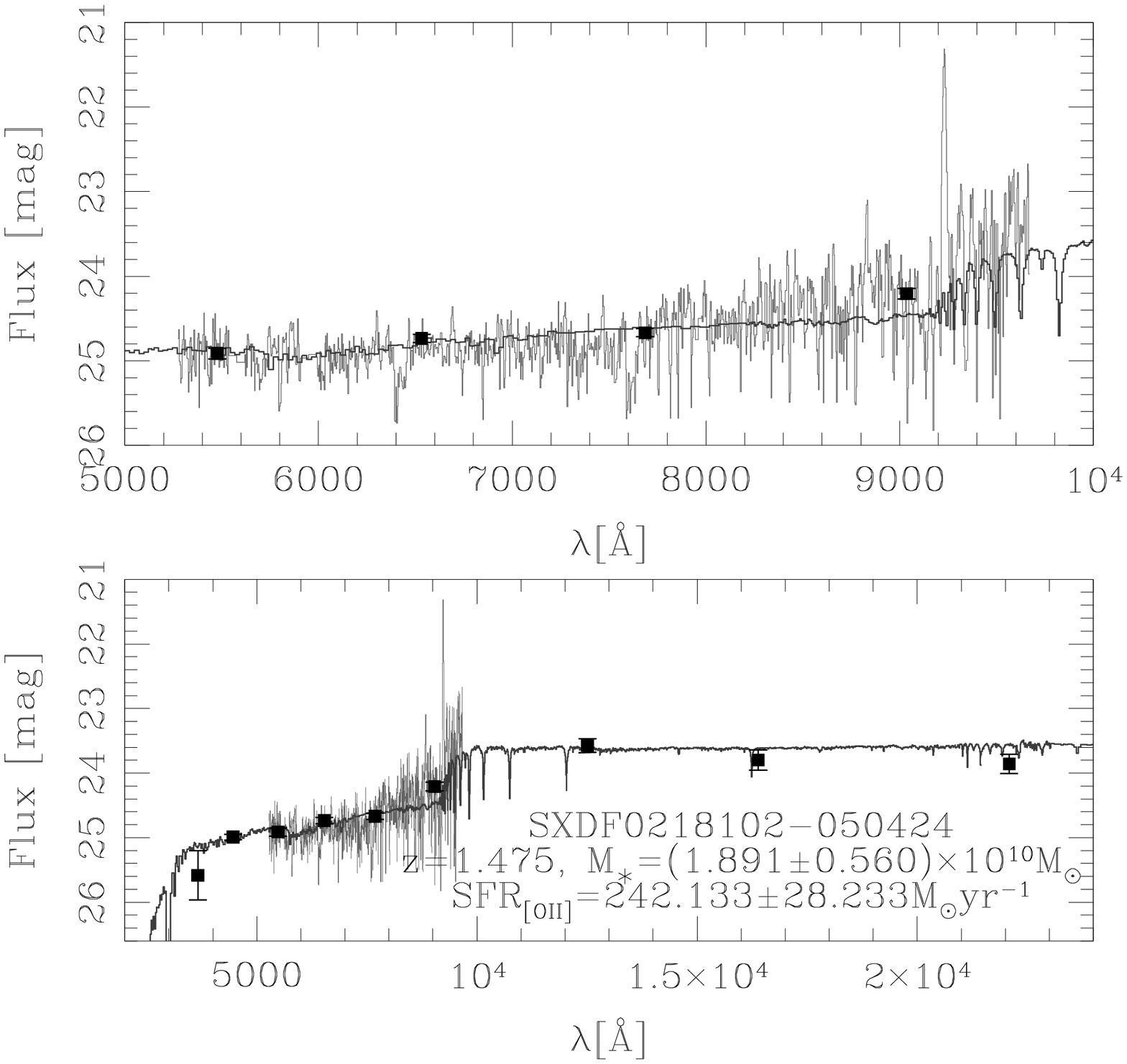}
\includegraphics[width=0.32\textwidth]{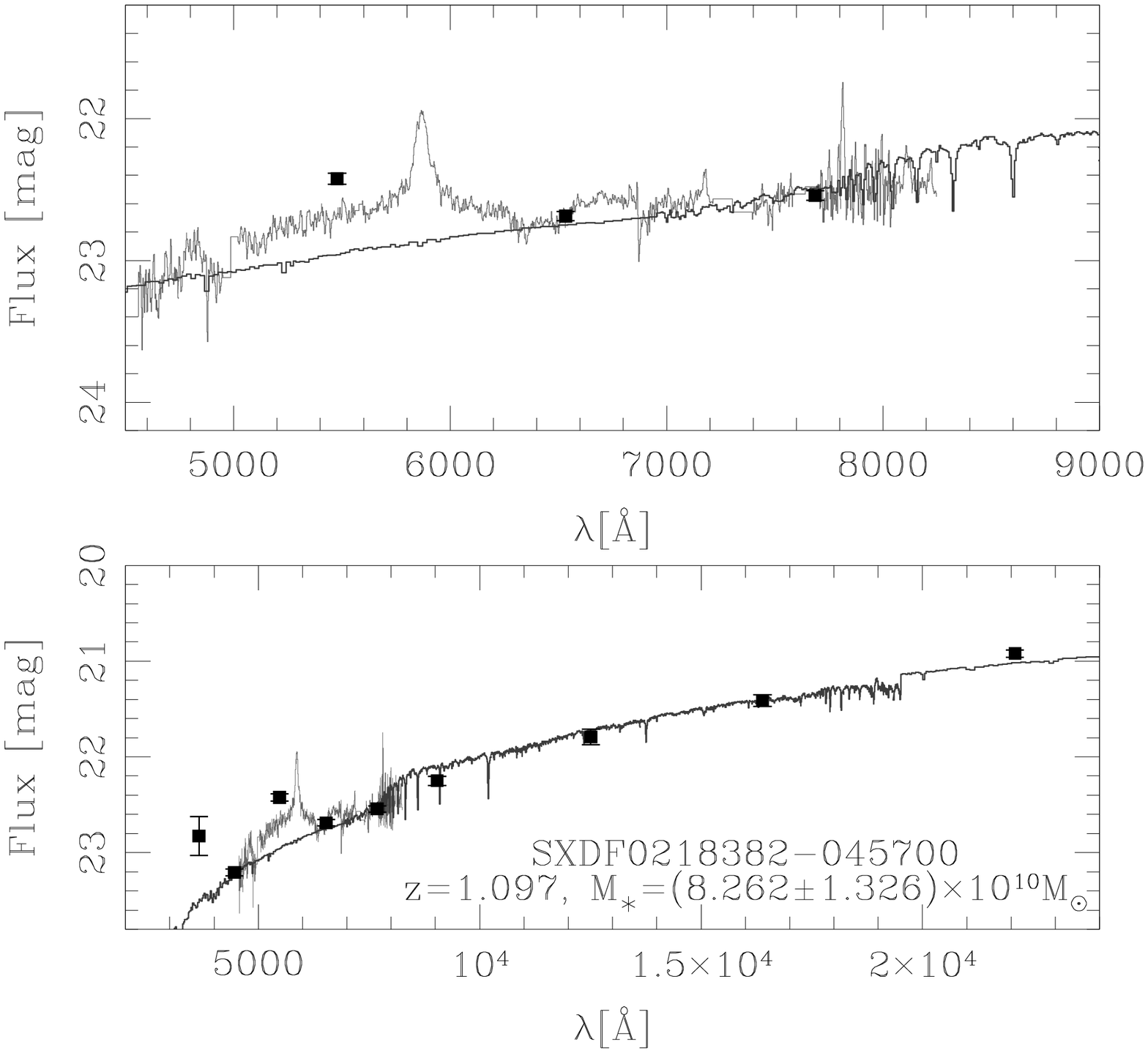}
\includegraphics[width=0.32\textwidth]{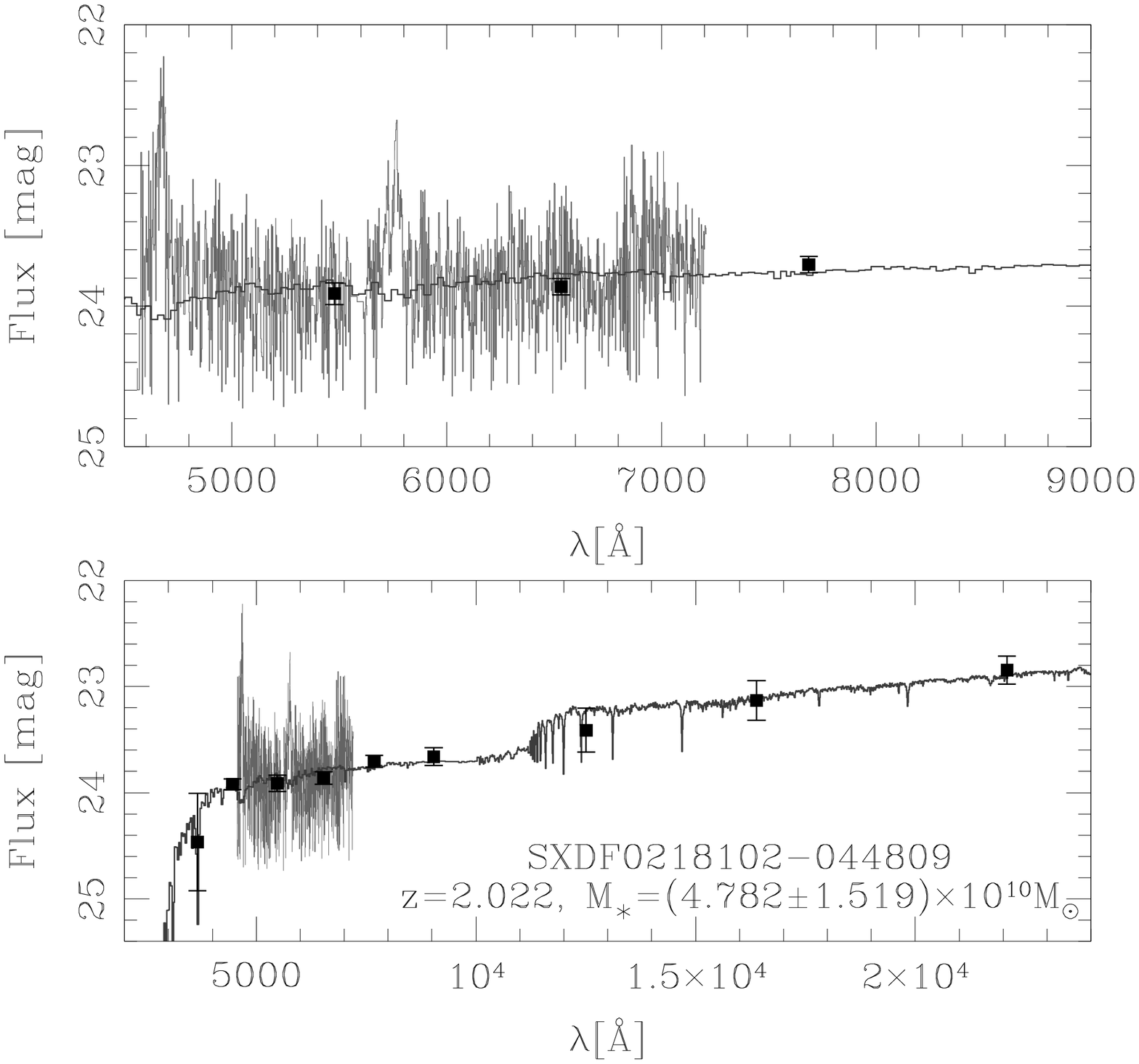}
\includegraphics[width=0.32\textwidth]{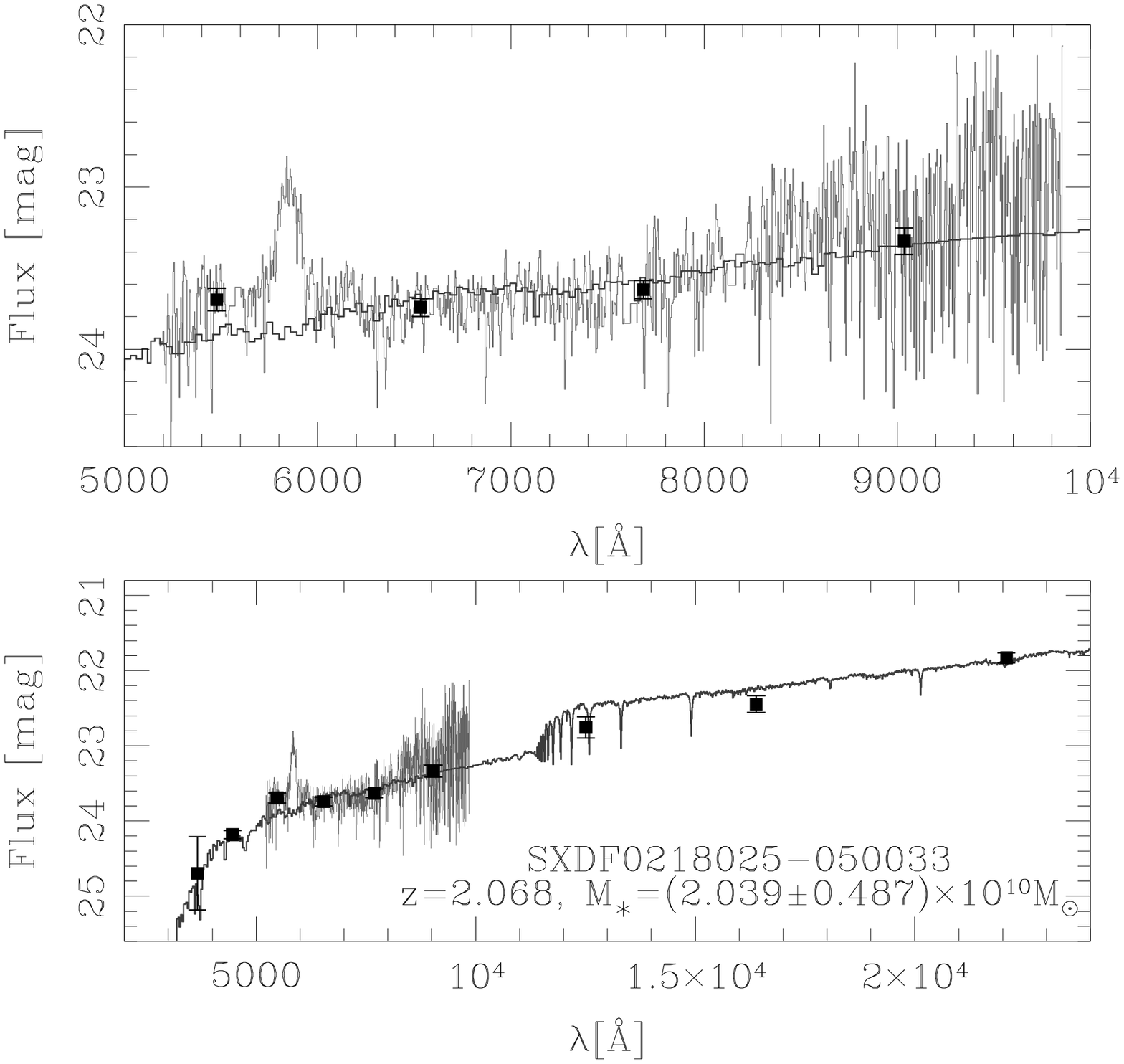}
\includegraphics[width=0.32\textwidth]{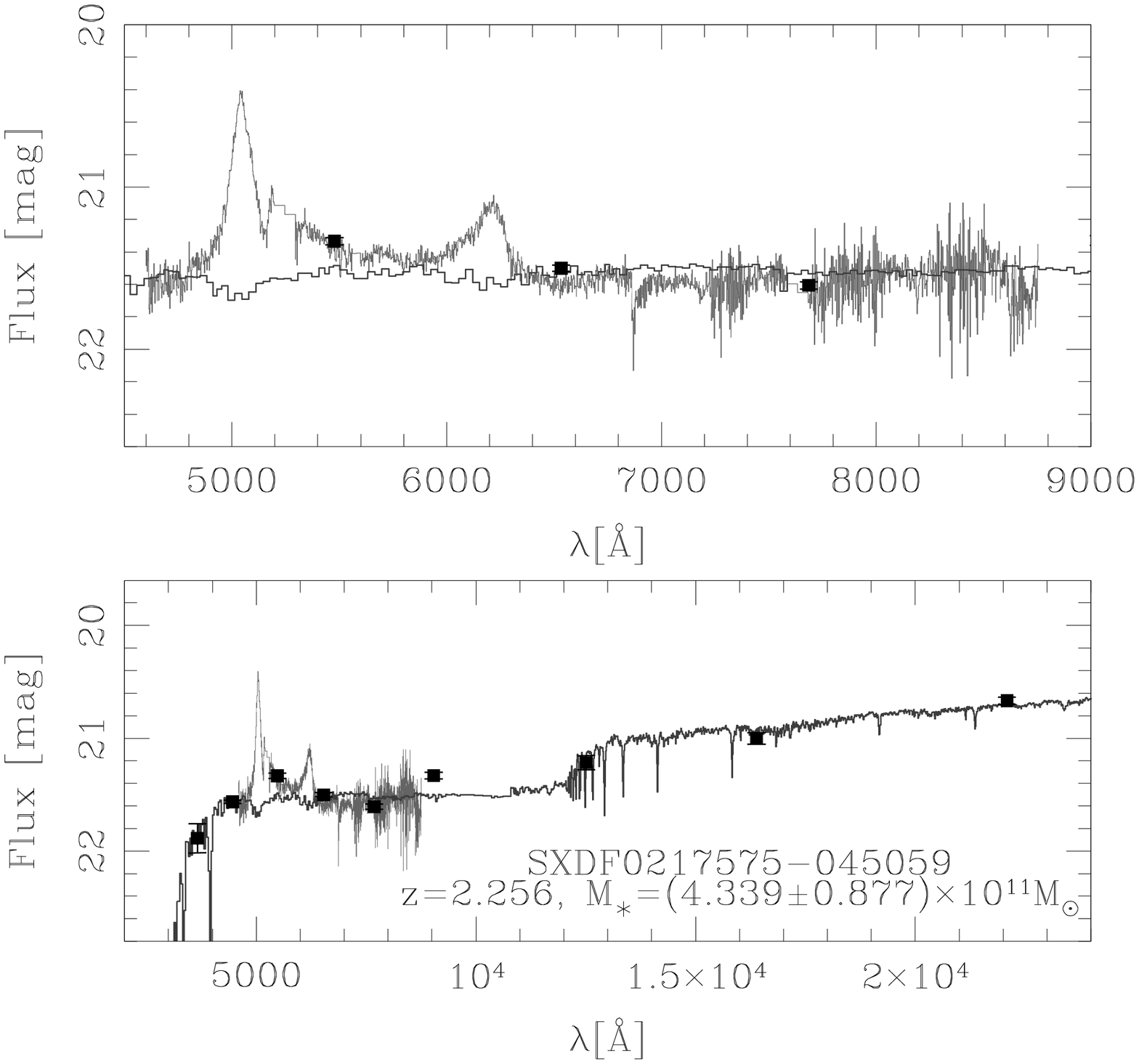}
\includegraphics[width=0.32\textwidth]{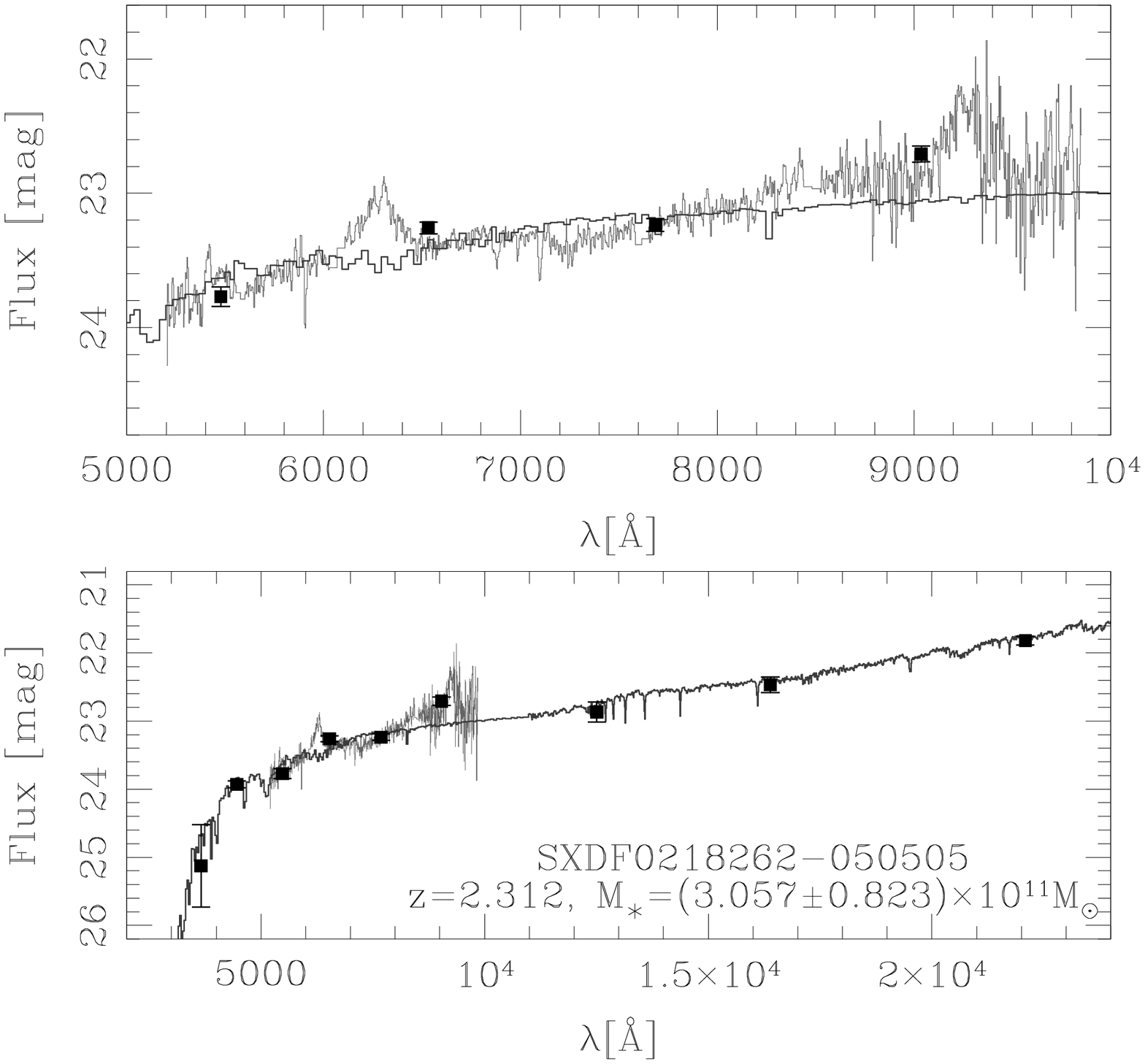}
\includegraphics[width=0.32\textwidth]{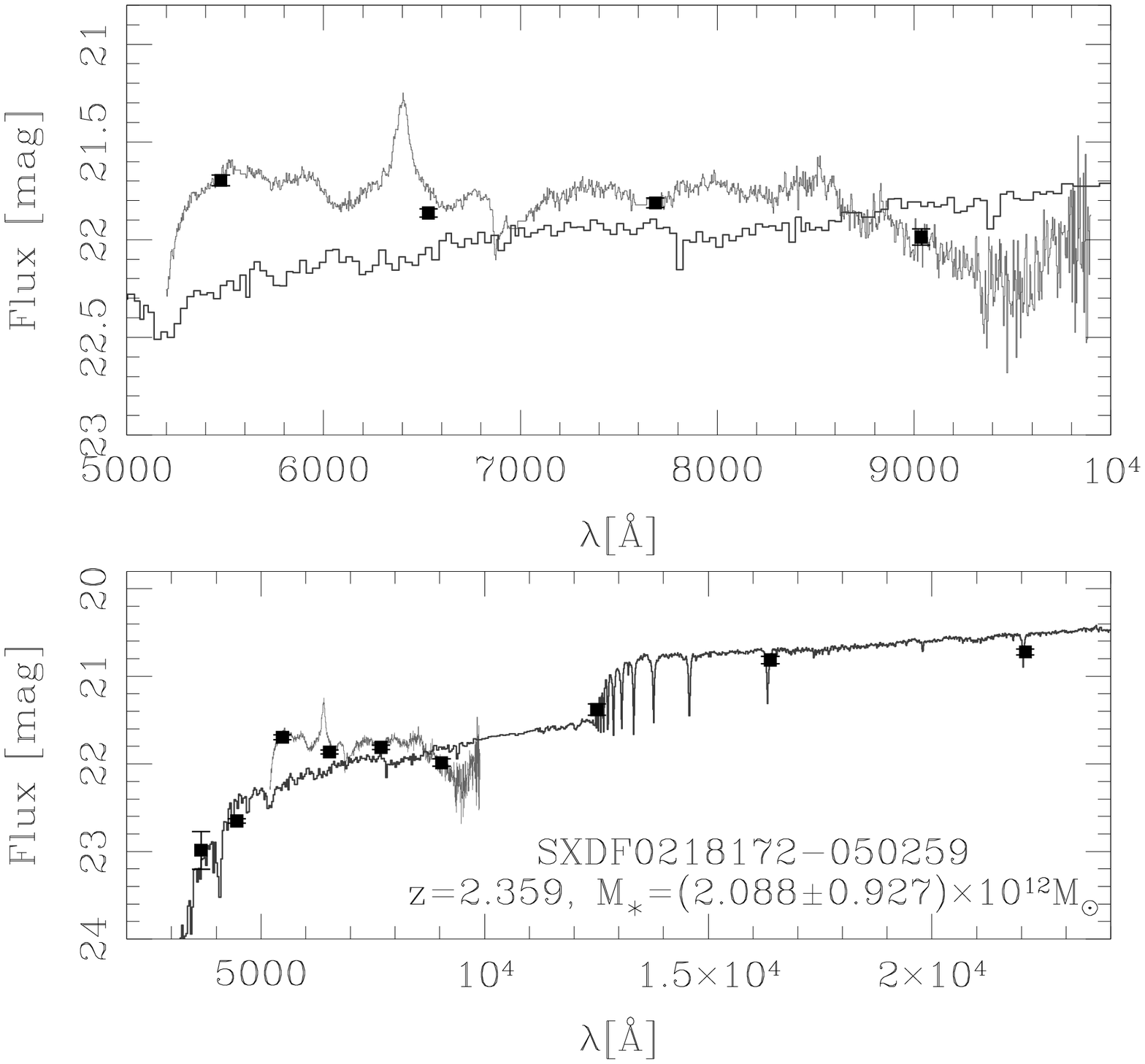}
\caption{\textit{- continued}}
\end{center}
\end{figure*}

\begin{figure*}
\addtocounter{figure}{-1}
\begin{center}
\includegraphics[width=0.32\textwidth]{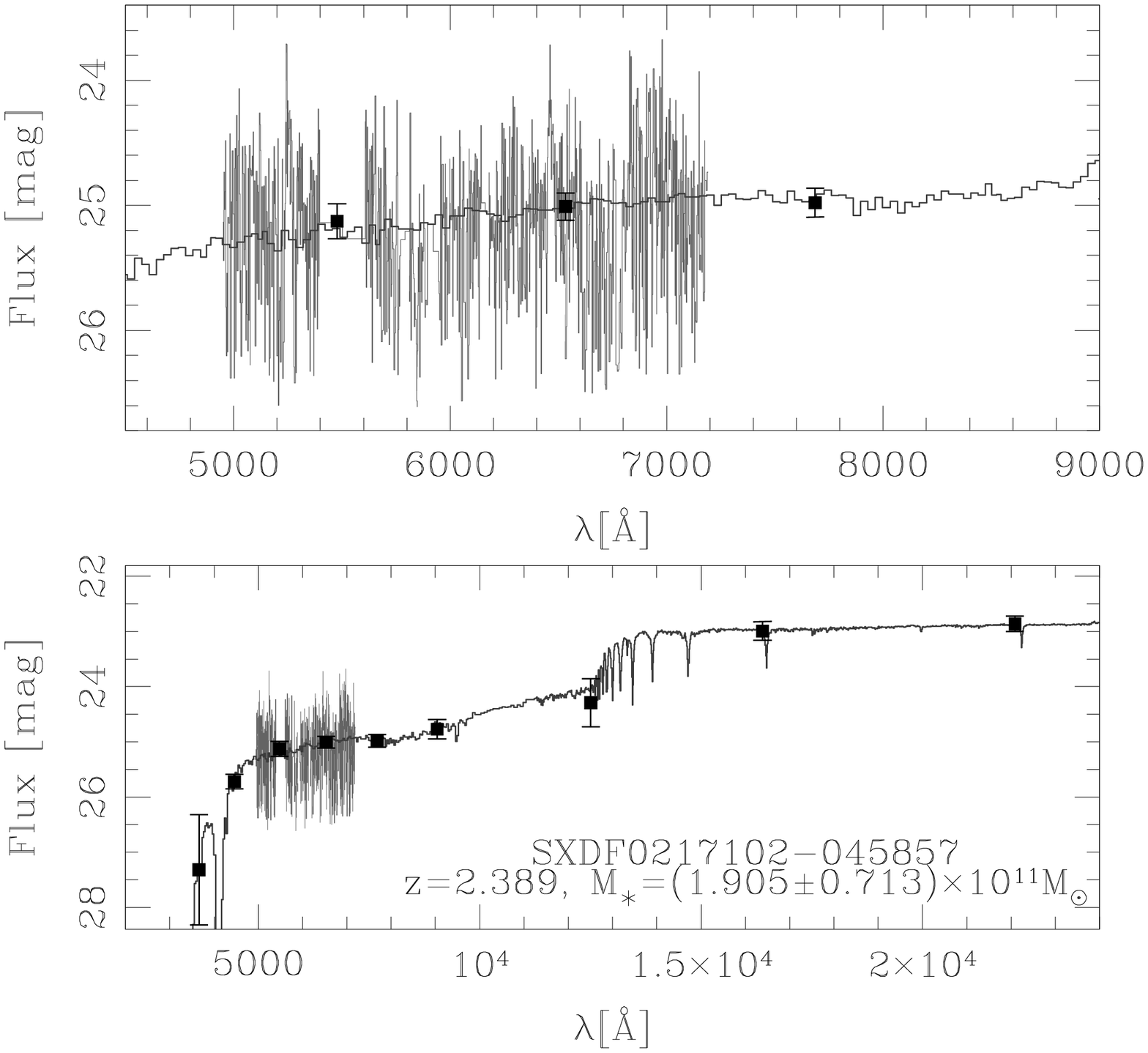}
\includegraphics[width=0.32\textwidth]{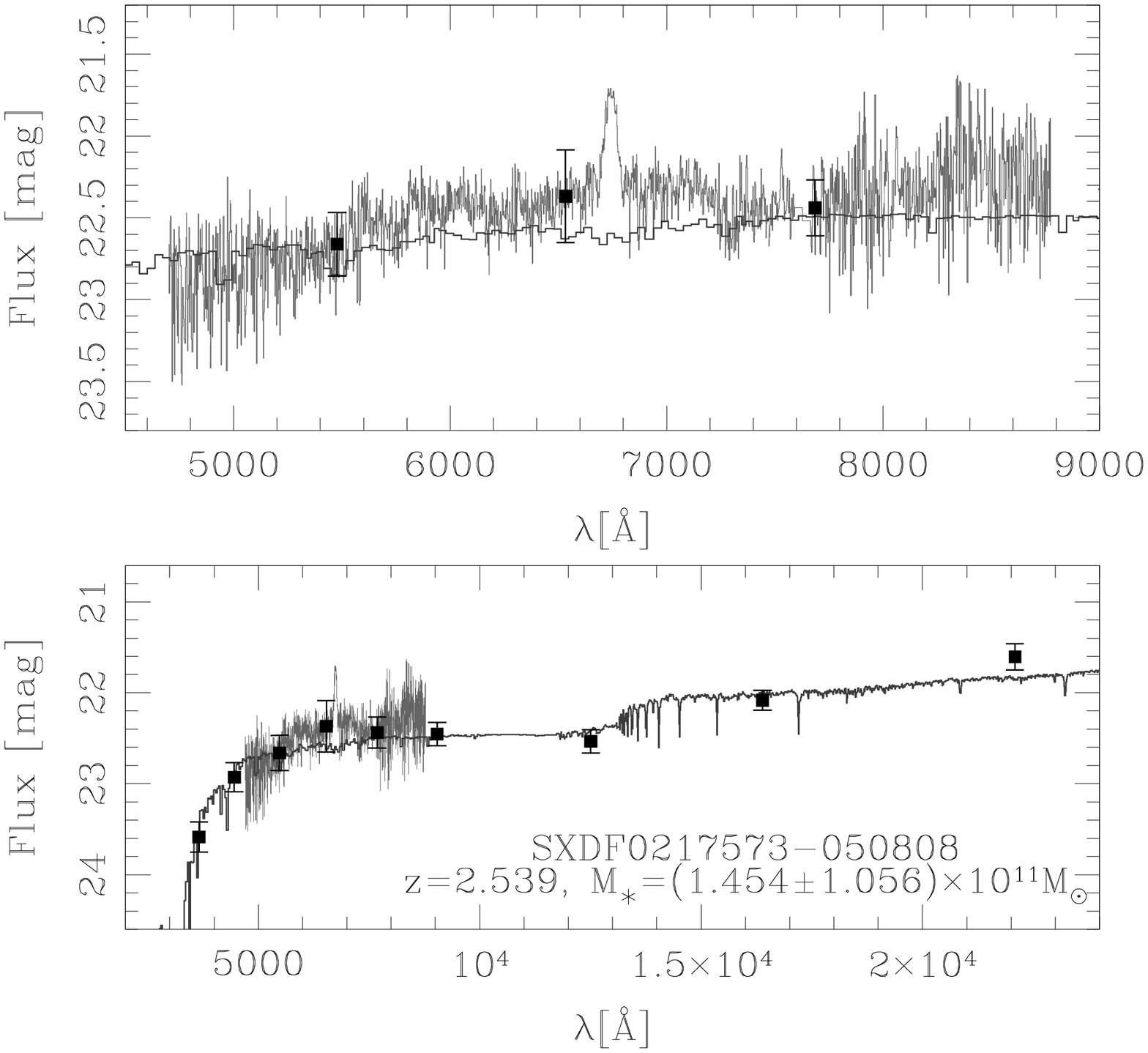}
\includegraphics[width=0.32\textwidth]{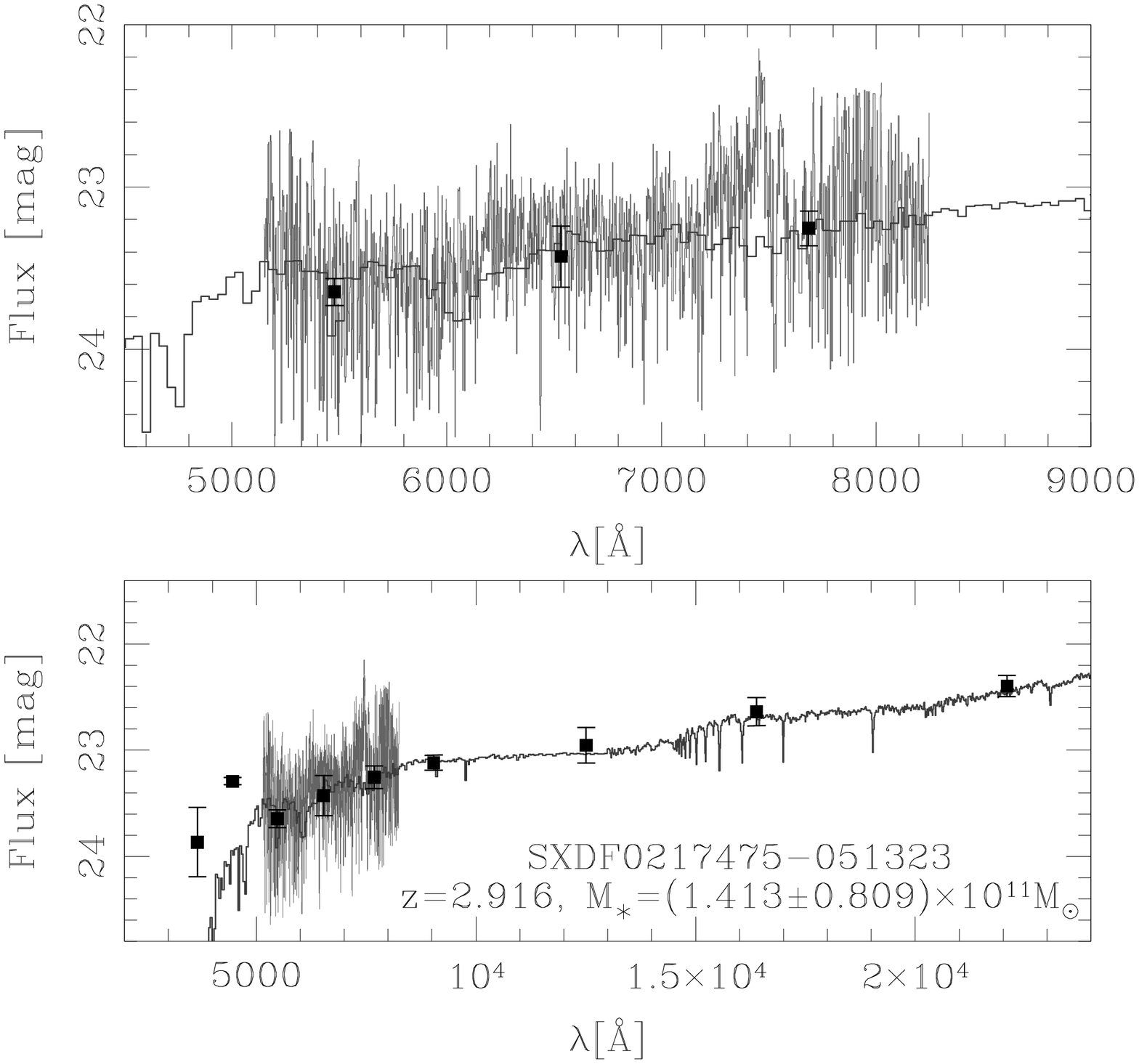}
\caption{\textit{- continued}}
\end{center}
\end{figure*}

\subsection{Spectral energy distribution fitting}\label{SED}

We generated a database of SED templates spanning the ages of [0.001Gyr; 13.5Gyr] and metallicities Z of [0.004; 0.05] with the GALAXEV code (\citeauthor{Bruzual}). A Salpeter initial mass function (IMF; Salpeter \citeyear{Salpeter}) and two types of star formation histories (SFHs) were assumed: passive evolution of stars (an instantaneous burst) and an exponentially declining SFR($t$) $\sim$ $e^{-t/\tau}$. We applied a dust reddening law (\citeauthor{Calzetti2}) to each template with $E(B-V)$ values spanning [0; 0.5] and a value of Rv=4.05 (i.e., for actively star-forming galaxies; \citeauthor{Calzetti2}). Then, with emission lines  masked, we used a least-squares $\chi^{2}$ fitting method to obtain the best template that matched the dominant stellar population.  The sample galaxies revealed values of total stellar masses in the range 10$^{9}-$10$^{11}M_{\odot}$ because we used these templates. In order to avoid the age$-$metallicity degeneracy found in SED fits, a constant metallicity of Z=0.02 was used since  it corresponds to the mean metallicity value found in galaxies with the same stellar mass range as our galaxies (\citeauthor{Gallazzi}), ignoring possible mass$-$metallicity evolution.

In this paper, all drawn properties of the galaxies are relative to the spectroscopy aperture, except for stellar masses. The aperture photometry parameters were used to be able to compare them directly to the properties derived from the spectra (SFRs, reddening) and to compute the color index and age, whereas total magnitudes were used to compute total stellar masses. We used a 90$\%$ confidence interval to determine the uncertainties of the derived properties. Figure \ref{seds} shows the spectrum and aperture photometry, together with the SED fitting for the sample galaxies. Each galaxy presents two panels: the lower panel has the whole SED fit, while the upper  panel shows only a zoom in the optical region.

\section{Properties of the sample}\label{phyprop}

\begin{figure}
\begin{center}
\includegraphics[width=0.40\textwidth]{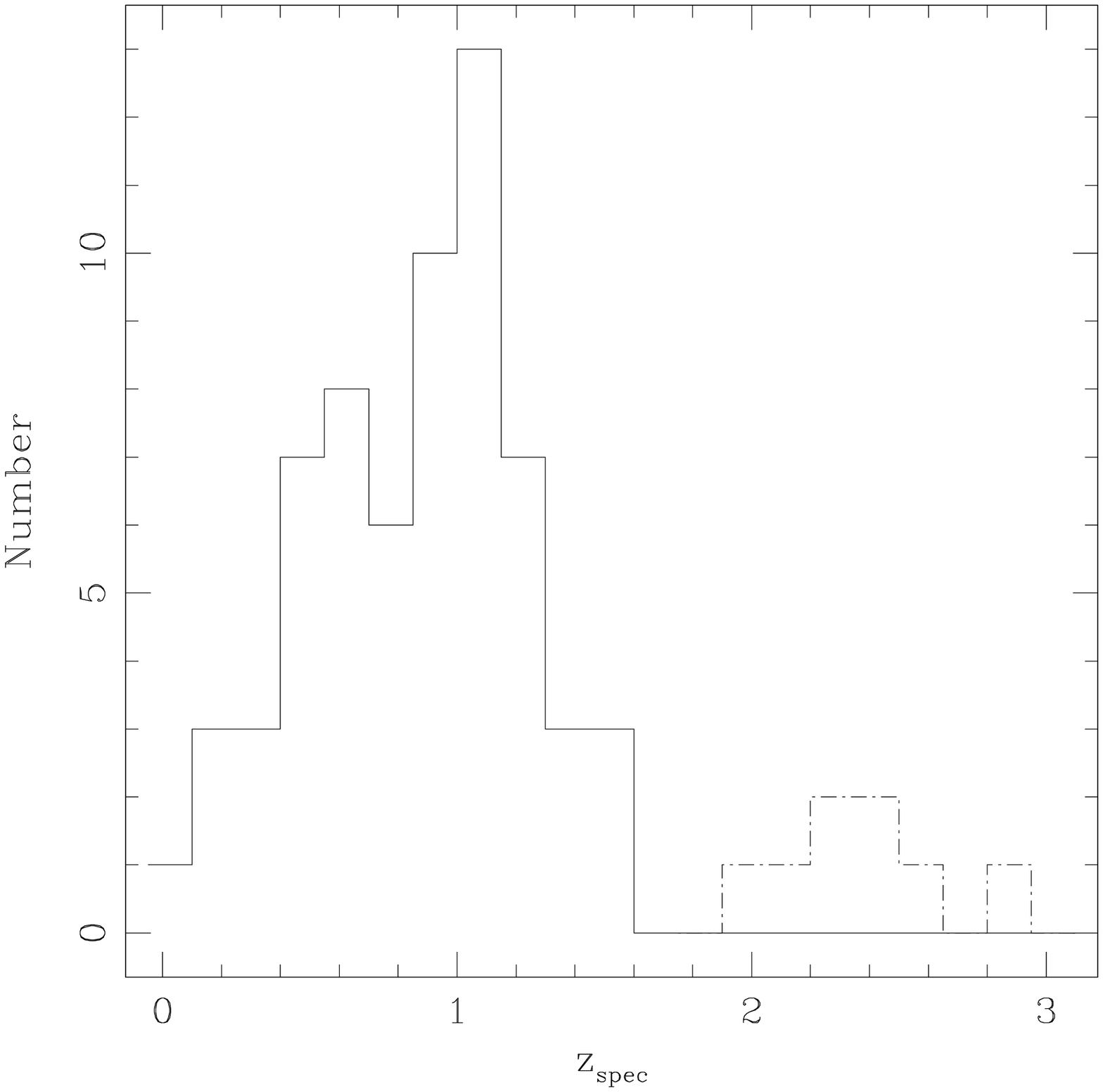}
\caption{Redshift distribution of galaxies presented in this paper. The solid line shows the redshift distribution for the narrow emission line galaxies, whereas the dash-dotted line shows the distribution for the broad emission line galaxies.}\label{redshift}
\end{center}
\end{figure}

The galaxies used in this paper, as mentioned in Section \ref{sample}, were only those that showed signs of star formation activity through the presence of at least one of the [OII]$\lambda$3727, H$\beta$ 4861 $\AA,{}$ or H$\alpha$ 6563 $\AA{}$ emission lines. Thereby, 64 galaxies achieved the required condition, with 45 spectra coming from the GMOS observations and 19 spectra from the IMACS observations. Additionally, we also selected galaxies that showed broad emission lines in their spectra to explore AGN activity at redshift $z$ $>$ 1.0. In this way, eight galaxies were added; five spectra were provided by the IMACS observations and three by the GMOS observations. Table \ref{spectra} summarizes the adopted names for the objects and the origin of the spectra, together with their coordinates, spectroscopic redshifts, and computed u$_{AB}$ magnitudes. Figure \ref{redshift} shows the redshift distribution of our galaxy sample. In this figure, the solid line shows the redshift distribution for the narrow emission line galaxies, whereas the dash-dotted line shows the distribution for the broad emission line galaxies. As can be seen, the broad emission line sample has the highest redshifts.

\begin{figure}
\begin{center}
\includegraphics[width=0.4\textwidth]{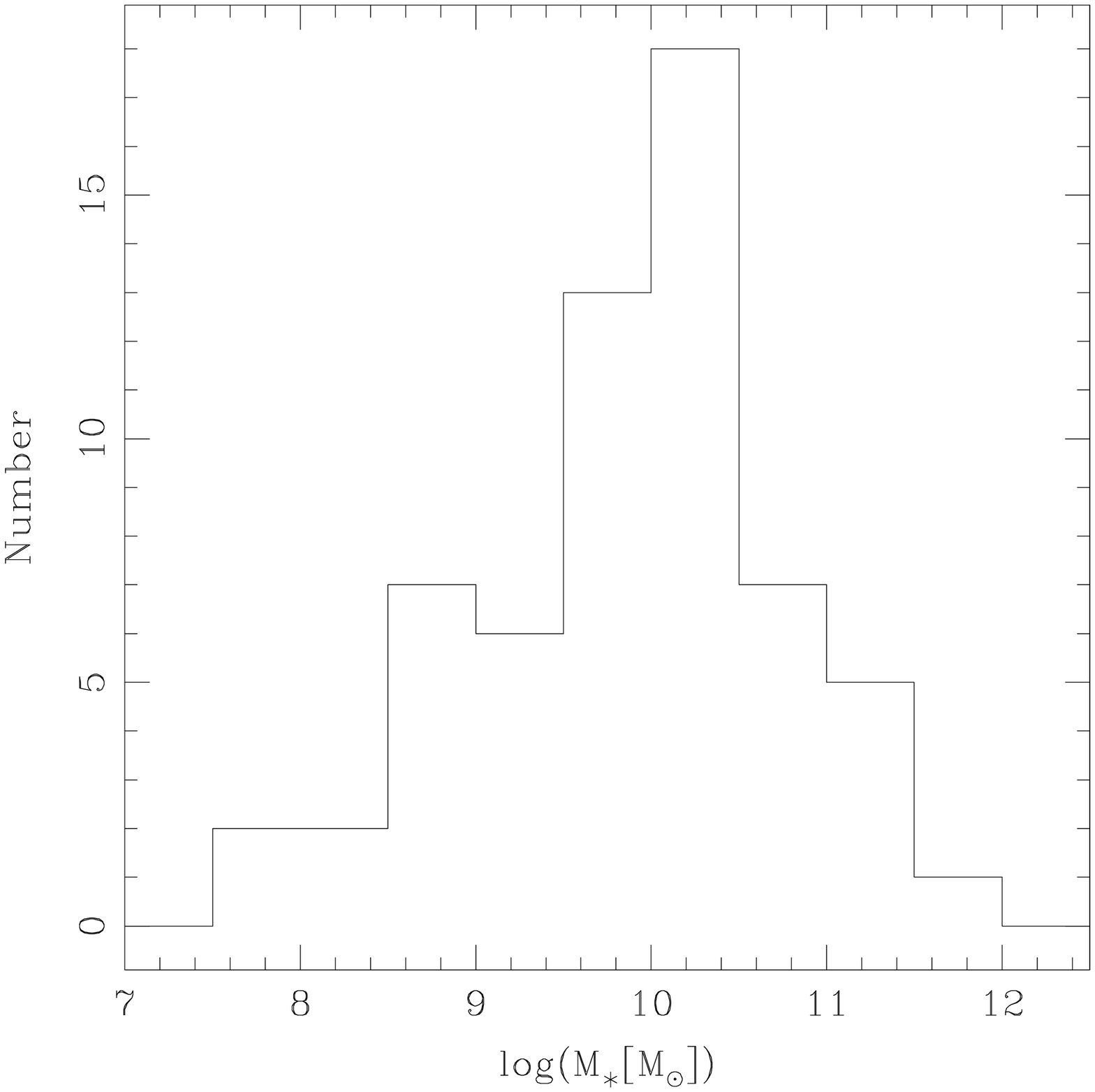}
\includegraphics[width=0.4\textwidth]{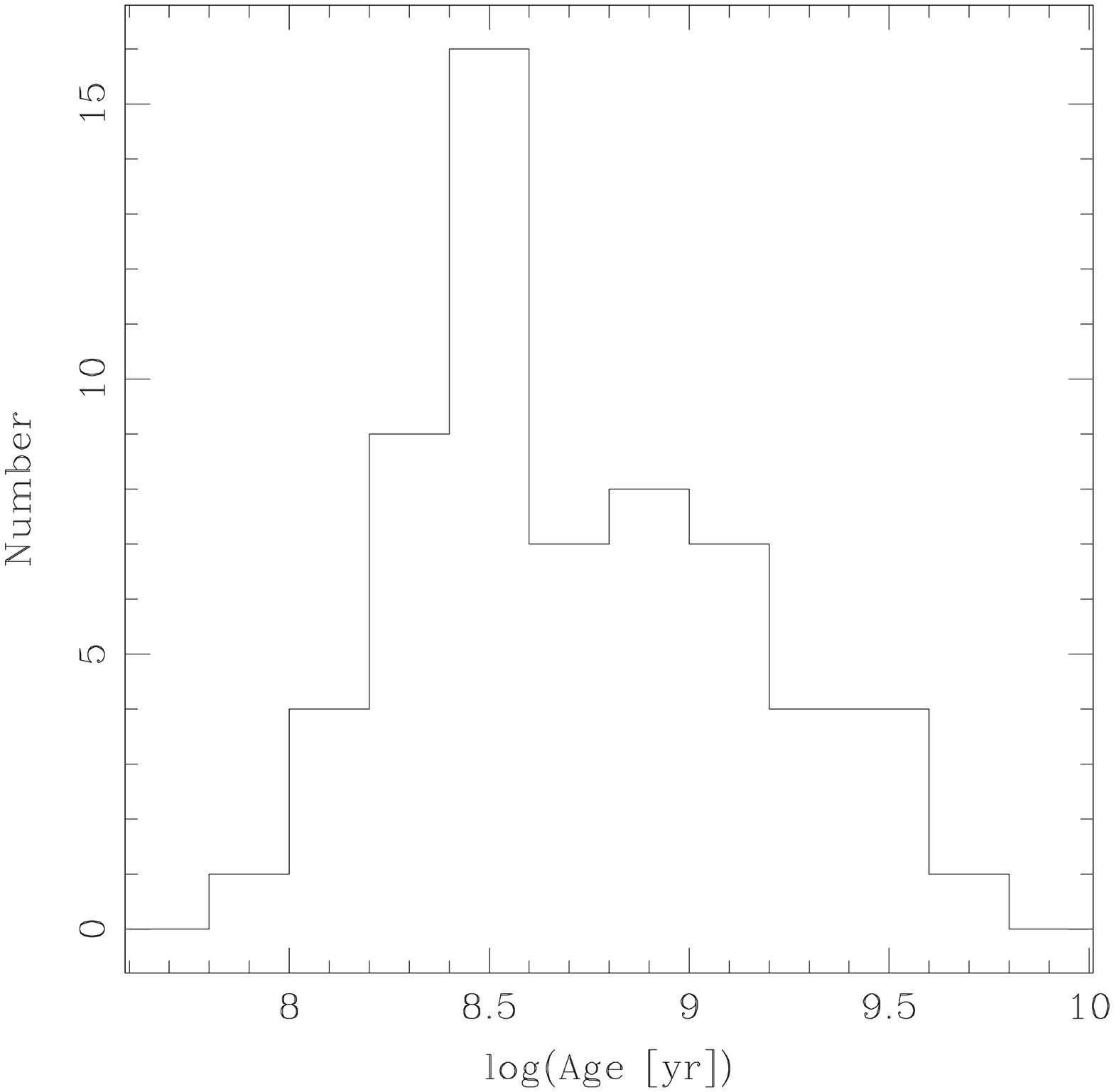}
\caption{Left panel: Total stellar mass distribution of galaxies presented in this paper. Right panel: Age distribution of galaxies. In both panels, the solid line shows the redshift distribution for the narrow emission line galaxies. Both physical properties were derived from SED fitting, using exponentially declining SFHs (Section \ref{SED}).}\label{prop}
\end{center}
\end{figure}

The total stellar masses were estimated  using the total magnitudes in all available photometric bands of the sample galaxies (Section \ref{SED}). The top panel of Figure \ref{prop} shows the stellar mass distribution obtained. It can be observed that the plot peaks at 10$^{10.25}$ $M_{\odot}$, with most of the galaxies (92$\%$) in the range 10$^{8.75}-$10$^{11.25}$ $M_{\odot}$. The ages of the galaxies were also estimated using the same $\chi^{2}$ statistic. The bottom panel of Figure \ref{prop} shows the age distribution of the sample with exponentially declining SFHs   and this plot peaking at 10$^{8.5}$ yr, with most of galaxies (71$\%$) having age values in the range 10$^{8.2}-$10$^{9.2}$ yr. It  is possible that some of our broad emission line galaxies were quasars rather than Seyfert 1 types, implying a continuum dominated by the AGN rather than by the host. Therefore, we have not included the properties of these galaxies in the analysis related to star formation. 

\begin{figure}
\begin{center}
\includegraphics[width=0.45\textwidth]{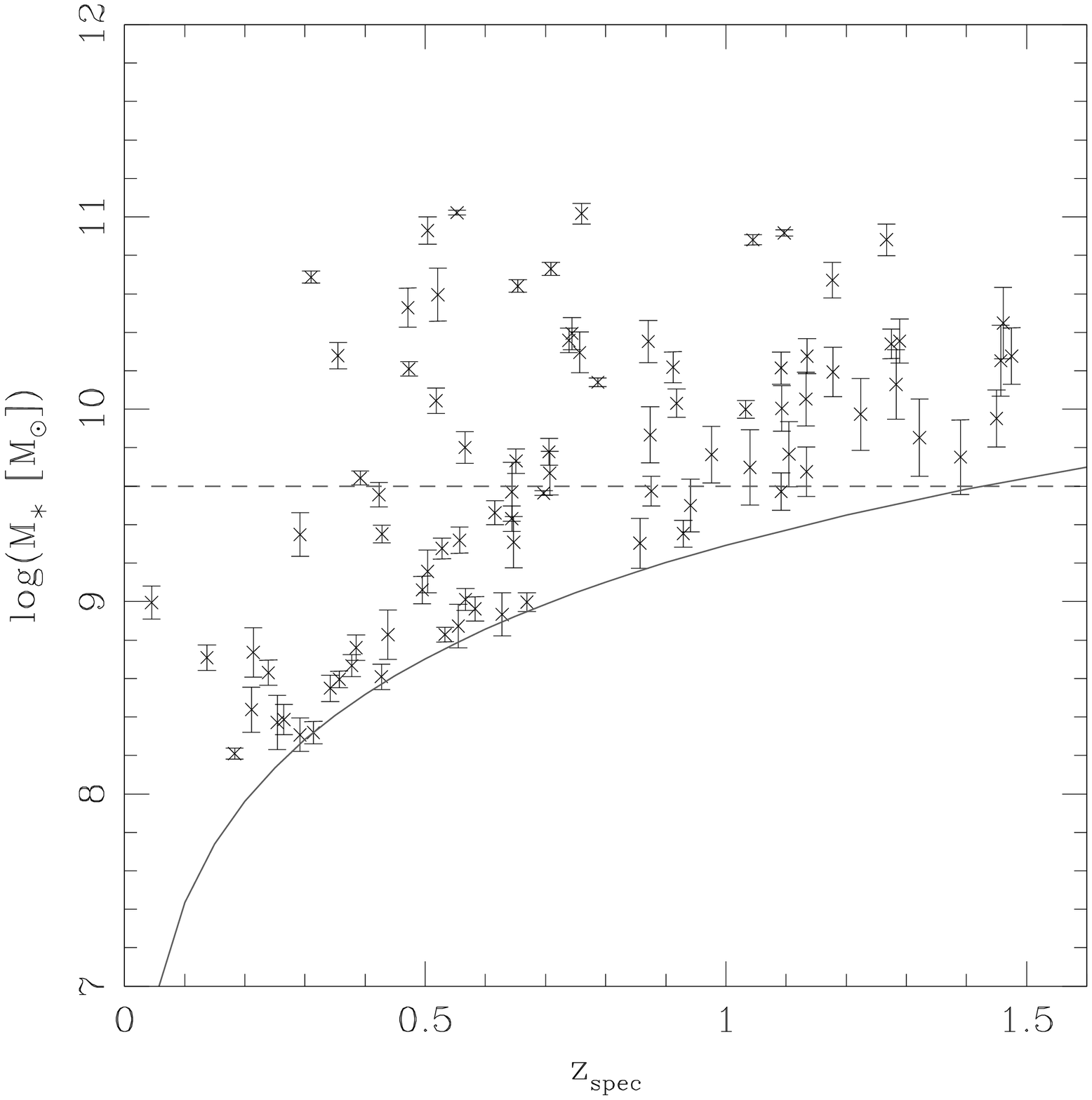}
\caption{Total stellar mass distribution of the galaxies. The solid line represents our mass limit computed with the $M/L$ ratio of a young starburst ($M/L_{V}=$ 0.02). The dashed line shows the stellar mass value at which the sample is well represented throughout the redshift range probed.}\label{massz}
\end{center}
\end{figure}

It is important for further analysis of the study of the stellar masses of our galaxies as a function of the redshift. In this way we can also determine the stellar mass limit of our survey. Figure \ref{massz} shows this diagram, where it can be seen that the galaxies with $M_{\ast}>$ 10$^{9.6}$ $M_{\odot}$ can be found throughout the redshift range probed by our survey (dashed line). The stellar mass limit of the survey is shown with a solid line. We have also included the galaxies presented in \citetalias{Diaz} with the purpose of showing the stellar mass range used in the analysis related to the evolution of SFR ($M_{\ast}>$ 10$^{9.6}$ $M_{\odot}$; section \ref{sfrevol}). 

\begin{figure}
\begin{center}
\includegraphics[width=0.45\textwidth]{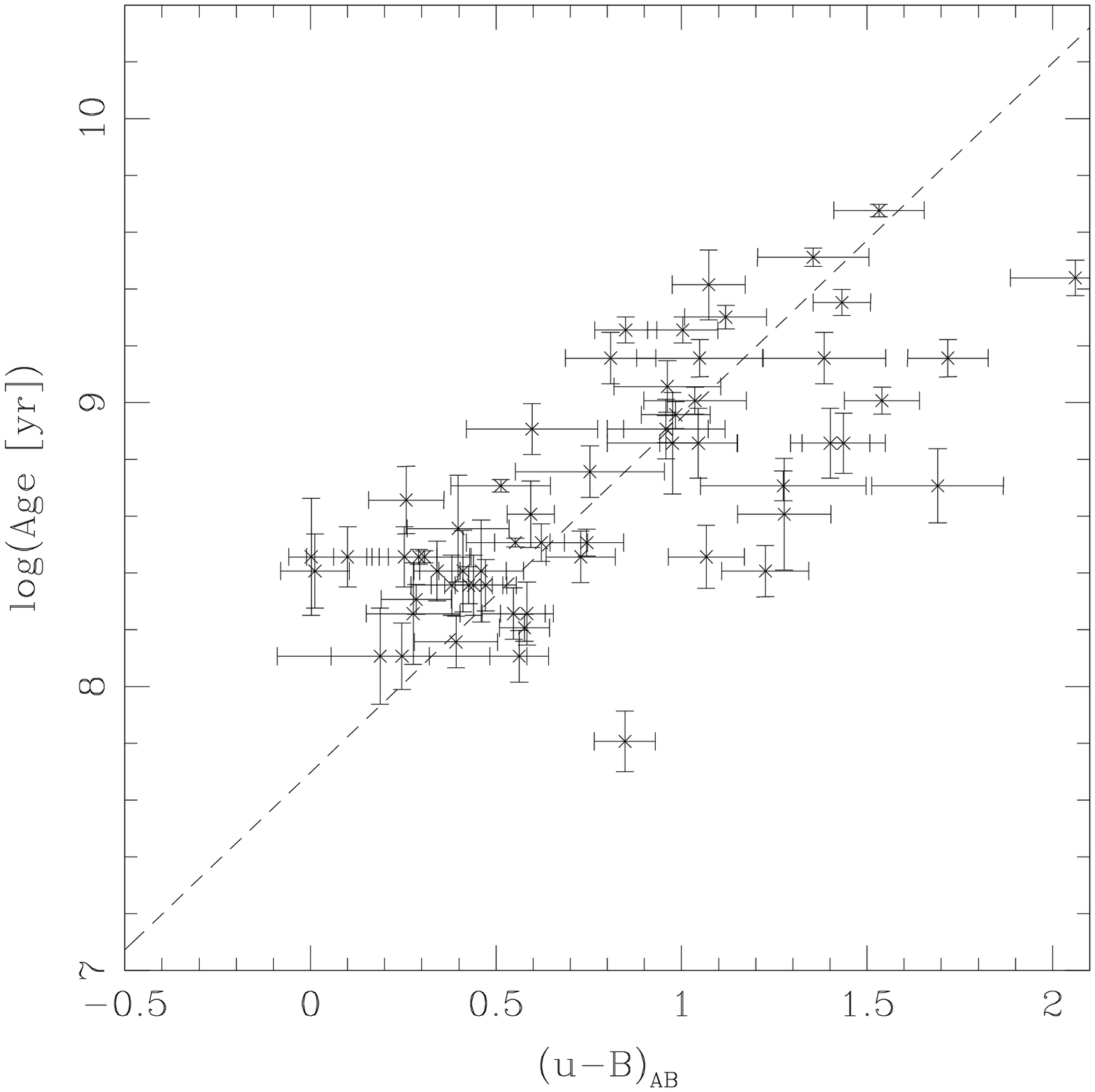}
\caption{Age versus rest-frame $(u-B)_{AB}$ color. The dashed line shows the linear fit between both quantities.}\label{colorage}
\end{center}
\end{figure}

Another useful property was the rest-frame color index $(u-B)_{AB}$, with figure \ref{colorage} showing the relation found between age and this color index. We applied a reddening correction and the $K-$correction to the observed-frame $(u-B)_{AB}$ color of each galaxy of the sample with the best SED model found. It can be observed that there is an evident correlation between both quantities, which reflects the capacity of this color index to be a good age indicator.  

\section{Contribution of active galactic nuclei }\label{nuc}

In this section, we explored whether there was a presence of AGN activity in the galaxies of our sample. Of these, eight galaxies showed broad emission lines in their optical spectra. For the redshift range that the bulk of our galaxies spanned, the available methods to detect AGN activity are mainly the use of AGN diagnostic diagrams in the mid-infrared (MIR) region and the detection of a counterpart in X-rays. We did not use the classical Baldwin, Phillips, and Terlevich (BPT; \citeyear{Baldwin}) diagnostic emission line diagrams because we could not locate any of the galaxies showing the necessary lines  in the BPT diagrams. 

The X-ray counterparts were searched for in public catalogs such as the one by Ueda et al. \citeyearpar{Ueda}, who observed with XMM reaching sensitivity limits of 6$\times$10$^{-16}$, 8$\times$10$^{-16}$, 3$\times$10$^{-15}$, and 5$\times$10$^{-15}$ erg cm$^{-2}$ s$^{-1}$ in the 0.5$-$2, 0.5$-$4.5, 2$-$10, and 4.5$-$10 keV bands. In addition, we used the X-ray catalog of Akiyama et al. (2015, accepted), who  summarized the results of the identification of the X-ray sources detected in the 0.5$-$2 and/or the 2$-$10 keV band within the five combined FOV of the SXDF. In these catalogs, we found a total of eight counterparts for the galaxies of our sample with four belonging to broad emission line galaxies, while the remaining belonged to narrow emission line galaxies. Two narrow emission line galaxies also showed the MgII $\lambda$2796, 2803 line in emission, thereby providing additional evidence to be considered as AGN hosts, while for the others two narrow emission line galaxies, this line was out of the spectral coverage due to its lower redshift ($z$ $<$ 0.78).

\begin{figure*}
\begin{center}
\includegraphics[width=0.49\textwidth]{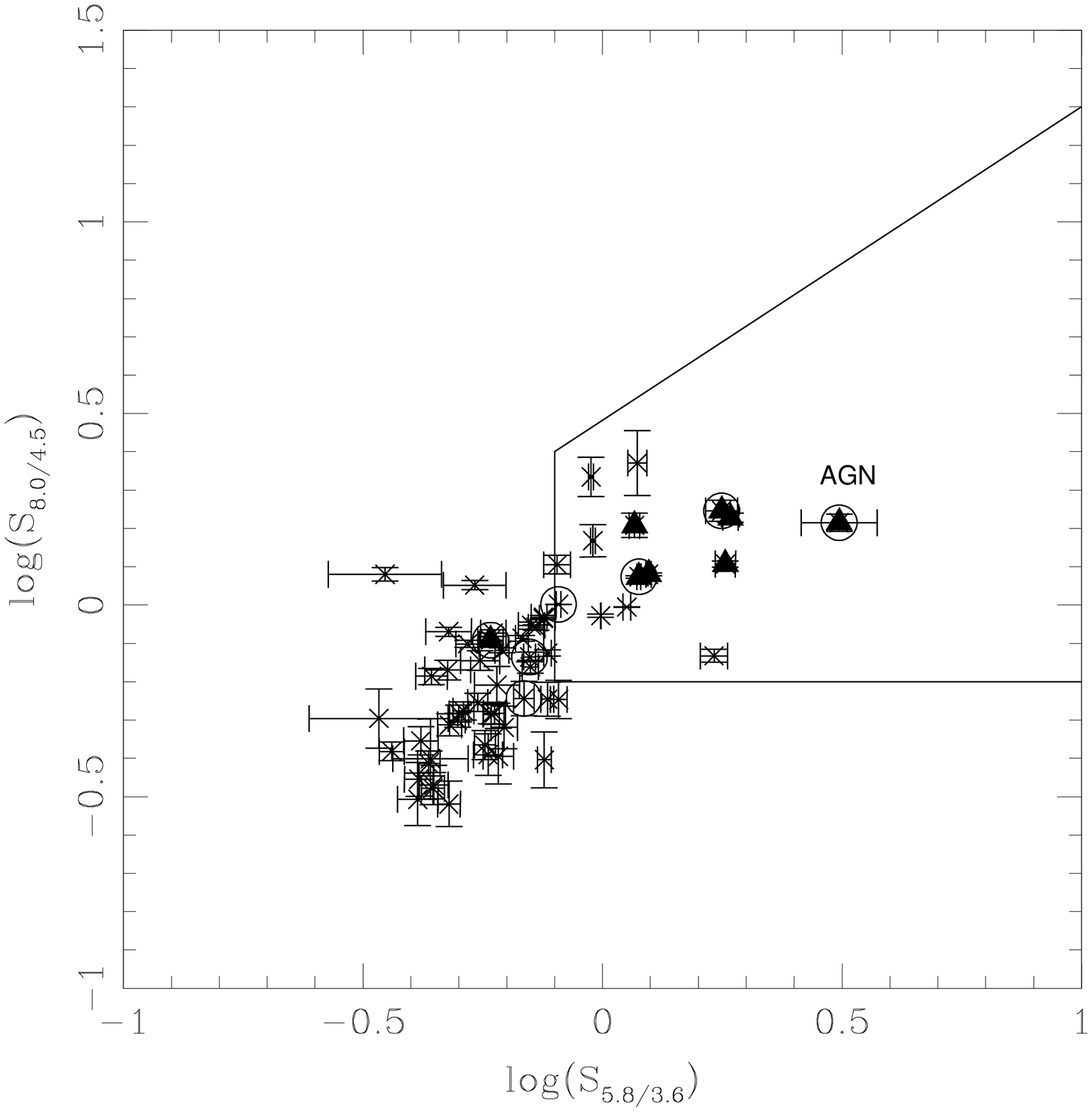}
\includegraphics[width=0.49\textwidth]{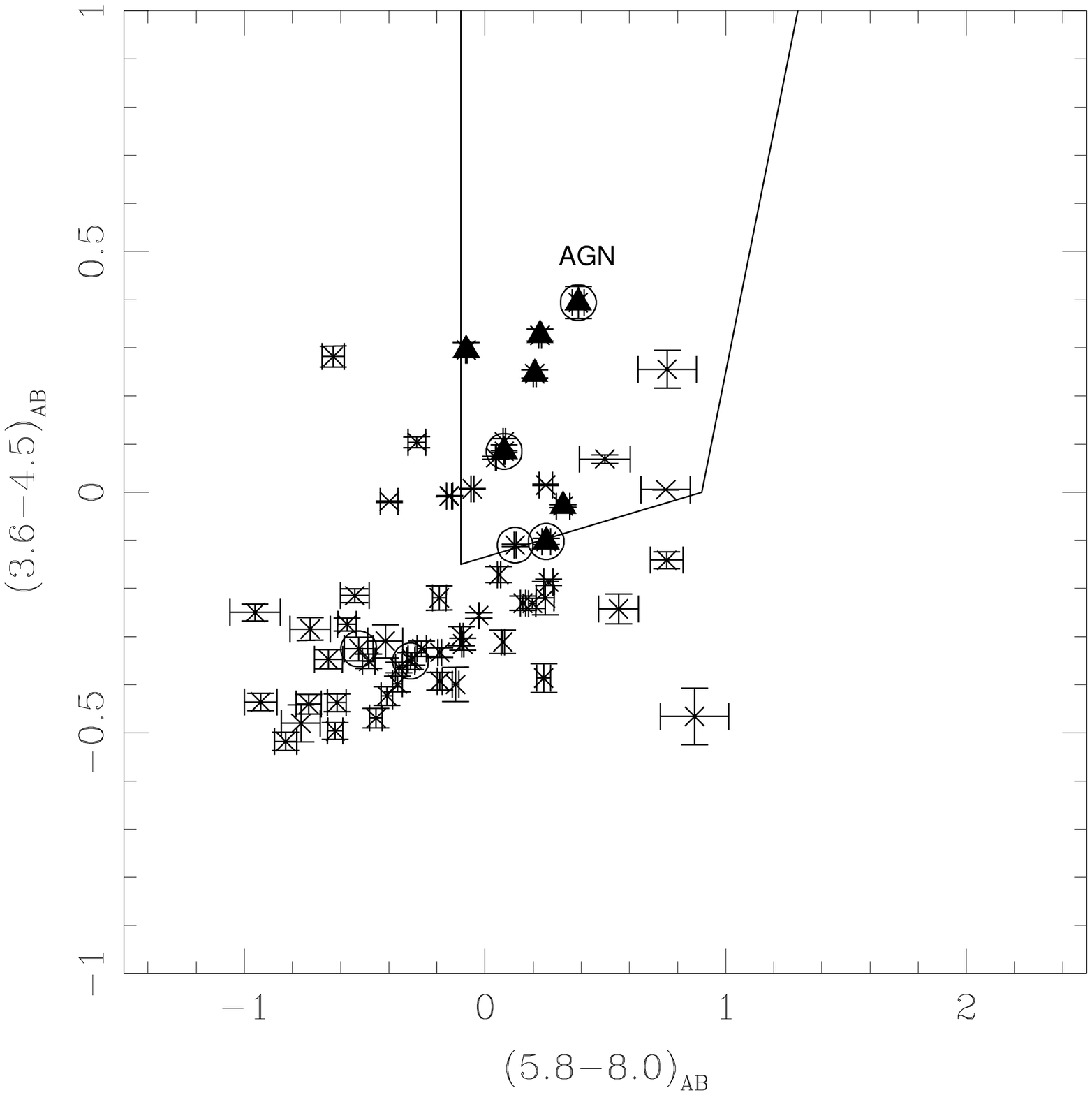}
\caption{Left panel: log($S_{8.0}/S_{4.5}$) vs log($S_{5.8}/S_{3.6}$) diagram. Right panel: (3.6$-$4.5$\mu$m) vs (5.8$-$8.0$\mu$m) diagram. Black solid lines show the limits adopted by Stern et al. \citeyearpar{Stern} and Lacy et al. \citeyearpar{Lacy} in the left and right panels,  respectively. The triangles show the galaxies with broad emission lines in their optical spectra, while the open circles show the galaxies with detection of a X-ray counterpart.}\label{iracd}
\end{center}
\end{figure*}

In addition to the X-ray counterparts, the MIR color$-$color diagrams were also used to investigate the existence of dusty AGNs (\citeauthor{Stern,Lacy}). Through their color indices, these diagrams are responsive to the power-law nature of the AGN continuum, which shows redder MIR colors than star-forming galaxies. There were 49 galaxies with available photometry in all the IRAC/\textit{Spitzer} bands with Figure \ref{iracd} showing the Lacy et al. \citeyearpar{Lacy} log($S_{8.0}/S_{4.5}$) versus log($S_{5.8}/S_{3.6}$) diagram (left panel) and the Stern et al. \citeyearpar{Stern} (3.6$-$4.5$\mu$m) versus (5.8$-$8.0$\mu$m) diagram (right panel). In this figure, the triangles show the galaxies with broad emission lines in their optical spectra, while the open circles show the galaxies with detection of an X-ray counterpart. It can be seen in both panels that all but one of the broad emission line galaxies were classified as AGN. In contrast, only two of the galaxies with X-ray counterparts and narrow emission lines in their spectra were not classified as AGN. This might imply that these  galaxies are X-ray star-forming or galaxies with SEDs in the MIR region dominated by the host rather than the AGN (\citeauthor{Donley}, \citetalias{Diaz}). Additionally, there were eight galaxies classified as AGN without an X-ray counterpart or AGN features in their optical spectra, implying they may be obscured AGNs (\citeauthor{Lacy2}, \citetalias{Diaz}). Therefore, all our AGN candidates were excluded from SFR analysis and their respective plots.

\section{Star formation rates}\label{starfor}

\begin{figure}
\begin{center}
\includegraphics[width=0.45\textwidth]{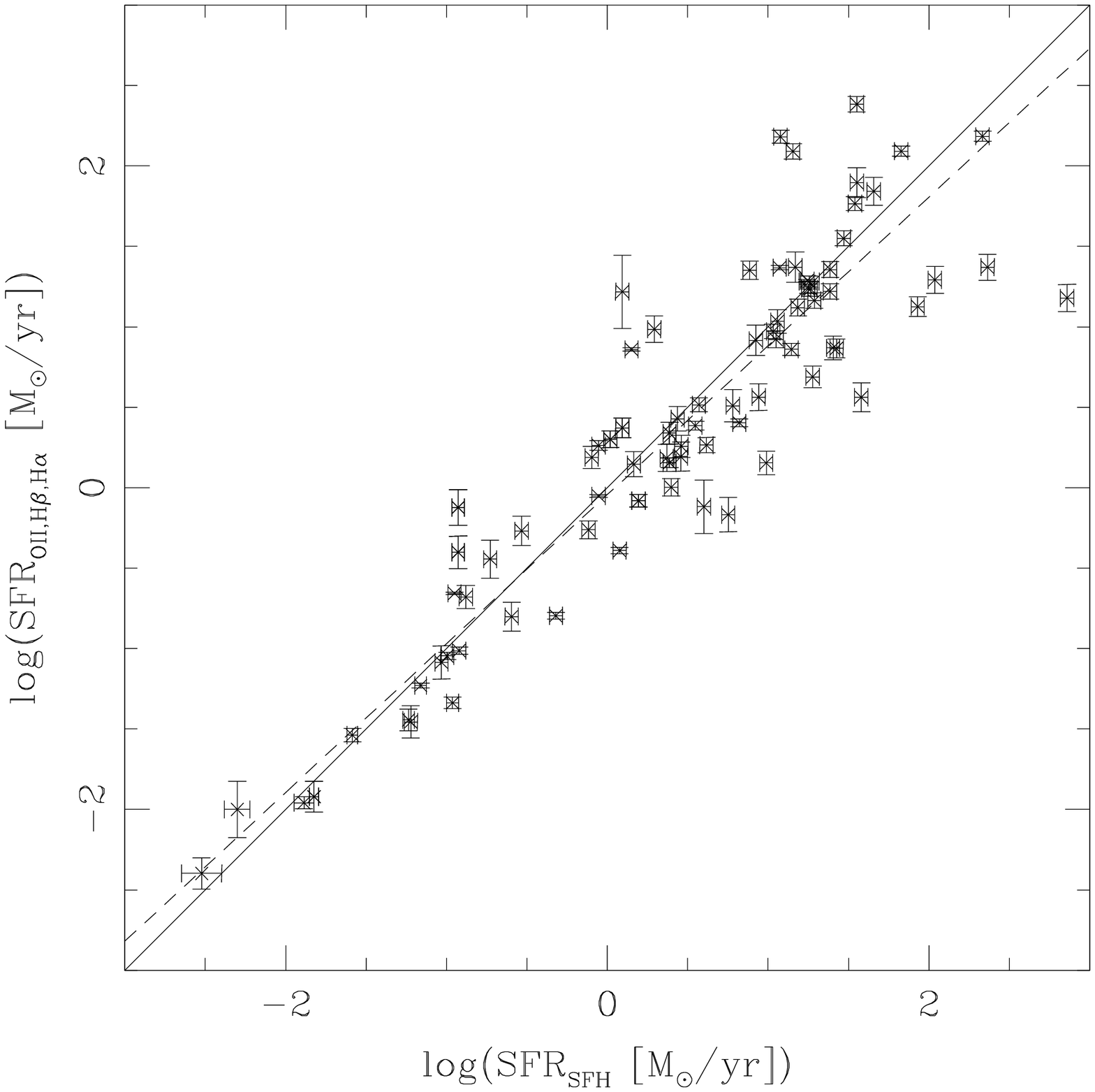}
\caption{Comparison between SFR$_{\text{[OII],H}\alpha,\text{H}\beta}$ versus SFR$_{\text{SFH}}$, obtained using exponentially declining SFHs. Simbols show the SFR values obtained for the spectroscopic sample presented in this work, and from the Díaz Tello et al. \citeyearpar{Diaz} sample. The solid line represents a 1:1 relation, while the dashed line shows the linear fit applied to the sample.}\label{sfr}
\end{center}
\end{figure}

The fluxes from one of the following emission lines were used to estimate the SFR from our sample galaxies, H$\alpha$ 6563 $\AA{}$, H$\beta$ 4861 $\AA,{}$ or [OII]$\lambda$3727. Particularly, we  used the Kennicutt et al. \citeyearpar{Kennicutt} formulas to estimate the SFR of H$\alpha$ and [OII] fluxes, after correcting by dust attenuation. However, we had six cases where the only emission line available was H$\beta$. Therefore, for the SFR$_{\text{H}\beta}$ computation we assumed an intrinsic flux ratio H$\alpha$/H$\beta$= 2.85. This is the typical value for a star-forming region (\citeauthor{Veilleux}). In this way, the following expression was obtained:

\begin{equation}
\text{SFR}_{\text{H}\beta} (M_{\odot} \text{yr}^{-1})=2.82 \times 10^{-40} d_{L}^{2}F_{\text{H}\beta} (\text{erg s}^{-1}),
\end{equation}
where $d_{L}$(cm) is the luminosity distance and $F_{\text{H}\beta}$(erg s$^{-1}$ cm$^{-2}$) is the flux measured in H$\beta$. Dust correction for both H$\alpha$ and H$\beta$ emission lines was carried out using the SED fits, as  described in Sec. \ref{SED}, and the $E(B-V)_{stellar}=0.44 E(B-V)_{nebular}$ relation of Calzetti et al. \citeyearpar{Calzetti2}. For [OII], the Kewley, Geller, and Jansen \citeyearpar{Kewley} reddening correction recipe was adopted.  The broad emission line galaxies were not included in subsequent analyses.

In addition, the SFR was determined in an independent way using the SFR values (from exponentially declining SFHs; Section \ref{SED}) derived from the fitted templates, where a 90$\%$ confidence interval was used to determine the uncertainty of these values. Figure \ref{sfr} shows the comparison between the SFR values obtained from the H$\alpha$, H$\beta,$ or [OII] lines versus the SFR values obtained from exponentially declining SFHs;  the crosses represent the galaxies presented in this work and from the \citetalias{Diaz} sample. The D13 sample galaxies were included to expand the current sample in the remaining analysis. There is no overlap between the \citetalias{Diaz} sample and the current sample. However, the mass distribution of the present sample was more massive than that from \citetalias{Diaz}, so the metallicity adopted was higher. A lower metallicity implies slightly higher SFR and SSFR values for each galaxy. Nevertheless, we observed differences that were not significant  in the presented results when the mass range was extended to lower masses. It can be seen that the methods used to derive SFRs gave very similar results, but with a slight trend for SFH models to produce higher SFR values at larger SFRs. The calculated \textit{rms} was 0.048.

\subsection{Evolution of the SFR}\label{sfrevol}

\begin{figure}
\begin{center}
\includegraphics[width=0.49\textwidth]{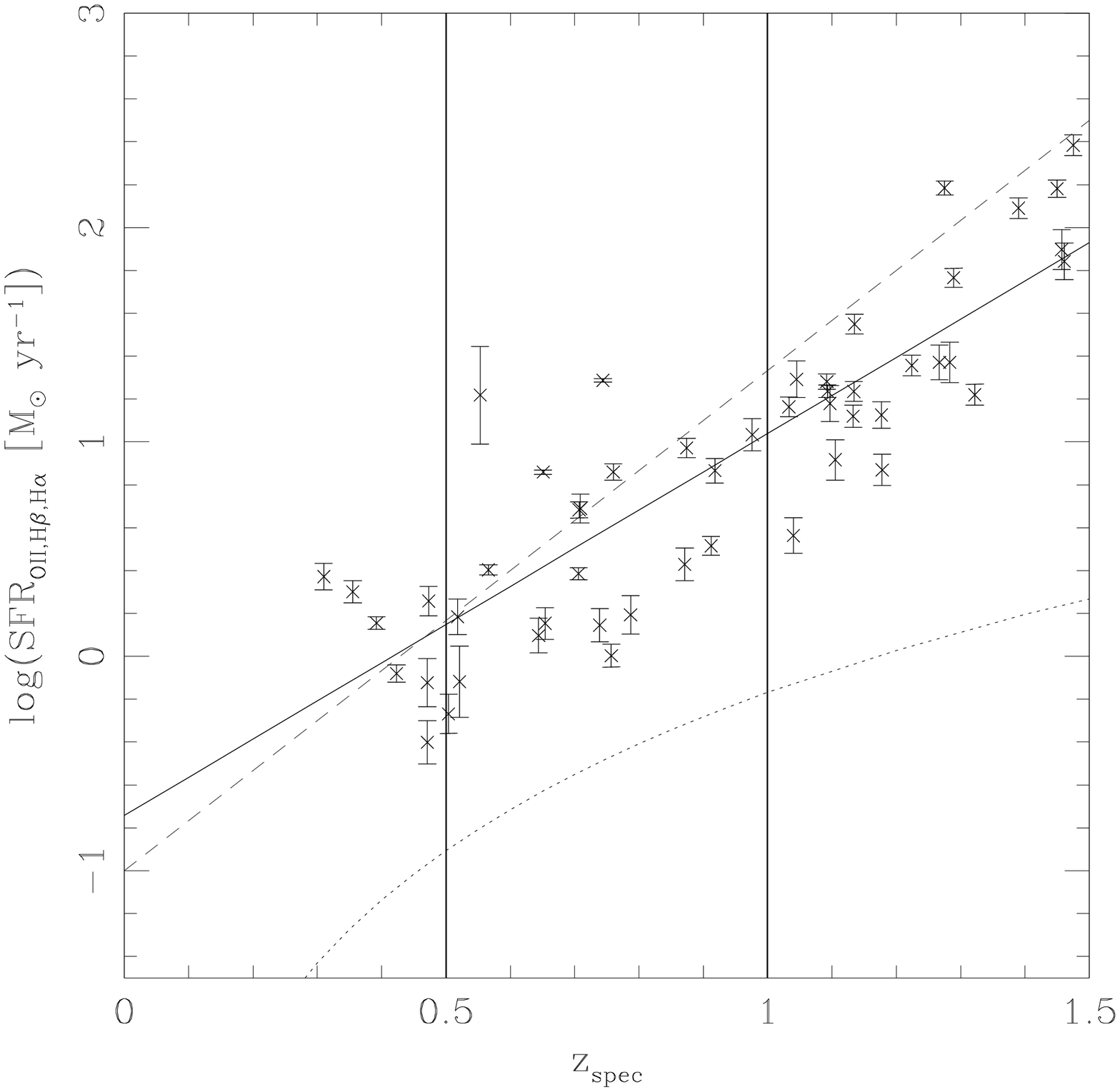}
\caption{Evolution of the SFR as function of the redshift. The points show the SFR values derived either from [OII], H$\beta,$ or H$\alpha$ lines in the mass limited sample. The solid line shows the linear fit applied to the sample, while the dashed line shows the trend found by Silverman et al \citeyearpar{Silverman} for zCOSMOS star-forming galaxies at 0.48 $<z<$ 1.05, extrapolated to the whole interval probed. The dotted line shows the limit SFR estimated for our survey, taking  dust reddening and Balmer absorption corrections in account. The vertical lines delimit the redshift intervals used for further analysis.}\label{sfrev}
\end{center}
\end{figure}

We next investigated how the SFR evolved as a function of redshift for our sample of galaxies and the possible implications for the star formation history of the universe (see figure \ref{sfrev}). The figure also shows the limit SFR observed for our survey (dotted line), for which we take reddening and Balmer absorption corrections  into account, as was described in Sec. (\ref{starfor}). As a result of the underlying stellar population, the Balmer absorption was estimated with the average value for the corresponding absorption lines measured from all SED spectra. It can be observed that the SFR values derived from the [OII] or H$\alpha$ lines in the sample evolved as a function of redshift;  the solid line shows the linear fit applied to the sample, while the dashed line indicates the trend found by Silverman et al. \citeyearpar{Silverman} for zCOSMOS star-forming galaxies at 0.48 $<$ $z$ $<$ 1.05 (vertical lines delimit the redshift intervals $z$ of 0.5). This evolution correlates with redshift. This trend is similar to that observed by Silverman et al. \citeyearpar{Silverman}, but with a steeper slope. We think this difference might be the result of a smaller cut value in the total stellar mass. From fig. \ref{massz} it can be observed that the cut we used was $M_{\ast}>$ 10$^{9.6}$ $M_{\odot}$, while that used by Silverman et al. \citeyearpar{Silverman} was $M_{\ast}>$ 10$^{10.6}$ $M_{\odot}$. The stellar mass cut used has excluded the galaxies for which their SFR values were obtained from H$\beta$. The cut has also mainly removed  galaxies with low SFR values (log(SFR)$\sim$0) in the redshift range probed, producing a notorious gap between the limit SFR curve and the observed SFR sequence. However, this cut was necessary to remove the stellar mass dependence on the SFR. Nevertheless, the aforementioned gap could also be the consequence of the parasite light problem mentioned in section \ref{sample}. The lights significantly affected the fainter spectra, which could have weak emission lines and therefore decrease the number of galaxies with low SFR values in the sample.

\begin{figure}
\begin{center}
\includegraphics[width=0.49\textwidth]{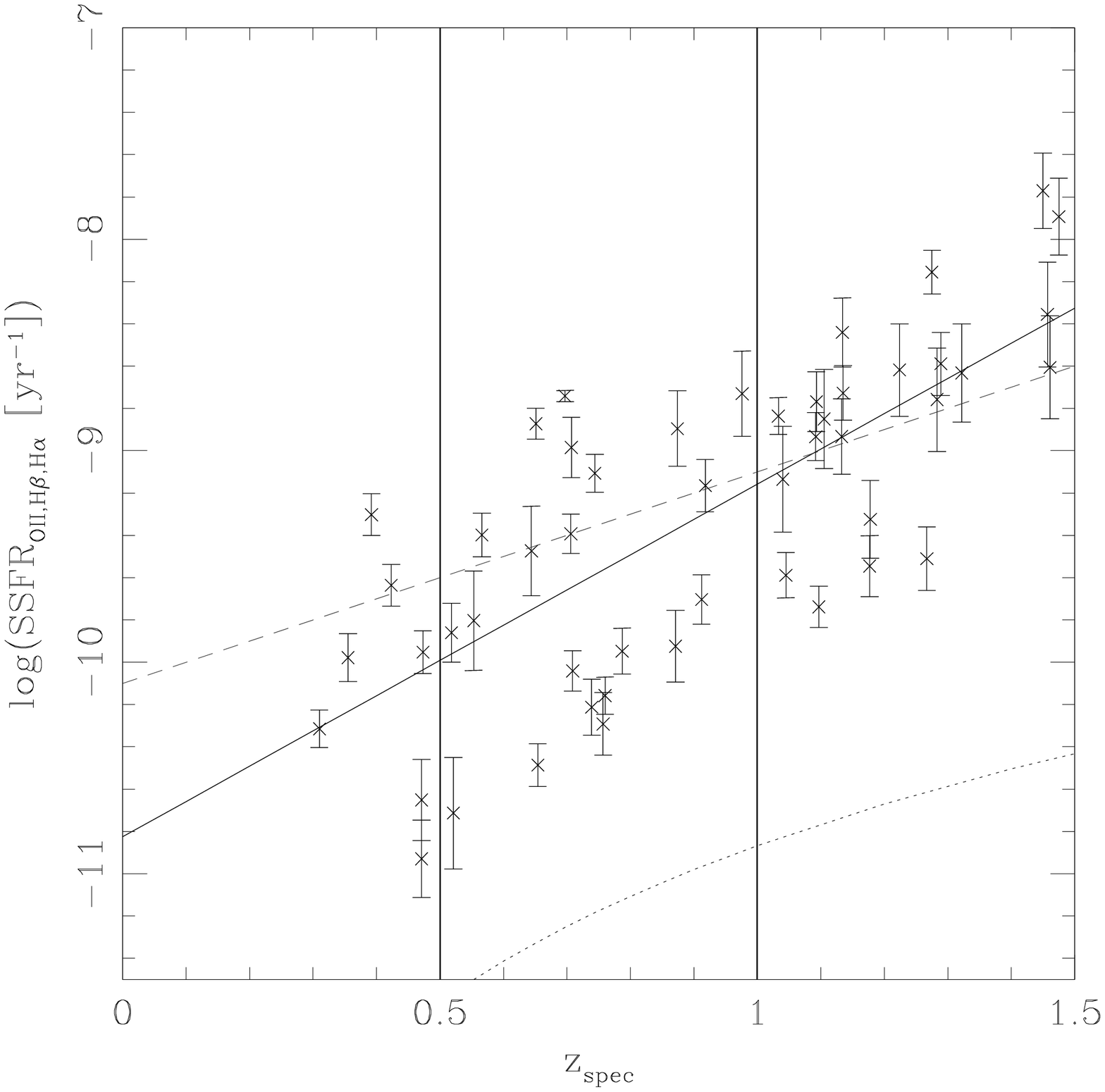}
\caption{Evolution of the SSFR as function of the redshift. The solid line shows the linear fit applied to the mass limited sample, while the dashed line shows the trend found by Juneau et al \citeyearpar{Juneau} for FIDEL infrared galaxies at 0.3$<z<$1.0, extrapolated to the whole interval probed. The dotted line shows the limit SSFR of our survey. The vertical lines delimit the redshift intervals used for further analysis.}\label{ssfrev}
\end{center}
\end{figure}

Another interesting parameter to study is the evolution of the specific SFR (SSFR), which is a measure of how efficient a galaxy is at forming stars relative to its total stellar mass. Figure \ref{ssfrev} shows this quantity as a function of redshift in the sample, with the solid line showing the linear fit applied to the data and the dashed line giving the relation found by Juneau et al \citeyearpar{Juneau} for FIDEL infrared galaxies at 0.3 $<z<$ 1.0. It can be observed, as in the case of the previous figure, that SSFR values increased with redshift. However, a spread is observed as a consequence of the different galaxy types selected by the survey, which included a range from young starburst (high SSFRs) to old star-forming (low SSFRs) galaxies. This spread is more notorious at the intermediate-redshift interval (0.5 $<z<$ 1.0), where the extended sample reaches a maximum. The figure also shows the limit SSFR observed in our survey (dotted line), taking in account the limit SFR shown in fig. \ref{sfrev} and the highest stellar mass value observed in our survey ($M_{\ast}$ $\sim 10^{11}$ $M_{\odot}$; fig. \ref{massz}). Furthermore, the correlation observed was similar to that found by Juneau et al. \citeyearpar{Juneau}, but our correlation had a shallower slope, possibly due to a different selection and thus to a different mean total stellar mass ($M_{\ast}$ $\sim$ 10$^{10.6}$ $M_{\odot}$).

\begin{table*}
\centering
\begin{minipage}{146mm}
\begin{footnotesize}
\caption{Spectroscopic sample obtained with IMACS and GMOS.}\label{spectra}
\begin{tabular}{c c c c c c c c}
\hline\hline
ID catalogue & Instrument & R.A. & Dec. & ${z}_{spec}$ & $u_{AB}^{a}$ & err $u$ & Comments$^{b}$\\
\hline
SXDF021726.5-045903 & IMACS & 02:17:26.529 & -04:59:03.47 & 0.094 & 25.10 & 0.08 & \\
SXDF021801.5-050504 & GMOS & 02:18:01.508 & -05:05:04.11 & 0.141 & 25.67 & 0.12 & \\
SXDF021812.5-050225 & GMOS & 02:18:12.596 & -05:02:25.35 & 0.137 & 23.36 & 0.03 & \\
SXDF021836.2-050442 & GMOS & 02:18:36.292 & -05:04:42.84 & 0.211 & 23.47 & 0.04 & \\
SXDF021852.0-045541 & IMACS & 02:18:52.040 & -04:55:41.44 & 0.314 & 25.25 & 0.09 & \\
SXDF021755.3-051143 & IMACS & 02:17:55.347 & -05:11:43.21 & 0.357 & 23.73 & 0.04 & \\
SXDF021812.8-050343 & GMOS & 02:18:12.898 & -05:03:43.90 & 0.385 & 23.68 & 0.04 & \\
SXDF021812.3-050307 & GMOS & 02:18:12.376 & -05:03:07.83 & 0.438 & 24.42 & 0.07 & \\
SXDF021834.3-045049 & IMACS & 02:18:34.332 & -04:50:49.62 & 0.473 & 22.42 & 0.02 & \\
SXDF021749.3-050713 & IMACS & 02.17:49.300 & -05:07:13.14 & 0.498 & 24.83 & 0.08 & \\
SXDF021739.4-045913 & IMACS & 02:17:39.430 & -04:59:13.48 & 0.504 & 24.48 & 0.07 & \\
SXDF021742.1-045628 & IMACS & 02:17:42.146 & -04:56:28.46 & 0.504 & 24.27 & 0.05 & 850$\mu$m source\\
SXDF021805.2-050437 & GMOS & 02:18:05.252 & -05:04:37.54 & 0.518 & 22.90 & 0.03 & \\
SXDF021827.5-050110 & GMOS & 02:18:27.501 & -05:01:10.15 & 0.521 & 24.25 & 0.06 & \\
SXDF021838.8-050344 & GMOS & 02:18:38.845 & -05:03:44.44 & 0.553 & 23.27 & 0.03 & 24$\mu$m source\\
SXDF021823.4-050116 & GMOS & 02:18:23.469 & -05:01:16.88 & 0.555 & 25.24 & 0.10 & \\
SXDF021803.1-050856 & IMACS & 02:18:03.140 & -05:08:56.12 & 0.566 & 23.68 & 0.04 & \\
SXDF021740.2-045500 & IMACS & 02:17:40.210 & -04:55:00.27 & 0.616 & 25.94 & 0.13 & \\
SXDF021806.9-045912 & GMOS & 02:18:06.904 & -04:59:12.46 & 0.620 & 25.97 & 0.15 & 24$\mu$m, 70$\mu$m, 850$\mu$m source\\
SXDF021840.6-050415 & GMOS & 02:18:40.692 & -05:04:15.84 & 0.628 & 26.46 & 0.17 & \\
SXDF021830.4-050016 & GMOS & 02:18:30.475 & -05:00:16.18 & 0.644 & 23.74 & 0.04 & \\
SXDF021810.3-044929 & IMACS & 02:18:10.320 & -04:49:29.51 & 0.654 & 25.63 & 0.12 & X-ray, 24$\mu$m, 70$\mu$m source\\
SXDF021735.8-045555 & IMACS & 02:17:35.851 & -04:55:55.49 & 0.709 & 25.21 & 0.09 & 24$\mu$m, 850$\mu$m source\\
SXDF021821.3-050049 & GMOS & 02:18:21.309 & -05:00:49.92 & 0.739 & 25.24 & 0.10 & \\
SXDF021731.8-045317 & IMACS & 02:17:31.814 & -04:53:17.21 & 0.754 & 25.00 & 0.05 & 24$\mu$m, 70$\mu$m source\\
SXDF021732.2-045743 & IMACS & 02:17:32.220 & -04:57:43.60 & 0.757 & - & - & \\
SXDF021745.7-051327 & IMACS & 02:17:45.728 & -05:13:27.49 & 0.760 & 23.77 & 0.03 & 24$\mu$m source\\
SXDF021721.4-050958 & IMACS & 02:17:21.400 & -05:09:58.02 & 0.787 & 23.25 & 0.03 & X-ray source\\
SXDF021727.0-045841 & GMOS & 02:17:27.069 & -04:58:41.44 & 0.857 & 25.79 & 0.12 & \\
SXDF021830.6-045951 & GMOS & 02:18:30.648 & -04:59:51.74 & 0.869 & 23.42 & 0.04 & 24$\mu$m source\\
SXDF021828.2-050005 & GMOS & 02:18:28.275 & -05:00:05.74 & 0.871 & 25.45 & 0.10 & \\
SXDF021837.8-050424 & GMOS & 02:18:37.826 & -05:04:24.01 & 0.874 & 24.96 & 0.08 & \\
SXDF021833.8-045941 & GMOS & 02:18:33.895 & -04:59:41.01 & 0.876 & 24.92 & 0.08 & \\
SXDF021846.0-050204 & IMACS & 02:18:46.020 & -05:02:04.21 & 0.912 & 24.18 & 0.06 & \\
SXDF021816.7-050104 & GMOS & 02:18:16.770 & -05:01:04.75 & 0.918 & 24.45 & 0.06 & \\
SXDF021806.0-045959 & GMOS & 02:18:06.008 & -04:59:59.87 & 0.920 & 25.61 & 0.11 & \\
SXDF021735.2-050412 & GMOS & 02:17:35.299 & -05:04:12.10 & 0.941 & 24.62 & 0.06 & \\
SXDF021823.0-050041 & GMOS & 02:18:23.001 & -05:00:41.13 & 0.976 & 24.95 & 0.07 & \\
SXDF021810.6-050337 & GMOS & 02:18:10.688 & -05:03:37.92 & 1.033 & 24.87 & 0.08 & \\
SXDF021804.9-050250 & GMOS & 02:18:04.978 & -05:02:50.66 & 1.040 & 23.69 & 0.04 & \\
SXDF021837.4-050432 & GMOS & 02:18:37.452 & -05:04:32.76 & 1.042 & 24.41 & 0.05 & \\
SXDF021817.3-045922 & IMACS & 02:18:17.380 & -04:59:22.63 & 1.045 & 23.45 & 0.04 & X-ray, 24$\mu$m, 70$\mu$m source\\
SXDF021728.9-045943 & GMOS & 02:17:28.977 & -04:59:43.54 & 1.091 & 23.51 & 0.04 & \\
SXDF021821.7-050038 & GMOS & 02:18:21.777 & -05:00:38.25 & 1.092 & 23.70 & 0.04 & \\
SXDF021726.4-045955 & GMOS & 02:17:26.482 & -04:59:55.74 & 1.092 & 25.41 & 0.09 & \\
SXDF021820.1-050056 & GMOS & 02:18:20.128 & -05:00:56.43 & 1.093 & 24.21 & 0.06 & \\
SXDF021739.1-050328 & GMOS & 02:17:39.118 & -05:03:28.72 & 1.105 & 24.89 & 0.07 & \\
SXDF021831.4-050452 & GMOS & 02:18:31.407 & -05:04:52.88 & 1.133 & 24.29 & 0.06 & \\
SXDF021829.4-050518 & GMOS & 02:18:29.481 & -05:05:18.66 & 1.134 & 24.65 & 0.06 & \\
SXDF021737.8-050412 & GMOS & 02:17:37.898 & -05:04:12.71 & 1.135 & 23.43 & 0.03 & \\
SXDF021734.3-045859 & GMOS & 02:17:34.377 & -04:58:59.69 & 1.177 & 24.85 & 0.07 & 24$\mu$m, 850$\mu$m source\\
SXDF021734.8-050026 & GMOS & 02:17:34.870 & -05:00:26.92 & 1.178 & 24.41 & 0.06 & \\
SXDF021745.3-050057 & GMOS & 02:17:45.307 & -05:00:57.12 & 1.224 & 23.78 & 0.04 & \\
SXDF021823.2-050416 & GMOS & 02:18:23.275 & -05:04:16.77 & 1.267 & 24.34 & 0.05 & 24$\mu$m source\\
SXDF021807.9-045934 & GMOS & 02:18:07.937 & -04:59:34.78 & 1.275 & 23.05 & 0.03 & \\
SXDF021733.8-050038 & GMOS & 02:17:33.823 & -05:00:38.51 & 1.283 & 24.11 & 0.05 & \\
SXDF021843.5-045936 & IMACS & 02:18:43.550 & -04:59:36.84 & 1.289 & 23.62 & 0.04 & \\
SXDF021827.8-050139 & GMOS & 02:18:27.879 & -05:01:39.67 & 1.322 & 24.31 & 0.05 & \\
SXDF021822.6-050508 & GMOS & 02:18:22.620 & -05:05:08.36 & 1.390 & 23.58 & 0.03 & \\
SXDF021802.2-045940 & GMOS & 02:18:02.235 & -04:59:40.07 & 1.450 & 23.57 & 0.03 & \\
SXDF021734.2-050107 & GMOS & 02:17:34.262 & -05:01:07.74 & 1.457 & 24.44 & 0.06 & \\
SXDF021732.5-045915 & GMOS & 02:17:32.519 & -04:59:15.53 & 1.461 & 24.20 & 0.05 & \\
SXDF021810.2-050424 & GMOS & 02:18:10.284 & -05:04:24.51 & 1.475 & 24.50 & 0.06 & \\
\hline
\end{tabular}
\end{footnotesize}
\end{minipage}
\end{table*}

\begin{table*}
\addtocounter{table}{-1}
\centering
\begin{minipage}{146mm}
\begin{footnotesize}
\caption{\textit{- continued}}
\begin{tabular}{c c c c c c c c}
\hline\hline
ID catalogue & Instrument & R.A. & Dec. & ${z}_{spec}$ & $u_{AB}$\tablefootmark{a} & err $u$ & Comments\tablefootmark{b}\\
\hline
SXDF021838.2-045700 & IMACS & 02:18:38.260 & -04:57:00.58 & 1.097 & 21.57 & 0.01 & X-ray, 24$\mu$m source\\
SXDF021810.2-044809 & IMACS & 02:18:10.200 & -04:48:09.01 & 2.022 & 23.32 & 0.03 & X-ray, 24$\mu$m source\\
SXDF021802.5-050033 & GMOS & 02:18:02.516 & -05:00:33.06 & 2.068 & 23.49 & 0.03 & X-ray, 24$\mu$m$-$160$\mu$m source\\
SXDF021757.5-045059 & IMACS & 02:17:57.520 & -04:50:59.63 & 2.256 & 20.52 & 0.01 & 24$\mu$m source\\
SXDF021826.2-050505 & GMOS & 02:18:26.212 & -05:05:05.80 & 2.312 & 23.61 & 0.04 & 24$\mu$m source\\
SXDF021817.2-050259 & GMOS & 02:18:17.204 & -05:02:59.08 & 2.359 & 21.72 & 0.01 & 24$\mu$m source\\
SXDF021710.2-045857 & IMACS & 02:17:10.200 & -04:58:57.54 & 2.389 & 26.74 & 0.19 & 24$\mu$m source\\
SXDF021757.3-050808 & IMACS & 02:17:57.300 & -05:08:08.66 & 2.539 & 22.15 & 0.05 & X-ray, 24$\mu$m, 70$\mu$m source\\
SXDF021747.5-051323 & IMACS & 02:17:47.590 & -05:13:23.75 & 2.916 & 22.69 & 0.02 & X-ray, 24$\mu$m source\\
\hline
\end{tabular}
\tablefoot{\\
\tablefoottext{a}{Measured observer frame $u$-band photometry in AB magnitudes.}\\
\tablefoottext{b}{Detection of a counterpart in X-ray, FIR, and/or submillimeter.}
}
\end{footnotesize}
\end{minipage}
\end{table*}

\subsection{ Evolution of the SFR$-$stellar mass relation}\label{evmsfr}

A strong correlation has been found between SFR and stellar mass both in local star-forming galaxies and at higher redshifts (\citeauthor{Brinchman,Elbaz}). Moreover, this correlation evolves with redshift, in the sense that at a given mass high redshift star-forming galaxies have on average higher SFR values than local galaxies (\citeauthor{Daddi2}, \citetalias{Diaz}; among others). However, this relation has been only studied using active star-forming galaxies. Instead, Drory $\&$ Alvarez \citeyearpar{Drory} investigated the relation between SFR and stellar mass in the FORS Deep Field at 0 $<z<$ 5, including massive star-forming galaxies, and found a break mass at which the SFR deviates from a power law and decreases with redshift. We use the phrase \textit{low SFR system} to define galaxies with old stellar populations and low levels of star formation activity. The parametric function used by these authors has the following shape:

\begin{equation}
\text{SFR(}M_{\ast})=\text{SFR}_{0} \left(\frac{M_{\ast}}{M_{0}}\right)^{\alpha} \exp \left(-\frac{M_{\ast}}{M_{0}}\right).
\end{equation}

\defcitealias{Drory}{D08}
\defcitealias{Brinchman}{B04}

For $M_{0}$, the mass at which the SFR begins to deviate from the power law, we adopted the values given by the relation found by Drory $\&$ Alvarez (\citeyear{Drory}; hereafter D08, $M_{0}(z)=$ 2.66$\times$10$^{10}$ (1+$z$)$^{2.13}$). This parameter was fixed since the number of shortage points in our sample with masses above 10$^{10.5}$ $M_{\odot}$ were not statistically sufficient. Our galaxies were separated into three groups with redshift ranges of [0.0: 0.5], [0.5: 1.0], and [1.0: 1.5] for which the adopted log($M_{0}$) values were 10.63 (z=0.25), 10.94 (z=0.75), and 11.17 (z=1.25), respectively. For comparison, Brinchman et al. (\citeyear{Brinchman}; hereafter B04) obtained for local galaxies (z$\sim$0.1) log($M_{0})\sim$10.50, which is very similar to that calculated from \citetalias{Drory}.

\begin{figure*}
\begin{center}
\includegraphics[width=0.6\textwidth]{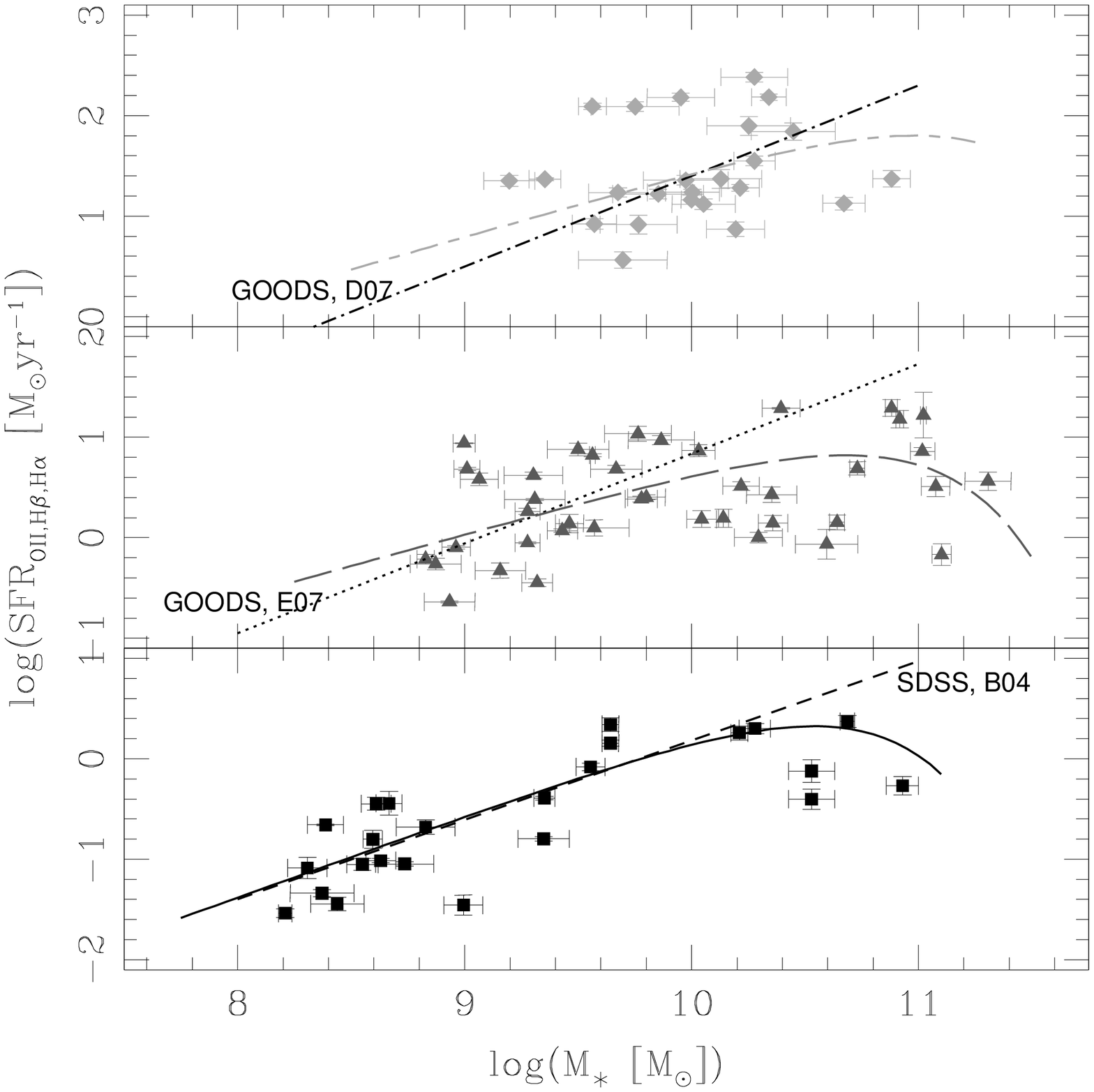}
\caption{SFR as function of total stellar mass. The bottom panel shows our low-redshift galaxies (black squares, 0 $<z<$ 0.5),  the medium panel shows our intermediate-redshift galaxies (gray triangles, 0.5 $<z<$ 1.0), while the top panel shows our high-redshift galaxies (light gray diamonds, 1.0 $<z<$ 1.5). The solid long-dashed, and short-long-dashed lines show the fitting function applied to each group. As a comparison, the trend observed in SDSS at redshift $z$ $<$ 0.2 (\citeauthor{Brinchman}) is indicated with a short-dashed line (bottom panel), the trend observed in GOODS by Elbaz et al. \citeyearpar{Elbaz} at redshift 0.8 $<$ $z$ $<$ 1.2 is shown with a dotted line (medium panel), while the trend found by Daddi et al. \citeyearpar{Daddi2} in GOODS at 1.4 $<$ $z$ $<$ 2.5 is indicated with a dash-dotted line (top panel).}\label{msfr}
\end{center}
\end{figure*}

\begin{table}
\centering
\begin{footnotesize}
\caption{\small{Fit parameters to the SFR$-$stellar mass relation.}}\label{sfrpar}
\begin{tabular}{c c c}
\hline\hline
$z$ & log(SFR$_{0}$ ($M_{\odot}$ yr$^{-1}$)) & $\alpha$ \\
\hline
0$-$0.5 & 0.75 $\pm$ 0.19 & 0.81 $\pm$ 0.10\\
0.5$-$1.0 & 1.22 $\pm$ 0.25 & 0.63 $\pm$ 0.19\\
1.0$-$1.5 & 2.20 $\pm$ 0.24 & 0.65 $\pm$ 0.20\\
\hline
\end{tabular}
\end{footnotesize}
\end{table}

\defcitealias{Elbaz}{E07}

Figure \ref{msfr} shows the SFR versus total stellar mass relation found in our data, split in three panels according to their redshift distribution as explained above. The limit of the SFR as a function of stellar mass for each redshift range is log(SFR)$=-$2.5 for log($M_{\ast})=$ 7.5 at low redshift, log(SFR)$=-$1 for log($M_{\ast})=$ 8.6 at intermediate redshift, and log(SFR)$=-$0.2 for log($M_{\ast})=$ 9.2 at high redshift. The parametric function applied to each group is represented by a solid line (low redshift, bottom panel), a long-dashed line (intermediate redshift, medium panel) and a short-long-dashed line (high redshift, top panel), respectively, with Table \ref{sfrpar} showing the best-fit values. As a comparison, the fits obtained for other star-forming galaxies in similar redshift intervals are also shown with a short-dashed line (SDSS galaxies at $z<$ 0.2, \citetalias{Brinchman}), a dotted line (GOODS galaxies at 0.8 $<z<$ 1.2; \citeauthor{Elbaz}, hereafter E07), and a dash-dotted line (GOODS galaxies at 1.4 $<z<$ 2.5, \citeauthor{Daddi2}), respectively. It can be observed, in agreement with \citetalias{Drory}, that galaxies with stellar masses higher than $M_{0}$ for each redshift group are located below the curve fit, hinting at a systematic offset of the star-forming sequence, which is consistent with the presence of low SFR systems. Furthermore, we found similar $\alpha$ slope values for the power-law component of the fitting function to those reported by \citetalias{Brinchman} (0.77) at low redshift,  to \citetalias{Elbaz} (0.90) and to Santini et al. (\citeyear{Santini}; 0.73) at intermediate redshift, as well as to those reported by \citetalias{Drory} (0.52$-$0.56) and by Santini et al. (\citeyear{Santini}; 0.65) at high redshift. 

\subsection{Evolution of the SSFR$-$stellar mass relation}

\begin{figure*}
\begin{center}
\includegraphics[width=0.6\textwidth]{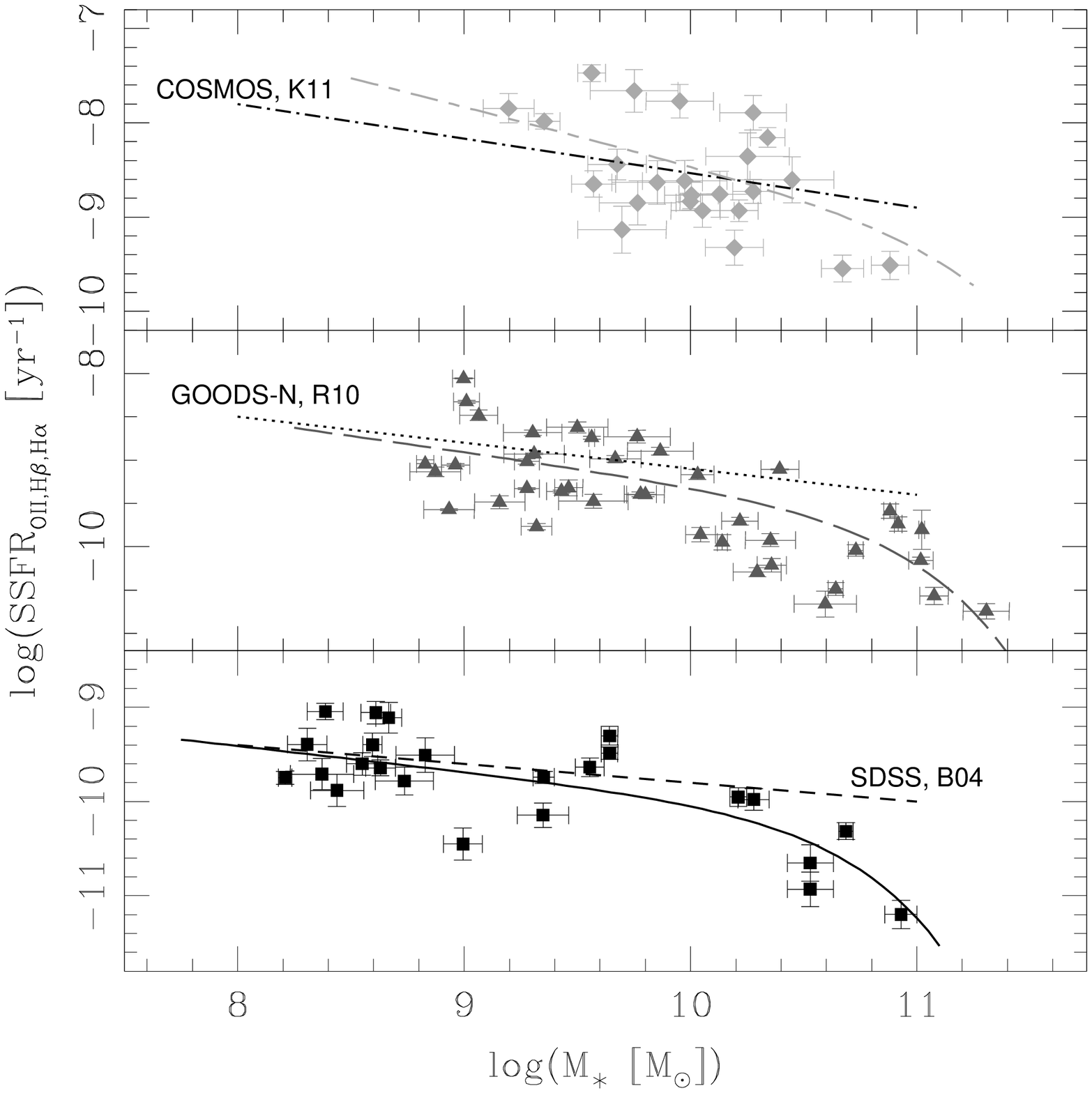}
\caption{SSFR as function of total stellar mass. As was shown in Figure \ref{msfr}, the bottom panel shows our low-redshift galaxies (black squares), the medium panel shows our intermediate-redshift galaxies (gray triangles), and the top panel shows our high-redshift galaxies (light gray diamonds). The solid, long-dashed, and short-long-dashed lines show the fitting function applied to each group. As a comparison, the trend observed in SDSS at redshift $z$ $<$ 0.2 (\citeauthor{Brinchman}) is shown with a short-dashed line (bottom panel), the trend observed in GOODS-N by Rodighiero et al. \citeyearpar{Rodighiero} at redshift 0.5 $<$ $z$ $<$ 1.0 is indicated with a dotted line (medium panel), while the trend found by Karim et al. \citeyearpar{Karim} in COSMOS at 1.6 $<$ $z$ $<$ 2.0 is shown with a dash-dotted line (top panel).}\label{mssfr}
\end{center}
\end{figure*}

Even though the SFR shows a positive correlation with stellar mass, if we look at the related quantity, the SFR per unit stellar mass, the specific SFR (SSFR) helps to interpret the degree of star formation activity of galaxies as a function of their mass. This is a standard practice in the literature (e.g., \citeauthor{Brinchman2,Bauer}) and we follow this practice  as well. We find there is an anticorrelation between the SSFR and stellar mass. Moreover, this anticorrelation also evolves with redshift in the sense that at a given stellar mass, high-redshift, star-forming galaxies have on average higher SSFR values than local galaxies (\citeauthor{Daddi2}, \citetalias{Diaz}, among others). In addition, in the local universe, Salim et al. \citeyearpar{Salim} found a clear sequence for star-forming galaxies with high and low levels of SFR in the SSFR versus stellar mass diagram. However, these authors did not determine the value of $M_{0}$, but rather the value of  $M_{\ast}$ from the Schechter function. Salim et al. \citeyearpar{Salim} indeed found AGNs as continuation of the star-forming sequence. However, Schiminovich et al. \citeyearpar{Schiminovich} reported that the bulk of AGNs are located in the zone where it is expected to find galaxies with residual star formation (SSFR $\sim-$11). Schiminovich et al. \citeyearpar{Schiminovich} also reported this trend occurring between the star-forming sequences and red sequences, finding that bulge-dominated galaxies near the characteristic transition mass ($\sim$ 10.5 $M_{\odot}$) could explain the growth rate of the non-star-forming population, taking into account reasonable assumptions regarding quenching timescales. 

Figure \ref{mssfr} shows the relation found between the SSFR and the total stellar mass for the sample galaxies, separated as shown previously. The limit SSFR as function of stellar mass for each redshift range is log(SSFR)$=-$13 for log($M_{\ast})=$ 7.5 at low redshift, log(SSFR)$=-$11.5 for log($M_{\ast})=$ 8.6 at intermediate redshift, and log(SSFR)$=-$10.7 for log($M_{\ast})=$ 9.2 at high redshift. We included in this plot the relation found by \citetalias{Brinchman} (dashed line) and those found by Rodighiero et al. (\citeyear{Rodighiero}; dotted line) and by Karim et al. (\citeyear{Karim}; dash-dotted line) in GOODS-N at redshift 0.5 $<$ $z$ $<$ 1.0, and also in COSMOS at 1.6 $<$ $z$ $<$ 2.0, respectively. As in the previous figure, we used the following modified version of the parametric function of \citetalias{Drory} to fit each group:

\begin{equation}
\text{SSFR(}M_{\ast})=\frac{\text{SFR}_{0}}{M_{\ast}} \left(\frac{M_{\ast}}{M_{0}}\right)^{\alpha} \exp \left(-\frac{M_{\ast}}{M_{0}}\right),
\end{equation}

where SSFR($M_{\ast}$)$=$SFR($M_{\ast}$)$/M_{\ast}$. After a little of algebraic manipulation, this equation can be expressed as follows:

\begin{equation}
\text{SSFR(}M_{\ast})=\text{SSFR}_{0} \left(\frac{M_{\ast}}{M_{0}}\right)^{\beta} \exp \left(-\frac{M_{\ast}}{M_{0}}\right),
\end{equation}

where SSFR$_{0}=$ SFR$_{0}$/$M_{0}$, and $\beta=$ $\alpha-$1.

\begin{table}
\centering
\begin{footnotesize}
\caption{\small{Fit parameters to the SSFR$-$stellar mass relation.}}\label{ssfrpar}
\begin{tabular}{c c c}
\hline\hline
$z$ & log(SSFR$_{0}$ (yr$^{-1}$)) & $\beta$ \\
\hline
0$-$0.5 & -10.12 $\pm$ 0.21 & $-$0.27 $\pm$ 0.11\\
0.5$-$1.0 & -9.60 $\pm$ 0.29 & $-$0.36 $\pm$ 0.22\\
1.0$-$1.5 & -9.16 $\pm$ 0.32 & $-$0.61 $\pm$ 0.32\\
\hline
\end{tabular} 
\end{footnotesize}
\end{table}

The same $M_{0}$ values used in the previous section, obtained from the fits reported by \citetalias{Drory}, were adopted. In Figure \ref{mssfr}, the solid line (bottom panel) represents the fitting function applied to the low-redshift group, the long-dashed line (medium panel) is the fit applied to the intermediate-redshift group and the short-long-dashed line (top panel) is the fit for the high-redshift group; Table \ref{ssfrpar} shows the values found for these fits. It can be observed that our galaxies qualitatively followed the trend given by the assumed value of $M_{0}$, with higher values occurring at higher redshifts. This is also qualitatively consistent with the results derived by Hopkins et al. \citeyearpar{Hopkins} from stellar mass functions (SMFs), in which early and late-type galaxies are separated by their SSFR values. Furthermore, it can be seen that the slope $\beta$ calculated for the low-redshift group is similar to that found by \citetalias{Brinchman} ($-$0.23 at $z<$ 0.2), while for the intermediate- and high-redshift groups $\beta$ is comparable with those found by Rodighiero et al. (\citeyear{Rodighiero}; $-$0.28 at 0.5 $<z<$ 1.0) and Karim et al. (\citeyear{Karim}; $-$0.30 at 1.2 $<z<$ 1.6). As in the previous figure, our data is consistent with the presence of low SFR systems.

\subsection{The evolution of the SSFR$-$color relation}

 Bauer et al. \citeyearpar{Bauer2} and Twite et al. \citeyearpar{Twite} reported that bluer galaxies, corresponding to galaxies of younger ages from \textbf{our} SED fitting analysis (Section \ref{phyprop}), had larger SSFR values. Moreover, \citetalias{Diaz} also showed that for a particular $(u-B)$ color, the high-redshift galaxies had on average higher SSFR values. Similarly, based on the correlation found between morphology and broadband color (\citeauthor{Driver}), Pannella et al \citeyearpar{Pannella} also reported that at higher redshift both early-type and late-type galaxies had on average higher SSFR values than their counterparts at low redshift. In this section, we explored the correlation between SSFR and color, but using a bigger sample than that utilized in \citetalias{Diaz}.  

\begin{figure}
\begin{center}
\includegraphics[width=0.49\textwidth]{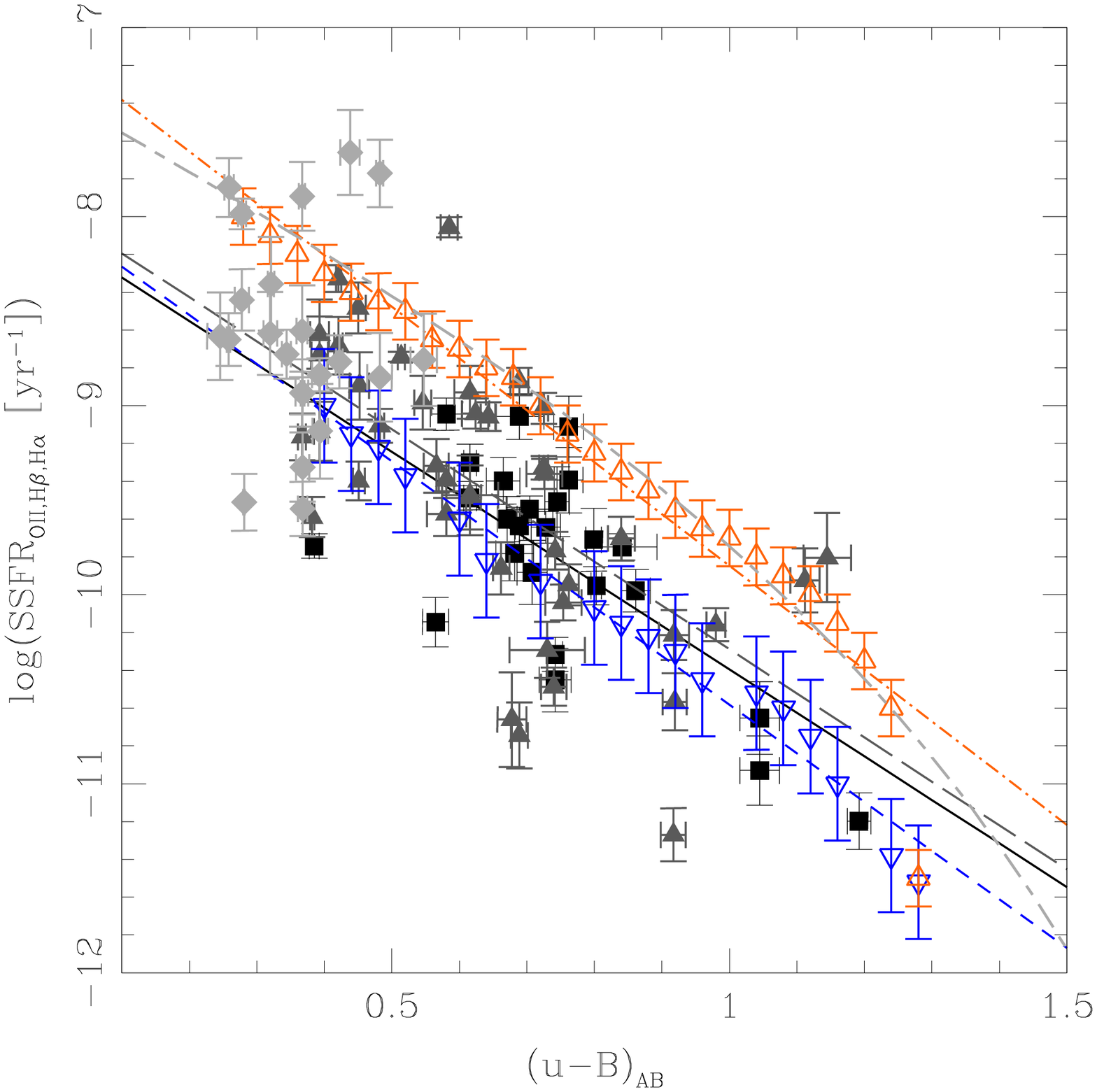}
\caption{SSFR as function of rest-frame color $(u-B)_{AB}$. As in Figure \ref{msfr}, black squares represent our galaxies  at 0 $<z<$ 0.5, gray triangles our galaxies at 0.5 $<z<$ 1.0, and light gray diamonds our galaxies at 1.0 $<z<$ 1.5. The solid and long-dashed lines show the linear fits used with each group. For comparison, we have included the trends found by using the Cooper et al. \citeyearpar{Cooper} data, with the blue empty inverted triangles showing galaxies at redshift 0.05 $<$ $z$ $<$ 0.1 from SDSS, and the red empty triangles indicating the DEEP2 sample at 0.75 $<$ $z$ $<$ 1.05. The blue dashed line and red dash-dotted lines show the linear fits estimated for the Cooper et al. \citeyearpar{Cooper} galaxies, while the gray short-long-dashed line is, as an example, the parametric curve proposed to fit these data.}\label{colssfr}
\end{center}
\end{figure}

\defcitealias{Cooper}{C08}

Figure \ref{colssfr} shows the SSFR as a function of rest-frame color $(u-B)_{AB}$, where we have discriminated our galaxies as in the previous sections and included the data used by Cooper et al. \citeyearpar{Cooper};  the blue empty inverted triangles represent SDSS galaxies at 0.05 $<$ $z$ $<$ 0.10 and the red empty triangles represent DEEP2 galaxies at 0.75 $<$ $z$ $<$ 1.05\footnote{Symbols represent the global trend observed, on average, for galaxies shown in figure 13 from Cooper et al. \citeyearpar{Cooper}.}. We obtained linear fit slopes of $-$2.38$\pm$0.48 and $-$2.21$\pm$0.34 for the low-redshift and intermediate-redshift groups, respectively.  These are different from those found by \citetalias{Diaz} ($\sim-$1.3$\pm$0.6) and from those estimated for the Bauer et al. (\citeyear{Bauer2}, $\sim-$1.2) and Twite et al. (\citeyear{Twite}; $\sim-$1.3) samples. However, our slopes are similar to those calculated for the Cooper et al. (\citeyear{Cooper}; hereafter C08) data ($\sim$ $-$2.57 for SDSS, and $-$2.74 for DEEP2 samples). It can also be observed a shift in the zero point of each subsample that indicates, on average, different levels of star formation in the redshift range probed. The differences between our data and \citetalias{Cooper} could be atributed to different stellar mass values for the bulk of the samples. In fact, the \citetalias{Cooper} data at high-redshift peak nearly at 10$^{9.6}$ M$_{\odot}$, while our data peaks at 10$^{10.25}$ M$_{\odot}$. Furthermore, from the \citetalias{Cooper} data, it can be seen that signs of a possible break color in the star-forming sequence might appear at the reddest colors.  

Based on the evolutionary sequence found in Section \ref{evmsfr}, we hypothesize that just as there is a break mass in the SFR-stellar mass relation, there might also exist a break color in the SSFR-color relation, thus revealing an evolutionary sequence between star-forming and low SFR systems. The following parametric function might describe this trend:

\begin{equation}
SSFR(u-B)= SSFR_{0} \left( \frac{u-B}{(u-B)_{0}} \right)^{\beta} \exp \left(-\frac{(u-B)}{(u-B)_{0}} \right),
\end{equation}

where $\beta$ is the slope of the power-law form, while $(u-B)_{0}$ is the color that may separate the location of low-SSFR early-type galaxies that are old and red from high-SSFR star-forming galaxies that are young and blue. The short-long-dashed curve shown in Figure \ref{colssfr} is an example of this  proposed parameterization. Evidence of this pattern has been previously observed by Salim et al. \citeyearpar{Salim2}. However, these authors used the index color $(g-r)$ instead, and did not fit a curve to their plot (figure 21, \citeauthor{Salim2}). No curve was fitted by \citetalias{Cooper} to their data either.

\begin{figure}
\begin{center}
\includegraphics[width=0.40\textwidth]{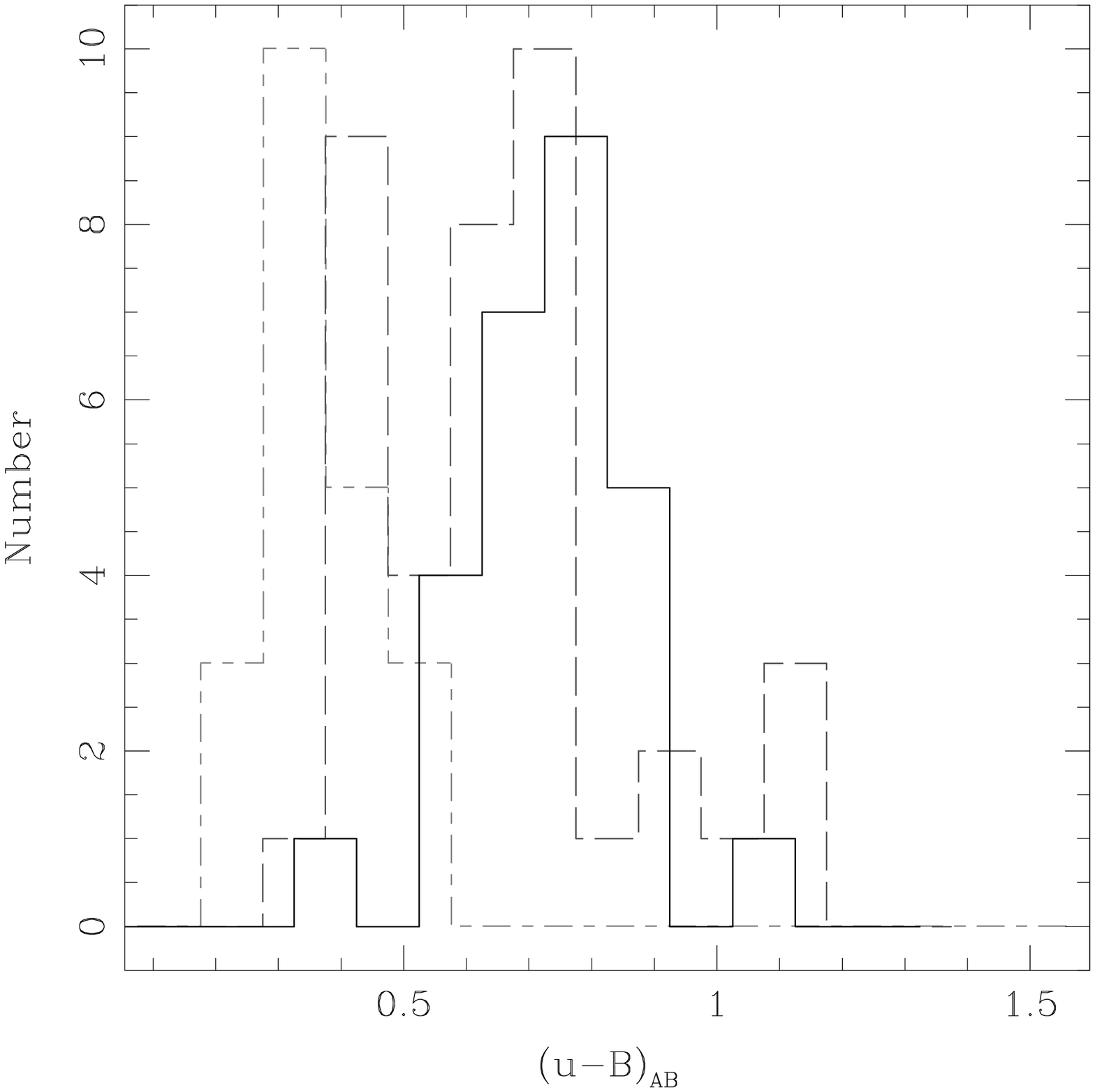}
\caption{Rest-frame color $(u-B)_{AB}$ distribution of the galaxies separated according to their redshift. As in Figure \ref{colssfr}, the solid line represents our galaxies at 0 $<z<$ 0.5, the long-dashed line our galaxies at 0.5 $<z<$ 1.0, while the short-long-dashed line shows our galaxies at 1.0 $<z<$ 1.5.}\label{histcol}
\end{center}
\end{figure}

It was not possible to calculate a slope for the high-redshift group because its color domain was very small. Figure \ref{histcol} shows a clearer view of this fact with the $(u-B)_{AB}$ color distribution of each galaxy group discriminated according to its redshift domain. It can be seen that the low-redshift group (0 $<z<$ 0.5, solid line) is mainly composed of galaxies with colors in the range [0.5; 1.0], while the intermediate-redshift group (0.5 $<z<$ 1.0, long-dashed line) is the only group\ that shows a more extended color range [0.3; 1.2]. In contrast, the high-redshift group (1.0 $<z<$ 1.5, short-long-dashed line) has a very small color domain [0.2; 0.6] and a linear fit is not used. This effect could in fact be due to the limiting SFR being higher at higher redshift. A similar interpretation can also be derived from Figure \ref{ssfrev}, assuming that the SSFR correlates fairly well with color. In this plot, the redshift interval was also divided in groups and showed that the intermediate-redshift sample had a more scattered range in SSFR values compared to the other groups.

\begin{figure}
\begin{center}
\includegraphics[width=0.49\textwidth]{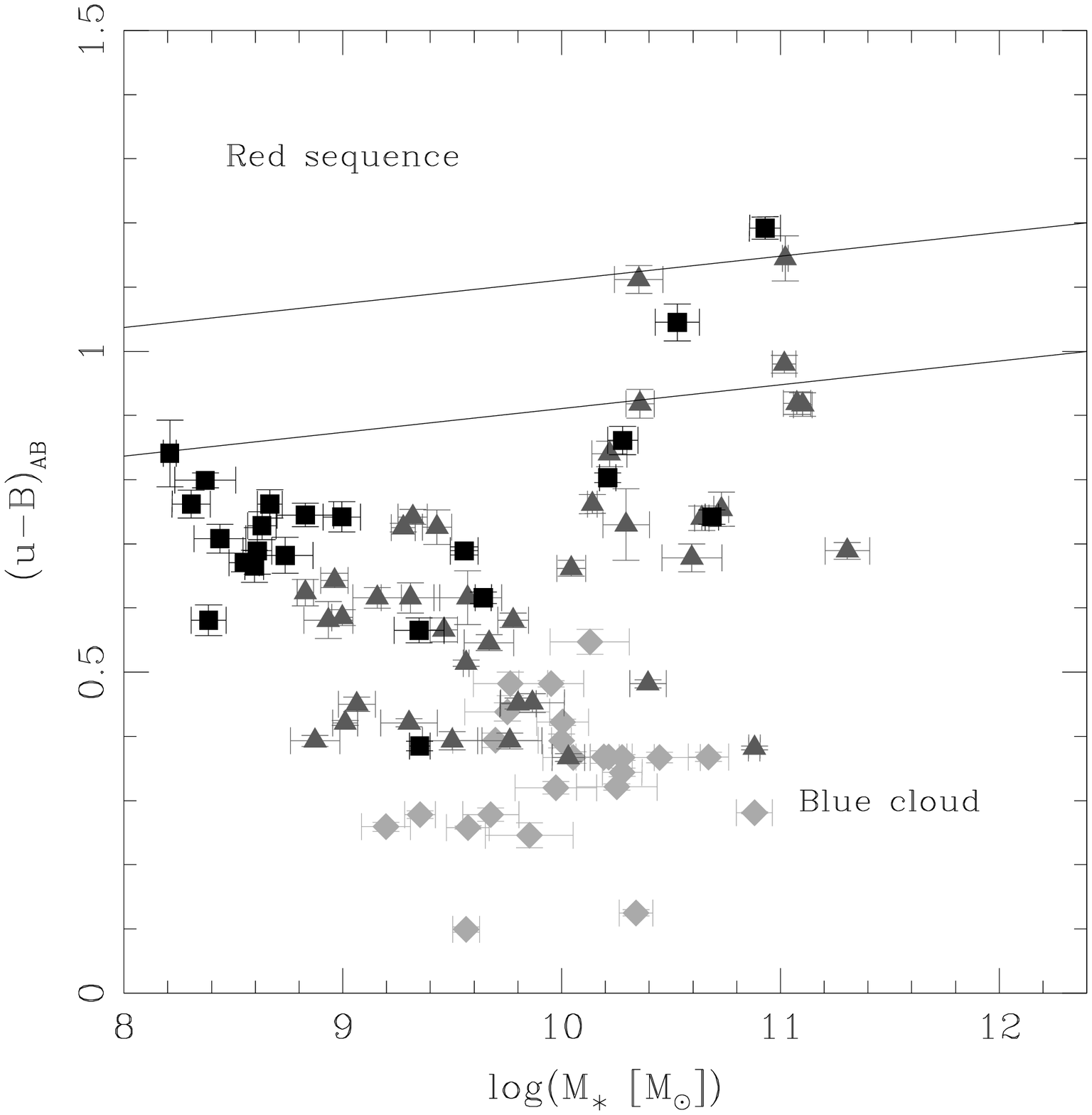}
\caption{Rest-frame color $(u-B)_{AB}$ versus total stellar mass. Symbols represent the same redshift domain as shown in Figure \ref{msfr}. The solid lines separate the regions where passive (upper), low SFR systems (middle), and star-forming (lower) galaxies could be located.}\label{colmass}
\end{center}
\end{figure}

A useful way to interpret the results found in the SFR$-$stellar mass, SSFR$-$stellar mass, and SSFR$-$color diagrams can be deduced from the rest-frame $(u-B)_{AB}$ color versus total stellar mass diagram. This is plotted in Figure \ref{colmass} , where the galaxies  are grouped in the same way as shown in the previous diagrams. It can be observed that the low-redshift and intermediate-redshift samples revealed a sequence along the stellar mass domain until $M_{\ast}\sim$ 10$^{10.4}$ $M_{\odot}$. Then, the galaxies with $M_{\ast}\gtrsim$ 10$^{10.4}$ $M_{\odot}$ and $(u-B) \gtrsim$ 0.9  leave the blue cloud to migrate to the transition region where the low SFR systems may be located. However, the high-redshift sample does not follow this trend. It can be observed that these galaxies are mainly blue  and show no signs of a transition toward redder colors. This fact opens a question about the observed trend in Fig. \ref{msfr}, where the break is again not noticeable for the high-redshift galaxies: Would be possible that the break for high-redshift galaxies is at higher stellar masses? The $M_{0}$ value found by \citetalias{Drory}, Bundy et al. \citeyearpar{Bundy} and Hopkins et al. \citeyearpar{Hopkins} was higher than 10$^{11}$ $M_{\odot}$. Larger volumes should be explored to answer this question. 
\section{Summary}\label{Sum}

Through their emission lines, we investigated  the physical properties of a sample of 64 Balmer break galaxies at redshift 0 $<z<$ 1.5 which showed signs of star formation. Also included in the study were eight Balmer break galaxies with broad emission lines  to explore the AGN contribution up to $z=$ 3. Most of the sample galaxies (92$\%$) have stellar masses ranging from 10$^{8.75}-$10$^{11.25}M_{\odot}$. This sample forms part of a spectroscopic survey pilot project in the SXDF out to z $\sim$ 3.0 with the purpose of investigating the decline of star formation in massive galaxies.  

Of the eight AGNs with broad emission lines, four of these had X-ray counterparts. In addition, there were four narrow emission line galaxies with X-ray counterparts. Of these four galaxies, two showed the MgII $\lambda$2796, 2803 line in emission, suggesting they have AGN activity, while  MgII was outside spectral coverage for the remaining two
galaxies.

According to the MIR diagnostic diagrams (\citeauthor{Lacy,Stern}), from 49 galaxies with available photometry in all the IRAC/\textit{Spitzer} bands, 16 galaxies (33$\%$) could host an AGN; seven of these objects were the AGNs with broad emission lines, and one was an X-ray source. The remaining eight sources could have hosted an obscured AGN or the nuclear star formation may have masked the optical AGN signatures.

The SFR was calculated for each galaxy, using: (1) [OII] $\lambda$3727, H$\beta,$ or H$\alpha$ luminosities; and (2) SED fitting with an exponentially declining SFH. Both methods gave very similar results with SFR values ranging between 0.01 and 300 $M_{\odot}$ yr$^{-1}$. The Díaz Tello et al. \citeyearpar{Diaz} sample galaxies were also included to improve statistically the results  in the analysis related to the SFR study.

The evolution in redshift of the SFR and SSFR was investigated, and it was found that both SFR and SSFR increased in redshift. These results were in agreement with those reported by Silverman et al. \citeyearpar{Silverman} and Juneau et al. \citeyearpar{Juneau}. The SFR$-$stellar mass and SSFR$-$stellar mass relations were also studied with the known correlation and anticorrelation found to exist between SFR and SSFR with the stellar mass, respectively. Signs of a break mass in the star-forming sequence were encountered, revealing the presence of low SFR systems;  this pattern was also reported by Salim et al \citeyearpar{Salim}, Schiminovich et al. \citeyearpar{Schiminovich}, and Drory $\&$ Alvarez \citeyearpar{Drory}.

The evolution with redshift of the SFR$-$stellar mass and SSFR$-$stellar mass relations was studied, which revealed that at a given mass high-redshift galaxies had on average higher SFR and SSFR values than local galaxies. A similar evolution was also previously reported in the evolution of the transition mass ($M_{tr}$) of the stellar mass function when it was separated in star-forming and passive galaxies (\citeauthor{Bundy,Hopkins}). Finally, we investigated the evolution with redshift of the SSFR$-(u-B)$ color relation, and it was found that at a given color high-redshift galaxies had on average higher SSFR values than local galaxies. Evidence of this trend was also reported by Cooper et al. \citeyearpar{Cooper} and Díaz Tello et al. \citeyearpar{Diaz}.

In summary, even though our extended sample was small, our results confirm and give more weight to the relations summarized above, but will benefit from larger number statistics.

\section*{Acknowledgments}

This research was partially supported by Consejo Nacional de Investigaciones Científicas E Técnicas (CONICET, Argentina), the Japan Society for the Promotion of Science (JSPS), and Dirección General de Asuntos del Personal Académico (DGAPA-UNAM, México). NP was supported by Fondecyt Regular and BASAL CATA. We are specially grateful to CONICET and DGAPA for the fellowship awarded to JD. We thank H. Muriel for his advice, which improved the analysis of this research, and native English speaker Paul Hobson for his revision of the manuscript. We also thank  the anonymous referee, who gave a detailed and constructive report that helped to improve the manuscript. 

This study is based on data principally collected using the Subaru telescope operated by the National Astronomical Observatory of Japan (NAOJ); the United Kingdom Infrared Telescope (UKIRT) operated by the Joint Astronomy Centre (JAC) on behalf of the Science and Technology Facilities Council of the U. K.; the \textit{Spitzer Space Telescope} operated by the Jet Propulsion Laboratory, California Institute of Technology (Caltech) under contract with NASA; the Blanco telescope of Cerro Tololo Inter-American Observatory (CTIO) operated by the Association of Universities Research in Astronomy (AURA), under cooperative agreement with the National Science Foundation as part of the National Optical Astronomy Observatories (NOAO); the Magellan telescope operated by the Carnegie Institution of Washington, Harvard University, Massachusetts Institute of Technology (MIT), University of Michigan, and University of Arizona; and the Gemini telescope operated by AURA, Inc., under a cooperative agreement with the NSF on behalf of the Gemini partnership: the National Science Foundation (United States), the Science and Technology Facilities Council (United Kingdom), the National Research Council (Canada), CONICYT (Chile), the Australian Research Council (Australia), Ministerio da Ciência e Tecnologia (Brazil), and Ministerio de Ciencia, Tecnología e Innovación Productiva (Argentina).

\begin{small}

\end{small}

\end{document}